\documentclass[a4paper,14pt,headings=chapterwithprefixline]{extreport}

\usepackage{iftex}
\newif\ifxetexorluatex   
\ifXeTeX
    \xetexorluatextrue
\else
    \ifLuaTeX
        \xetexorluatextrue
    \else
        \xetexorluatexfalse
    \fi
\fi

\usepackage{pdflscape}                              
\usepackage{geometry}                               

\usepackage{amsthm,amsfonts,amsmath,amssymb,amscd}  
\usepackage{mathtools}                              

\newlength{\curtextsize}
\newlength{\bigtextsize}

\ifxetexorluatex
    \usepackage{polyglossia}                        
\else
    \RequirePDFTeX                                  
    \usepackage{cmap}                               
    \usepackage[T2A]{fontenc}                       
    \usepackage[utf8]{inputenc}                     
    \usepackage[ukrainian,main=english]{babel}            
    \IfFileExists{pscyr.sty}{\usepackage{pscyr}}{}  
\fi

\usepackage[normalem]{ulem}  

\usepackage{indentfirst}                            

\usepackage[dvipsnames,usenames]{color}
\usepackage{colortbl}

\usepackage{longtable}                              
\usepackage{multirow,makecell,array}                
\usepackage{booktabs}                               

\usepackage{soulutf8}                               
\usepackage{icomma}                                 

\usepackage{enumitem} 

\usepackage{graphicx}                               

\usepackage{enumitem}

\usepackage{caption}                                
\usepackage{subcaption}                             

\usepackage[onehalfspacing]{setspace}               

\usepackage[figure,table]{totalcount}               
\usepackage{totcount}                               
\usepackage{totpages}                               

\usepackage{hyperref}

\ifxetexorluatex
    \usepackage{cleveref}                           
\else
    \usepackage[ukrainian]{cleveref}                  
\fi
\creflabelformat{equation}{#2#1#3}                  

\usepackage{fancyhdr}

\usepackage{calc}               

\usepackage{titlesec}           

\usepackage{tocloft}

\usepackage{chngcntr}           

\usepackage{tabularx,tabulary}  

\usepackage{fancyvrb}
\usepackage{listings}

\usepackage{floatrow}

\usepackage{textcomp}

\usepackage{ifthen}                 
\newcounter{intvl}
\newcounter{otstup}
\newcounter{contnumeq}
\newcounter{contnumfig}
\newcounter{contnumtab}
\newcounter{pgnum}
\newcounter{bibliosel}
\newcounter{chapstyle}
\newcounter{headingdelim}
\newcounter{headingalign}
\newcounter{headingsize}
\newcounter{tabcap}
\newcounter{tablaba}
\newcounter{tabtita}

\definecolor{linkcolor}{rgb}{0,0,0} 
\definecolor{citecolor}{rgb}{0,0,0} 
\definecolor{urlcolor}{rgb}{0,0,0} 

\renewcommand{\chaptertitlename}{CHAPTER}
\renewcommand{\appendixname}{APPENDIX}
\newcommand{\authorbibtitle}{List of the candidate's publications related to the dissertation}
\newcommand{\fullbibtitle}{BIBLIOGRAPHY} 

\newcommand{\thesisAuthor}             
{%
    \texorpdfstring{
        \todo{Багрова Ольга Миколаївна}
    }{%
        Багрова, Ольга Миколаївна
    }%
}
\newcommand{\thesisUdk}                
{\todo{xxx.xxx}}
\newcommand{\thesisTitle}              
{\texorpdfstring{\todo{\MakeUppercase{Electromechanical phenomena in normal and superconducting nanostructures based on a movable quantum dot}}}{Electromechanical phenomena in normal and superconducting nanostructures based on a movable quantum dot}}
\newcommand{\thesisSpecialtyNumber}    
{\texorpdfstring{\todo{XX.XX.XX}}{XX.XX.XX}}
\newcommand{\thesisSpecialtyTitle}     
{\texorpdfstring{\todo{Название специальности}}{Название специальности}}
\newcommand{\thesisDegree}             
{\todo{доктора філософії}}
\newcommand{\thesisCity}               
{\todo{Харків}}
\newcommand{\thesisYear}               
{\todo{2022}}
\newcommand{\thesisOrganization}       
{\todo{Національна академія наук України \\ Фізико-технічний інститут низьких температур \\ імені Б.І.~Вєркіна}}

\newcommand{\thesisInOrganization}       
{\todo{учреждении, в~котором выполнялась данная диссертационная работа}}

\newcommand{\supervisorFio}            
{\todo{Фамилия Имя Отчество}}
\newcommand{\supervisorRegalia}        
{\todo{уч. степень, уч. звание}}

\newcommand{\opponentOneFio}           
{\todo{Фамилия Имя Отчество}}
\newcommand{\opponentOneRegalia}       
{\todo{доктор физико-математических наук, профессор}}
\newcommand{\opponentOneJobPlace}      
{\todo{Не очень длинное название для места работы}}
\newcommand{\opponentOneJobPost}       
{\todo{старший научный сотрудник}}

\newcommand{\opponentTwoFio}           
{\todo{Фамилия Имя Отчество}}
\newcommand{\opponentTwoRegalia}       
{\todo{кандидат физико-математических наук}}
\newcommand{\opponentTwoJobPlace}      
{\todo{Основное место работы c длинным длинным длинным длинным названием}}
\newcommand{\opponentTwoJobPost}       
{\todo{старший научный сотрудник}}

\newcommand{\leadingOrganizationTitle} 
{\todo{Федеральное государственное бюджетное образовательное учреждение высшего профессионального образования с~длинным длинным длинным длинным названием}}

\newcommand{\defenseDate}              
{\todo{DD mmmmmmmm YYYY~г.~в~XX часов}}
\newcommand{\defenseCouncilNumber}     
{\todo{NN}}
\newcommand{\defenseCouncilTitle}      
{\todo{Название учреждения}}
\newcommand{\defenseCouncilAddress}    
{\todo{Адрес}}

\newcommand{\defenseSecretaryFio}      
{\todo{Фамилия Имя Отчество}}
\newcommand{\defenseSecretaryRegalia}  
{\todo{д-р~физ.-мат. наук}}            

\newcommand{\synopsisLibrary}          
{\todo{Название библиотеки}}
\newcommand{\synopsisDate}             
{\todo{DD mmmmmmmm YYYY года}}

\newcommand{\keywords}
{}      
\ifxetexorluatex
    \setmainlanguage[babelshorthands=true]{english}  
    \setotherlanguage{ukrainian}                       
    \ifXeTeX
        \defaultfontfeatures{Ligatures=TeX,Mapping=tex-text}
    \else
        \defaultfontfeatures{Ligatures=TeX}
    \fi
    \newfontfamily\cyrillicfont{Times New Roman}
    \newfontfamily\cyrillicfontsf{Arial}
    \newfontfamily\cyrillicfonttt{Courier New}
\else
    \IfFileExists{pscyr.sty}{}{}
\fi

\DeclareCaptionLabelSeparator*{emdash}{~ }             
\ifthenelse{\equal{\thetabcap}{0}}{%
    \newcommand{\tabcapalign}{\raggedright}  
}

\ifthenelse{\equal{\thetablaba}{0} \AND \equal{\thetabcap}{1}}{%
    \newcommand{\tabcapalign}{\raggedright}  
}

\ifthenelse{\equal{\thetablaba}{1} \AND \equal{\thetabcap}{1}}{%
    \newcommand{\tabcapalign}{\centering}    
}

\ifthenelse{\equal{\thetablaba}{2} \AND \equal{\thetabcap}{1}}{%
    \newcommand{\tabcapalign}{\raggedleft}   
}

\ifthenelse{\equal{\thetabtita}{0} \AND \equal{\thetabcap}{1}}{%
    \newcommand{\tabtitalign}{\raggedright}  
}

\ifthenelse{\equal{\thetabtita}{1} \AND \equal{\thetabcap}{1}}{%
    \newcommand{\tabtitalign}{\centering}    
}

\ifthenelse{\equal{\thetabtita}{2} \AND \equal{\thetabcap}{1}}{%
    \newcommand{\tabtitalign}{\raggedleft}   
}

\DeclareCaptionFormat{tablenocaption}{\tabcapalign #1\strut}        
\ifthenelse{\equal{\thetabcap}{0}}{%
    \DeclareCaptionFormat{tablecaption}{\tabcapalign #1#2#3}
    \captionsetup[table]{labelsep=emdash}                       
}{%
    \DeclareCaptionFormat{tablecaption}{\tabcapalign #1#2\par
        \tabtitalign{#3}}                                       
    \captionsetup[table]{labelsep=space}                        
}
\DeclareCaptionLabelFormat{continued}{Продолжение таблицы~#2}

\AddEnumerateCounter{\Asbuk}{\@Asbuk}{\CYRM} 
\AddEnumerateCounter{\asbuk}{\@asbuk}{\cyrm}

\ifLuaTeX
    \hypersetup{
        unicode,                
    }
\fi

\hypersetup{
    linktocpage=true,           
    plainpages=false,           
    colorlinks,                 
    linkcolor={linkcolor},      
    citecolor={citecolor},      
    urlcolor={urlcolor},        
    pdftitle={\thesisTitle},    
    pdfauthor={\thesisAuthor},  
    pdfsubject={\thesisSpecialtyNumber\ \thesisSpecialtyTitle},      
    pdfkeywords={\keywords},    
    pdflang={english},
}

\DeclareRobustCommand{\todo}{\textcolor{black}}       
\renewcommand{\labelitemi}{\normalfont\bfseries{--}} 
\setlist{nosep,
    labelindent=\parindent,leftmargin=*
}
\graphicspath{{images/}{Dissertation/images/}}         

\newlength{\otstuplen}
\ifthenelse{\equal{\theheadingalign}{0}}{
    \newcommand{\hdngalign}{\filcenter}                
}{%
    \newcommand{\hdngalign}{\filright}                 
} 

\cftsetrmarg{2.55em plus1fil}                       

\ifthenelse{\theheadingdelim > 0}{%
    \renewcommand\cftchapaftersnum{\quad }   
}{%
\renewcommand\cftchapaftersnum{.\ }     
}
\ifthenelse{\theheadingdelim > 1}{%
    \renewcommand\cftsecaftersnum{.\ }    
    \renewcommand\cftsubsecaftersnum{.\ } 
}{%
\renewcommand\cftsecaftersnum{\quad}      
\renewcommand\cftsubsecaftersnum{\quad}   
}

\ifthenelse{\equal{\thepgnum}{1}}{%
    \addtocontents{toc}{~\hfill{P.}\par}
}

\ifnum\curtextsize>\bigtextsize     
\else
\fi

\makeatletter
\let\ps@plain\ps@fancy              
\makeatother
\titleformat{\chapter}[display]                                
        {\filcenter\fontsize{14pt}{16pt}\selectfont\bfseries}
        {\chaptertitlename~\thechapter\cftchapaftersnum}                       
       {1em}
       {} 

\titleformat{\section}[block]                                
        {\hdngalign\fontsize{14pt}{16pt}\selectfont\bfseries}%
        {\thesection\cftsecaftersnum}                        
        {0em}
        {}%

\titleformat{\subsection}[block]                             
        {\hdngalign\fontsize{14pt}{16pt}\selectfont\bfseries}%
        {\thesubsection\cftsubsecaftersnum}                  
        {0em}
        {}%

\addtocontents{toc}{\def\protect\cftchappresnum{\chaptertitlename{} }%
\setlength{\cftchapnumwidth}{\widthof{\cftchapfont\chaptername~WW\cftchapaftersnum}}%
}

\beforetitleunit=\curtextsize
\aftertitleunit=\curtextsize

\titlespacing{\chapter}{\theotstup\parindent}{-1.7em}{*\theintvl}       
\titlespacing{\section}{\theotstup\parindent}{*\theintvl}{*\theintvl}
\titlespacing{\subsection}{\theotstup\parindent}{*\theintvl}{*\theintvl}
\titlespacing{\subsubsection}{\theotstup\parindent}{*\theintvl}{*\theintvl}

\ifthenelse{\equal{\theheadingsize}{1}}{
    \setlength{\cftbeforetoctitleskip}{-1.2\curtextsize}        
    \setlength{\cftbeforeloftitleskip}{-1.4\curtextsize}        
    \setlength{\cftbeforelottitleskip}{-1.4\curtextsize}        
    \sectionformat{\chapter}{
        format=\hdngalign\Large\bfseries, 
        top-=0.4em,                       
    }
    \sectionformat{\section}{
        format=\hdngalign\large\bfseries, 
    }
}

\ifthenelse{\equal{\theheadingsize}{1}\AND \curtextsize < \bigtextsize}{
    \sectionformat{\chapter}{
        top-=0.2em, 
    }
}

\ifthenelse{\equal{\thecontnumeq}{1}}{%
    \counterwithout{equation}{chapter} 
}
\ifthenelse{\equal{\thecontnumfig}{1}}{%
    \counterwithout{figure}{chapter}   
}
\ifthenelse{\equal{\thecontnumtab}{1}}{%
    \counterwithout{table}{chapter}    
}

\makeatletter
\def\formbytotal#1#2#3#4#5{%
    \newcount\@c
    \@c\totvalue{#1}\relax
    \newcount\@last
    \newcount\@pnul
    \@last\@c\relax
    \divide\@last 10
    \@pnul\@last\relax
    \divide\@pnul 10
    \multiply\@pnul-10
    \advance\@pnul\@last
    \multiply\@last-10
    \advance\@last\@c
    \total{#1}~#2%
    \ifnum\@pnul=1#5\else%
    \ifcase\@last#5\or#3\or#4\or#4\or#4\else#5\fi
    \fi
}
\makeatother

\AtBeginDocument{
    \regtotcounter{totalcount@figure}
    \regtotcounter{totalcount@table}       
    \regtotcounter{TotPages}               
}

\def\zz{\ifx\[$\else\aftergroup\zzz\fi}
\def\zzz{\setbox0\lastbox
\dimen0\dimexpr\extrarowheight + \ht0-\dp0\relax
\setbox0\hbox{\raise-.5\dimen0\box0}%
\ht0=\dimexpr\ht0+\extrarowheight\relax
\dp0=\dimexpr\dp0+\extrarowheight\relax 
\box0
}

\newlength{\twless}
\newlength{\lmarg}
    

\DefineVerbatimEnvironment
{Verb}{Verbatim}
{fontsize=\fontsize{12pt}{14pt}\selectfont}

\RawFloats[figure,table]            

\DeclareNewFloatType{ListingEnv}{
    placement=htb,
    within=chapter,
    fileext=lol,
    name=Листинг,
}

\ifthenelse{\equal{\thebibliosel}{0}}{%

\usepackage{natbib}                                   
\usepackage{multibib}

\bibliographystyle{apsrev4-1}

\newtotcounter{citenum}
\def\oldcite{}
\let\oldcite=\bibcite
\def\bibcite{\stepcounter{citenum}\oldcite}
}{

\usepackage[%
backend=biber,
bibencoding=utf8,
sorting=none,
style=numeric-comp,
language=autobib,
autolang=other,
clearlang=true,
defernumbers=true,
sortcites=true,
url=false,
]{biblatex}

\renewbibmacro{in:}{} 

\DeclareFieldFormat[article,periodical]{volume}{\mkbibbold{#1}} 

\DeclareSourcemap{ 
    \maps{
        \map{
            \step[fieldsource=language, fieldset=langid, origfieldval, final]
            \step[fieldset=language, null]
        }
        \map{
            \step[fieldsource=numpages, fieldset=pagetotal, origfieldval, final]
            \step[fieldset=pagestotal, null]
        }
        \map[overwrite]{
            \step[fieldset=issn, null]
        }
        \map[overwrite]{
            \step[fieldsource=abstract]
            \step[fieldset=abstract,null]
        }
        \map[overwrite]{ 
            \step[fieldsource=urldate,
            match=\regexp{([0-9]{2})\.([0-9]{2})\.([0-9]{4})},
            replace={$3-$2-$1$4}, 
            final]
        }
        \map[overwrite]{ 
            \perdatasource{biblio/otherref.bib}
            \step[fieldset=keywords, fieldvalue={otherref,bibliofull}]
       }
        \map[overwrite]{ 
            \perdatasource{biblio/myref.bib}
            \step[fieldset=keywords, fieldvalue={myref,bibliofull}]
        }
        \map[overwrite]{
            \step[typesource=online, fieldsource=howpublished, fieldset=organization, origfieldval, final]
            \step[fieldset=howpublished, null]
        }
        \map[overwrite]{
           \step[fieldsource=media,
           match={eresource},
            final]
            \step[fieldset=media, null]
        }
    }
}

\addbibresource{biblio/otherref.bib}

\addbibresource{biblio/myref.bib}

\newtotcounter{citenum}
\makeatletter
\defbibenvironment{counter} 
  {\setcounter{citenum}{0}%
  \renewcommand{\blx@driver}[1]{}%
  } 
  {} 
  {\stepcounter{citenum}} 
\makeatother
\defbibheading{counter}{}

\newtotcounter{citeauthor}
\makeatletter
\defbibenvironment{countauthor} 
{\setcounter{citeauthor}{0}%
    \renewcommand{\blx@driver}[1]{}%
} 
{} 
{\stepcounter{citeauthor}} 
\makeatother
\defbibheading{countauthor}{}

}

\begin{document}

\renewcommand{\alsoname}{див. так.}
\renewcommand{\seename}{див.}
\renewcommand{\headtoname}{вх.}
\renewcommand{\ccname}{вих.}
\renewcommand{\enclname}{вкл.}
\renewcommand{\pagename}{P.}
\renewcommand{\partname}{Part}
\renewcommand{\abstractname}{Abstract}
\renewcommand{\contentsname}{CONTENTS} 
\renewcommand{\figurename}{\textit{Fig.}} 
\renewcommand{\tablename}{Табл.} 
\renewcommand{\indexname}{Предметний вказівник}
\renewcommand{\listfigurename}{Список рисунків}
\renewcommand{\listtablename}{Список таблиць}
\renewcommand{\refname}{\fullbibtitle}
\renewcommand{\bibname}{\fullbibtitle}

\renewcommand{\subfigurename}{\textit{}}                   


\thispagestyle{empty}%
\begin{center}%
\MakeUppercase{\textbf{National Academy of Sciences of Ukraine\\ B.~V\MakeLowercase{erkin} Institute for Low Temperature Physics and Engineering}}
\end{center}%

\begin{center}%
\MakeUppercase{National Academy of Sciences of Ukraine\\ B.~V\MakeLowercase{erkin} Institute for Low Temperature Physics and Engineering}
\end{center}%

\begin{flushright}%
Qualification scientific \\ work printed as manuscript

\vspace{0pt plus6fill} 
\begin{center}%
{\textbf{Bahrova M. Olha} }
\end{center}%
\text {UDK 538.93}
\end{flushright}%
\begin{center}%
\textbf{DISSERTATION}
\end{center}%
\vspace{0pt plus1fill} 
\begin{center}%
\textbf {\large \MakeUppercase{Electromechanical phenomena\\ in normal and superconducting nanostructures \\based on a movable quantum dot }}

\vspace{0pt plus2fill} 
{
104 --- "Physics and Astronomy"\\ 
10~--- "Natural Sciences"
}

\vspace{0pt plus2fill} 
Submitted for obtaining the Doctor of Philosophy degree

\vspace{0pt plus4fill} 
{\small The dissertation contains the results of own research. The ideas, results, and texts of the other authors are referred accordingly $\underset{\text{(candidate's signature)}}{\underline{\hspace{0.16\textwidth}}}$ O.M.~Bahrova}
\end{center}%
\vspace{0pt plus4fill} 
\begin{flushright}%
\textbf{Supervisor:~Kulinich I. Sergei,\\ Candidate of Physico-Mathematical Sciences,\\ Senior Researcher}
\end{flushright}%
%
%
\vspace{0pt plus4fill} 
\begin{center}%
{Kharkiv~--~2023}
\end{center}%
\newpage
\chapter*{\MakeUppercase{\textbf{abstract}}}
\textbf{Bahrova O.M. \MakeUppercase{Electromechanical phenomena in normal and superconducting nanostructures based on a movable quantum dot}.~--- Manuscript.}

Dissertation for a Doctor of Philosophy degree on speciality 104~-- Physics and Astronomy.~--- B. Verkin Institute for Low Temperature Physics and Engineering, NAS of Ukraine, Kharkiv, 2023. 

The dissertation is devoted to the study of new fundamental phenomena which emerge due to electromechanical coupling in mesoscopic systems based on movable quantum dot. 

In the \textbf{introduction} it is briefly justified the relevance of the dissertation topic, defined
the purpose and main tasks of the research, as well as objects, subject and research
methods. The scientific novelty and practical value of the obtained results are
formulated. The information about the publications, the personal applicant’s
contribution and the approbation of the results of the dissertation are discussed. The information
about the structure and volume of the dissertation is also given.

The \textbf{chapter 1} is devoted to the review and analysis of the literature related to the topic of the dissertation. The main phenomena which arise in the electron transport through a single-electron transistor, are considered. Namely, Landauer-B{\"u}ttiker approach and the Coulomb blockade of electron tunneling are introduced. The \textbf{subsection 1.1.2} is devoted to the polaronic effects in transport of electrons in molecular transistors. In particular, the origin of the Franck-Condon (polaronic) blockade and polaronic narrowing of the energy level width are discussed as well as non-monotonic temperature dependence of the differential conductance. In addition, in the last part of the \textbf{subsection 1.1.2} a special case of a non-equilibrium vibron subsystem is briefly considered. 

In contrast to the first part of the \textbf{chapter 1}, where influence of the mechanical vibrations of a quantum dot on the electron transport is discussed, in the further parts we alternatively take into account the evolution of the mechanical subsystem under an impact of the tunneling of electrons. Thus, in the \textbf{section 1.2} the concept of a driven qubit  and Landau-Zener-St{\"u}ckelberg-Majorana formula for transition probability are introduced. Also, some protocols for quantum error correction codes and its importance in the further consideration are discussed.  
In the \textbf{section 1.3} nature of the mechanical instability phenomenon and key results are considered.

The \textbf{chapter 2} is devoted to the derivation and analysis of polaronic effects which emerge due to the \textit{non-equilibrium} coherent vibron subsystem. 

In the \textbf{section 2.1} a model device is introduced. A single-molecule transistor consists of a big molecule which is placed between two bulk electrodes biased by a constant voltage. The quantum dot which models the molecule, undergoes quantum oscillations in the direction perpendicular to the electron transport flow. It also gated by the gate voltage in order to control the energy of a single-electron level in the quantum dot. 

In the \textbf{section 2.2} Hamiltonian of the system under consideration is presented and equations for the density matrix of the electronic subsystem are obtained. 

In the \textbf{section 2.3} an analytical expression for the electric current through the single-molecular transistor is derived. In the \textbf{section 2.4} results of numerical calculations for the current-voltage characteristics (\textit{I-V} curves) are presented and analysed. The correspondence between the current-voltage curves obtained for the assumption of the vibron subsystem being in coherent (\textit{non-equilibrium}) state and Franck-Condon steps for equilibrated vibrons are drawn. It is demonstrated that in contrast to the Franck-Condon theory, in our case of coherent vibrons steps in the current-voltage characteristics are completely non-regular. Moreover, for the vibrons being in coherent state, the current saturates at much lower bias voltages. This can be effective in experiments which require working in a regime out of the polaronic blockade, i.e., maximal currents.

In the \textbf{section 2.5} a quite simple analytical formula for the electric current is found. The approximation gives high-precision agreement with the main results.

The \textbf{chapter 3} is devoted to the obtaining and analysis of entanglement between electronic and mechanical degrees of freedom in a superconducting nanoelectromechanical device. 

In the \textbf{section 3.1} a model of the nanoelectromechanical device under consideration is introduced. 
The system consists of a superconducting nanowire suspended over two superconducting leads. The nanowire which is treated as a charge qubit (Cooper pair box), undergoes bending vibrations in the perpendicular to the nanowire axis direction. Furthermore, the nanowire is capacitively coupled to the gate electrodes which allow one to control the difference between the energy levels of the qubit. Also, the superconducting phase difference between the electrodes can be tuned by the constant bias voltage applied to them as a result of non-stationary (ac) Josephson effect.
The Hamiltonian of the system is derived. 

In the \textbf{section 3.2} time evolution of the pure state of the system is found. It is demonstrated that initial pure state evolves into the state represented by entanglement between the two qubit
states and two coherent states of the mechanical resonator. 

In the \textbf{section 3.3}, which represents the main result of this chapter, we propose and derive a specific bias voltage manipulation protocol which results in the formation of entanglement between two states of the charge qubit and two Schr{\"o}dinger-cat states (superposition of two coherent states) starting from the initial pure state. The considered protocol due to its simplicity can effectively be implemented in experiments with encoding quantum information from the electronic qubit states to coherent (cat states, in particular) of a nanomechanical resonator. Moreover, the cat states due to its structure are not sensitive to errors. Thus, the proposed scheme does not require additional quantum error correction protocols. 

In the \textbf{section 3.4} the entanglement (von Neumann) entropy is considered in order to quantitatively analyse the entanglement between charge states of the qubit and coherent states of the nanomechanical resonator. Further, in the \textbf{section 3.5} time evolution of the mechanical subsystem is discussed. A clear justification of presence of the entanglement is presented by analysing corresponding Wigner functions.

In the \textbf{section 3.6} an experimentally feasible method for the detection of 
signatures of the entanglement by measuring average current is discussed. 

The \textbf{chapter 4} is devoted to the derivation and analysis of nanomechanical phenomena which arise due to proximity effect in the following hybrid nanoelectromechanical device. The system under consideration involves a carbon nanotube suspended above a trench in a normal metal electrode and positioned in a gap between two superconducting leads. Moreover, the nanotube undergoes bending vibrations in between two superconducting electrodes in such a way that the bending of the nanotube moves it closer to one electrode and further away from the other. It results in the position-dependent tunneling amplitudes. In addition, due to the presence of superconducting phase difference between the leads, the off-diagonal order parameter of the quantum dot emerges as a result of superconducting proximity effect. Lastly, the bias voltage applied to the normal electrode induces directed electron dynamics in the system. 

In the \textbf{section 4.1} the semi-classical approach within the density matrix approximation is used to obtain and analyse the regime of mechanically \textit{unstable} states.

In the \textbf{subsection 4.1.1} the model of the considered nanoelectromechanical device and Hamiltonian are introduced. In the \textbf{subsection 4.1.2} the density matrix approximation is considered. The system of equation for the density matrix elements together with the second-order nonlinear differential equation for the quantum dot displacement is derived. Additionally, in the \textbf{subsection 4.1.3} the Green function formalism is used to find the quantum dot order parameter induced by superconducting proximity effect. 

In the \textbf{subsection 4.1.4} the consideration within an adiabatic limit allow one to simplify the problem to one strongly nonlinear differential equation (which is the central one in this chapter) for the displacement and analytically analyse it by using a simple linearization method as in \textbf{subsection 4.1.5}. Furthermore, in the \textbf{subsection 4.1.6} the Krylov-Bogoliubov method of averaging is used to find an approximate solution and analyse regimes in which the nanoelectromechanical system under consideration can operate. Two states of mechanical subsystem are discussed. In particular, it is demonstrated that in the mechanically \textit{unstable} regime the limit cycles of self-sustained oscillations occur. Moreover, the self-saturation effect takes place. In the \textbf{subsection 4.1.7} the main results are generalized to the case of asymmetric tunnel contacts and the influence of a thermodynamic environment. 

In the \textbf{subsection 4.1.8} a possibility to experimentally detect the mechanical instability in the system due to electric current measurements is discussed. It is demonstrated that the device can operate in  
transistor and diode regimes. 

In the \textbf{subsection 4.1.9} we discuss numerically calculated time evolution of the considered system in the diabatic limit which cannot be done analytically. 

In the \textbf{section 4.2} quantum-mechanical fluctuations are taken into account. It is demonstrated that we can achieve ground-state cooling regime as a result of the superconducting proximity effect. 

In the \textbf{subsection 4.2.1} Hamiltonian of the nanoelectromechanical under consideration is introduced.
In the \textbf{subsection 4.2.2} the system of equations that describes dynamics in the stationary regime is derived and analysed by using the Wigner function representation. 

In the \textbf{subsection 4.2.3} the regime of cooling of nanomechanical vibrations is discussed. 

In the \textbf{subsection 4.2.4} the electric current through the system is discussed. It is demonstrated that the cooling of the mechanical vibrations and ground-state cooling, particularly, can be experimentally explored via electric current measurements.

\textbf{Keywords:} Quantum dot (QD), nanoelectromechanical system (NEMS), molecular transistor, coherent state, proximity effect, qubit.     
\chapter*{\MakeUppercase{\textbf{Анотація}}}

\textbf{Багрова О.М. \MakeUppercase{Електромеханічні явища в нормальних та надпровідних наноструктурах на основі рухомої квантової точки}.~--- Рукопис.}

Дисертація на здобуття наукового ступеня доктора філософії за спеціальністю 104 – фізика та астрономія. – Фізико-технічний інститут низьких температур імені Б.І. Вєркіна НАН України, Харків, 2023.

Дисертація присвячена вивченню нових фундаментальних явищ, які виникають внаслідок електромеханічного зв'язку в мезоскопічних системах на основі рухомої квантової точки. 

У \textbf{вступі} коротко обґрунтовано актуальність теми дисертації, визначено
мету та основні завдання дослідження, а також об'єкт, предмет і методи дослідження. Сформульовано наукову новизну та практичне значення отриманих результатів.
Наведено відомості про публікації, особистий внесок здобувача та апробацію результатів дисертації.
Також приведено відомості про структуру та обсяг дисертаційної роботи.

\textbf{Розділ 1} присвячено огляду та аналізу літератури за темою дисертації. Розглянуто основні явища, які виникають при транспортуванні електронів через одноелектронний транзистор. Зокрема, введено підхід Ландауера-Бюттікера та поняття кулонівської блокади тунелювання електронів. \textbf{Пункт 1.1.2} присвячено розгляду поляронних ефектів у транспорті електронів у молекулярних транзисторах. Зокрема, обговорюється походження блокади Франка-Кондона (поляронної) і поляронного звуження ширини енергетичного рівня, а також немонотонна температурна залежність диференціальної провідності. Окрім того, в останній частині \textbf{пункту 1.1.2} коротко розглянуто окремий випадок нерівноважної вібронної підсистеми. 

На відміну від першої частини \textbf{розділу 1}, де обговорюється вплив механічних коливань квантової точки на транспорт електронів, на противагу цьому в подальших частинах розглядається еволюція механічної підсистеми під впливом тунелювання електронів. Так, у \textbf{підрозділі 1.2} введено поняття керованого кубіта та формулу Ландау-Зенера-Штукельберга-Майорани для ймовірності переходу. Крім того, обговорено деякі протоколи для квантових кодів корекції помилок та їх важливість для подальшого розгляду.  
У \textbf{підрозділі 1.3} розглянуто природу явища механічної нестійкості.

\textbf{Розділ 2} присвячено розгляду та аналізу поляронних ефектів, які виникають завдяки \textit{нерівноважній} когерентній вібронній підсистемі. 

У \textbf{підрозділі 2.1} представлено модель системи, що розглядається. Одномолекулярний транзистор складається з макромолекули, яку розміщено між двома об'ємними електродами, до яких прикладено постійну тягнучу напругу. Квантова точка, яка моделює молекулу, зазнає квантових коливань у напрямку, перпендикулярному до напрямку переносу електронів. За допомогою напруги на затворі виникає можливість керування енергією одноелектронного рівня квантової точки. 

У \textbf{підрозділі 2.2} представлено гамільтоніан досліджуваної системи та отримано рівняння для матриці густини електронної підсистеми. 

У \textbf{підрозділі 2.3} отримано аналітичний вираз для електричного струму через одномолекулярний транзистор. У \textbf{підрозділі 2.4} наведено та проаналізовано результати чисельних розрахунків вольт-амперних (\textit{I-V}) характеристик (ВАХ). Встановлено відповідність між ВАХ, отриманими в припущенні, що вібронна підсистема перебуває в когерентному (\textit{нерівноважному}) стані, та франк-кондонівськими сходинками для вібронів у рівноважному стані. Показано, що на відміну від теорії Франка-Кондона, у випадку когерентних вібронів сходинки на вольт-амперних характеристиках є нерегулярними. Більш того, для когерентного стану вібронів струм насичення виникає при значно менших тягнучих напругах. Останній факт може бути вирішальним в експериментах, які вимагають роботи в режимі зняття поляронної блокади, тобто максимальних струмів.

У \textbf{підрозділі 2.5} знайдено аналітичну формулу для електричного струму. Наближення дає гарне узгодження з основними чисельними результатами.

\textbf{Розділ 3} присвячено отриманню та аналізу заплутаності, яка виникає між електронними та механічними ступенями свободи в надпровідному наноелектромеханічному пристрої. 

У \textbf{підрозділі 3.1} представлено модель наноелектромеханічного пристрою, що розглядається. 
Система складається з надпровідного нанодроту, що підвішений між двома надпровідними електродами. Нанодріт, який розглядається як зарядовий кубіт (сховище куперівських пар), зазнає згинальних коливань у напрямку, перпендикулярному до осі нанодроту. Окрім того, нанодріт з'єднаний з електродами затвора за допомогою ємнісного зв'язку, що дозволяє керувати відстанню між енергетичними рівнями кубіта. До того ж, різниця фаз між надпровідними електродами може бути підлаштована постійною тягнучою напругою, що прикладена до них, як результат нестаціонарного ефекту Джозефсона.
Також представлено гамільтоніан системи.

У \textbf{підрозділі 3.2} знайдено часову еволюцію чистого стану системи. Показано, що початковий чистий стан еволюціонує до стану, представленого заплутаністю між двома станами кубіта та двома когерентними станами механічного резонатора. 

У \textbf{підрозділі 3.3}, який представляє основний результат цього розділу, запропоновано і виведено специфічний протокол маніпуляції тягнучою напругою, який призводить до утворення заплутаності між двома станами зарядового кубіта і двома станами типу "Schr{\"o}dinger cat"$~$ (суперпозиція двох когерентних станів), починаючи з чистого стану. Розглянутий протокол завдяки своїй простоті може бути ефективно реалізований в експериментах з кодуванням квантової інформації з електронних станів кубіта до когерентних (зокрема, так званих "cat states") наномеханічного резонатора. До того ж, "cat states"$~$  завдяки своїй структурі не чутливі до виникнення помилок. Таким чином, запропонована схема не потребує додаткових протоколів корекції квантових помилок. 

У \textbf{підрозділі 3.4} розглянуто ентропію заплутаності (фон Неймана) з метою кількісного аналізу заплутаності між зарядовими станами кубіта і когерентними станами наномеханічного резонатора. Далі в \textbf{підрозділі 3.5} обговорено часову еволюцію механічної підсистеми. Чітке обґрунтування наявності заплутаності представлено шляхом аналізу відповідних функцій Вігнера.

У \textbf{підрозділі 3.6} описано чіткий метод для експериментального виявлення заплутаності шляхом вимірювання середнього струму. 

\textbf{Розділ 4} присвячено розгляду та аналізу наномеханічних явищ, які виникають завдяки ефекту близькості в наступному гібридному наноелектромеханічному пристрої. Система, що розглядається, включає вуглецеву нанотрубку, підвішену над канавкою в звичайному металевому електроді і розміщену в проміжку між двома надпровідними електродами. Крім того, нанотрубка зазнає згинальних коливань між двома надпровідними електродами таким чином, що згинання нанотрубки переміщує її ближче до одного електрода і далі від іншого. Це призводить до залежних від положення амплітуд тунелювання. До того ж, завдяки наявності різниці фаз між надпровідними електродами, недіагональний параметр порядку квантової точки виникає в результаті надпровідного ефекту близькості. На додаток, тягнуча напруга, що прикладена до нормального електрода, спричиняє направлену динаміку електронів у системі. 

У \textbf{підрозділі 4.1} напівкласичний підхід в рамках наближення матриці густини використано для отримання та аналізу режиму \textit{нестійких} станів механічної підсистеми.

У \textbf{пункті 4.1.1} вводиться модель досліджуваного наноелектромеханічного пристрою та його гамільтоніан. У \textbf{пункті 4.1.2} розглянуто наближення матриці густини. Виведено систему рівнянь для елементів матриці густини разом з нелінійним диференціальним рівнянням другого порядку для координати квантової точки. Крім того, в \textbf{пункті 4.1.3} формалізм функцій Гріна використано для знаходження параметра порядку квантової точки, який виникає за рахунок надпровідного ефекту близьості.

У \textbf{пункті 4.1.4} розгляд в рамках адіабатичного режиму дозволяє спростити задачу до одного нелінійного диференціального рівняння (яке є центральним у цьому розділі) для координати квантової точки і аналітично проаналізувати його за допомогою методу лінеаризації, як у \textbf{пункті 4.1.5}. Крім того, в \textbf{пункті 4.1.6} використано метод усереднення Крилова-Боголюбова для знаходження наближеного розв'язку та аналізу режимів, в яких може працювати наноелектромеханічна система, що розглядається. Розглянуто два стани механічної підсистеми. Зокрема, показано, що в механічно \textit{нестійкому} режимі виникають граничні цикли самопідтримних коливань. До того ж, має місце ефект самонасичення. У \textbf{пункті 4.1.7} основні результати узагальнено на випадок несиметричних тунельних контактів і впливу термодинамічного оточення. 

У \textbf{пункті 4.1.8} обговорюється можливість експериментального виявлення механічної нестійкості в системі за допомогою вимірювання електричного струму. Продемонстровано, що дана система може працювати в  
транзисторному та діодному режимах. 

У \textbf{пункті 4.1.9} обговорюється чисельно розрахована еволюція розглянутої системи в діабатичній границі, що не може бути зроблено аналітично. 

У \textbf{підрозділі 4.2} враховано вплив квантово-механічних флуктуацій. Продемонстровано, що можна досягти режиму охолодження до основного стану в результаті ефекту близькості. 

У \textbf{пункті 4.2.1} введено гамільтоніан наноелектромеханічної системи, що розглядається.
У \textbf{пункті 4.2.2} виведено систему рівнянь, яка описує динаміку в стаціонарному режимі, і проаналізовано її за допомогою представлення функцій Вігнера. 

У \textbf{пункті 4.2.3} розглянуто режим охолодження наномеханічних коливань. 

У \textbf{пункті 4.2.4} розглянуто електричний струм через систему. Показано, що охолодження механічних коливань і, зокрема, охолодження до основного стану можна експериментально дослідити за допомогою вимірювання електричного струму.

\textbf{Ключові слова:} квантова точка, наноелектромеханічна система, молекулярний транзистор, когерентний стан, ефект близькості, кубіт.    
\chapter*{\authorbibtitle}

The main results of the dissertation are published in 10 scientific works, 4 research papers in leading special scientific journals among them~[1-4]:
\begin{enumerate}
    \item \textbf{O.M. Bahrova}, S.I. Kulinich, I.V. Krive, Polaronic effects induced by non-equilibrium vibrons in a single-molecule transistor, \textit{Low Temp. Phys.}~\textbf{46}, No. 7, 671, (2020) [\textit{Fiz. Nizk. Temp.}, \textbf{46}, 799 (2020)],  DOI: 10.1063/10.0001362
    \item \textbf{O.M. Bahrova}, L.Y. Gorelik, S.I. Kulinich, Entanglement between charge qubit states and coherent states of nanomechanical resonator generated by ac Josephson effect, \textit{Low Temp. Phys.,} \textbf{47}, No. 4, 287, (2021) [\textit{Fiz. Nizk. Temp.}, \textbf{47}, 315 (2021)], DOI: 10.1063/10.0003739
    \item \textbf{O.M. Bahrova}, L.Y. Gorelik, S.I. Kulinich, R.I. Shekhter, H.C. Park, Nanomechanics driven by the superconducting proximity effect, \textit{New J. Phys.}, \textbf{24}, 033008 (2022), DOI:  10.1088/1367-2630/ac5758
    \item \textbf{O.M. Bahrova}, L.Y. Gorelik, S.I. Kulinich, R.I. Shekhter, H.C. Park, Cooling of nanomechanical vibrations by Andreev injection, \textit{Low Temp. Phys.}, \textbf{48}, No. 6, 476 (2022) [\textit{Fiz. Nizk. Temp.}, \textbf{48}, 535 (2022)], DOI: 10.1063/10.0010443
    %
    %
    %
    \item \textbf{O.M. Bahrova}, I.V. Krive, How to control transport of spin-polarized electrons via magnetic field in a molecular transistor, Physics and Scientific\&Technological progress: student scientific conference, p.3, (2018).
    \item \textbf{O. M. Bahrova}, S. I. Kulinich, I. V. Krive, Polaronic effects induced by coherent vibrons in a single-molecule transistor, I International Advanced Study Conference Condensed matter \& Low Temperature Physics, June 8-14, 2020, Ukraine, Kharkiv, Abstracts, p. 183, (2020).
    \item A.D. Shkop, \textbf{O.M. Bahrova}, Coulomb and vibration effects in spin-polarized current through a single-molecule transistor, XI Conference of Young Scientists “Problems of Theoretical Physics”, December 21-23, 2020, Ukraine, Kyiv, Abstracts, p.15-16, (2020).
    \item \textbf{O.M. Bahrova}, L.Y. Gorelik, S.I. Kulinich, Schrödinger-cat states generation via mechanical vibrations entangled with a charge qubit, II International Advanced Study Conference Condensed matter \& Low Temperature Physics, June 6–12, 2021, Ukraine, Kharkiv, Abstracts, p.201, (2021).
    \item \textbf{O.M. Bahrova}, L.Y. Gorelik, S.I. Kulinich, H.C. Park, R.I. Shekhter, Self-sustained mechanical oscillations promoted by superconducting proximity effect, The International Symposium on Novel maTerials and quantum Technologies, December 14–17, 2021, Abstracts, p.134, (2021).
    %
    %
    \item \textbf{O.M. Bahrova}, L.Y. Gorelik, S.I. Kulinich, H.C. Park, R.I. Shekhter, Nanomechanics provoked by Andreev injection, 29th International Conference on Low Temperature Physics, August 18-24, 2022, Abstracts, p.1554 {\&} 1771, (2022).
\end{enumerate}


\tableofcontents
\chapter*{INTRODUCTION}							
\addcontentsline{toc}{chapter}{INTRODUCTION}	

\textbf{Justification of the relevance of the research topic.}
Nanotechnology is in the front of modern science nowadays. State-of-the-art technology
allows one to manipulate with molecular orbitals of a single molecule and to molecule-based
transistors of a high quality. Single-molecule transistors (SMTs) mostly studied in experiments are
a macromolecule (fullerenes or carbon nanotubes)-based device, where a molecule is tunnel
coupled to source and drain electrodes and capacitively coupled to a gate electrode.
 As a result of coupling of the mechanical (vibronic) and electronic degrees of freedom, the transport properties of such a nano-scale transistor are drastically changed.  The main new effect caused by vibrons is the appearance of inelastic
channels of electron tunneling in single-electron transistor. For strong electron-vibron interaction
the current at low voltages is strongly suppressed (Franck-Condon or polaronic blockade) and the lifting of this
blockade by bias voltage or by temperature leads to step-like \textit{I-V} characteristics and nonmonotonic
temperature dependence of conductance. If the vibration excitations (vibrons)
of the central part of the transistor are coupled to a heat bath and the vibron relaxation time is much smaller than the
characteristic time of electron tunneling, the vibron subsystem is in equilibrium. The electron
transport through molecular transistors with equilibrated vibrons is usually considered. However, it is not the case when the coupling of vibron subsystem to the environment is weak.

On the other hand, nanoelectromechanical systems (NEMS) provide a promising platform for investigations into the
quantum mechanical interplay between mechanical and electronic subsystems.
One of the most important phenomena providing the foundation of NEMS functionality
is the generation of self-sustained mechanical oscillations by a constant value current flow.
Nevertheless, nanoelectromechanical systems promise to manipulate the mechanical motion of a nano-object using electronic dynamics. There are many approaches to control nanomechanical performance providing a number of new functionalities of nano-device operations, in particular, pumping or cooling of the mechanical subsystem. One of the main approaches exploits the electronic flow through a nanosystem induced by either the bias voltage or temperature drop between two electronic reservoirs connected by a quantum dot (QD).

In general, there are several types of interaction between the electronic and mechanical subsystems. The most common cause of this interaction is due to localization of the electron charge or spin. However,  incorporating of superconducting elements into NEMS allows one to use the coupling based on delocalization of Cooper pairs, as a foundation for the electro-mechanical performance.

The above range of unresolved issues related to the study of nanoelectromechanical systems and polaronic effects in single-electron transistors determines the \textbf{relevance} of the topic of this dissertation.

\textbf{Relation to research programs, plans, topics.}

The dissertation was performed at the B. Verkin Institute for Low Temperatures Physics and Engineering of the National Academy of Sciences of Ukraine within the framework of the thematic plan of the B. Verkin ILTPE of NASU on department topics: "Theoretical studies of collective phenomena in quantum condensed structures and nanomaterials" (registration number 0117U002292, code 1.4.10.26.4, the period of execution is 2017-2021), "Theoretical studies of quantum phenomena in complex low-dimensional condensed matter" (registration number 0122U001505, code 1.4.10.26.5, the period of execution is 2022-2026).
Part of the dissertation work was carried out at the Center for Theoretical Physics of Complex Systems, Institute for Basic Science, Daejeon, Republic of Korea, as part of the projects "Condensed matter theory at nanoscale" (IBS-R024-D1) and "Disorder and chaos in low-dimensional systems" (IBS Young Scientist Fellowship (IBS-R024-Y3-2021)).

\textbf{Goal and tasks of the research.}

\textit{The goal} 
of the dissertation work is a theoretical description of quantum effects in electron transport in nanoelectromechanical systems and molecular transistors.

To achieve this goal, it was necessary to solve the following \textit{tasks}:
\begin{itemize}
    \item to study the electron transport through a single-molecule transistor for the case when the mechanical subsystem is in a non-equilibrium state, in particular, the coherent state;
    \item to obtain the time evolution of a superconducting nanoelectromechanical system based on a carbon nanotube;
    \item to study the nature of entanglement between the charge states of the qubit and coherent states of the nanomechanical oscillator; 
    \item to study the dynamics of a hybrid nanoelectromechanical device which arise due to the superconducting proximity effect;
    \item to obtain regions of the mechanical instability of such a system;
    \item to study the influence of quantum fluctuations on the steady state of the hybrid nanomechanical system based on a carbon nanotube.
\end{itemize}

\textbf{Object of research} of the dissertation is quantum transport of electrons in nanostructures based on a movable quantum dot.

\textbf{Subjects of research} are 
mechanical instability and tunneling processes in nanoelectromechanical systems, in particular, molecular transistors.

\textbf{Research methods.}
The results of the dissertation were obtained using the methods of theoretical condensed matter physics. The density matrix method and perturbation theory were used to analytically find the regions of mechanical instability and study polaron effects in nanoelectromechanical systems. Also, to find the effects associated with coherent oscillations in a molecular transistor, numerical calculations were performed (solving the system of differential equations by the Runge-Kutta method).

\textbf{Scientific novelty of the obtained results.}

\begin{enumerate}
    \item \textit{For the first time}, electron transport through a single-molecule transistor was studied for the case when the mechanical subsystem is in a non-equilibrium coherent state, in particular, the current-voltage characteristics of such a transistor were obtained;
    \item \textit{For the first time}, possibility of generating quantum entanglement between the charge states of the qubit and coherent states of the nanomechanical resonator using the protocol of bias  voltage manipulation was shown;
    \item  Quantum dynamics of a hydrid nanoelectromechanical system based on a carbon nanotube which emerge due to the superconducting proximity effect was studied \textit{for the first time};
    \item \textit{For the first time}, regions of instability for the nanoelectromechanical system based on a carbon nanotube were found and the self-saturation phenomenon was obtained, which arise as a result of delocalization of Cooper pairs due to the proximity effect;
    \item \textit{For the first time}, the effect of ground-state cooling of nanomechanical vibrations for a nanoelectromechanical system where the electromechanical coupling is of quantum origin --- arising from the proximity effect --- was theoretically obtained.
\end{enumerate}

\textbf{Practical significance of the results.}

The results of the research presented in this dissertation are of fundamental importance, as they deepen and extend the knowledge of electron transport in nanoelectromechanical systems. The effects predicted in the work, such as self-sustained nanomechanical oscillations and cooling to the ground state of ones, can be detected in experiments. The obtained protocols for bias voltage manipulation can be used to encode quantum information between qubit and nano-resonator states within a single system. Based on the study of electron transport in a single-electron transistor, the basic element of which is a movable quantum dot, more efficient molecular transistors can be created.

\textbf{The candidate's contribution.}
In all works that were co-authored and included in the dissertation, the author performed all analytical calculations, participated in the discussion of the results and wrote the articles. Thus, the personal contribution of the candidate to the solution of the theoretical problems discussed in the dissertation is decisive.

\textbf{Approbation of results of the dissertation.}
The main results which this dissertation includes were presented on the following 7 international scientific conferences:
\renewcommand{\labelitemi}{$\bullet$}
\begin{itemize}
    \item Physics and Scientific\&Technological progress: student scientific conference (Kharkiv, Ukraine, April 10-12, 2018);
    \item I International Advanced Study Conference Condensed matter \& Low Temperature Physics (Kharkiv, Ukraine, June 8-14, 2020);
    \item XI Conference of Young Scientists “Problems of Theoretical Physics” (Kyiv, Ukraine (online), December 21-23, 2020);
    \item II International Advanced Study Conference Condensed matter \& Low Temperature Physics (Kharkiv, Ukraine, June 6–12, 2021);
    \item The International Symposium on Novel maTerials and quantum Technologies, (Kanagawa, Japan (online), December 14–17, 2021);
    \item Quantum Thermodynamics Conference 2022, (Belfast, United Kingdom (online), June 27-July 1, 2022);
    \item 29th International Conference on Low Temperature Physics, (Sapporo, Japan (online), August 18-24, 2022);
\end{itemize}

\textbf{Publications.} 
Results which this dissertation based on have been published in 4 research papers~\cite{Bahrova2020,Bahrova2021,Bahrova2022,Bahrova2022b} and 6 conference abstracts~\cite{BahrovaC2018,BahrovaC2020,ShkopC2020,BahrovaC2021,BahrovaC2021b,BahrovaC2022b}.

\textbf{Structure of the dissertation.} The dissertation consists of abstracts, introduction, review chapter, three original chapters with figures, conclusions, and bibliography. The total length of the dissertation is \formbytotal{TotPages}{}{}{} pages. It contains \formbytotal{totalcount@figure}{}{}{} figures and bibliography with 210 sources in 21 pages.


\chapter{EFFECTS OF ELECTROMECHANICAL COUPLING IN NANOSCALE SYSTEMS}\label{chapt1}

In this chapter we briefly review main effects emerged in the electron tunneling in mesoscopic devices.

\section{Electron transport in single-electron transistors.} \label{sect1_1}

In contrast to a conventional transistor, a single-electron transistor (SET) exhibits some non-trivial features due to relevance of quantum effects. Main phenomena that arise in the electron transport in nanoelectrical and nonoelectromechanical devices are Coulomb blockade, vibration and shuttle effects, Kondo and Luttinger liquid effects. In this section we start with a simple model of a SET and proceed to more complex through the consideration, so that we leave behind vibration effects in the first subsection,~\ref{subsect1_1_1}.

\begin{figure}
\centering
\includegraphics[width=0.75\columnwidth]{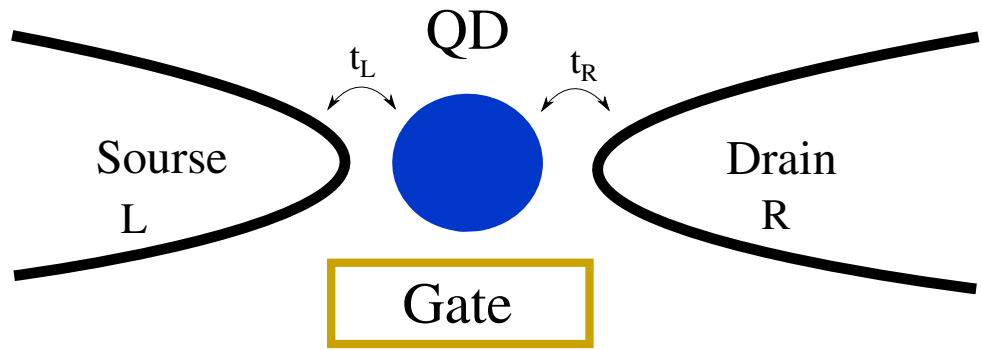}
\caption{\textit{Schematic representation of a single-electron transistor. A quantum dot is placed between two bulk electrodes (source and drain) and tunnel couples ($t_{L,R}$) to them. The gate voltage is set to manipulate the energy of the QD.}} \label{fig:fig1_1}
\end{figure}

A single-electron transistor can be viewed in a simplest model as a quantum dot (QD) placed between two bulk source and drain electrodes and gated by a third electrode, see Fig.~\ref{fig:fig1_1}. The central part of the system, i.e., quantum dot, is a zero-dimensional mesoscopic structure with a discrete energy spectrum. It can be represented by a metallic grain (island), quantum nanowire (including a carbon nanotube (CNT)) or a big molecule (like fullerene one). In the latter case they are usually called by molecular transistors, see subsection~\ref{subsect1_1_2}. There are enough comprehend reviews and textbooks on the topic, see, e.g., Refs.~\cite{Krive2010,Devoret1992,Datta1997,Nazarov2013,Ryndyk2019}. The QD is coupled to bulk (with non-interacting electrons) electrodes by quantum tunneling processes (tunnel). It means that this system can be considered as one-dimensional double-barrier one (with ballistic transport inside) which is connected to the reservoirs of electrons. The case when the energy of tunneling electrons is within the energy window of the tunneling width of the resonant energy level inside the structure, is refereed to the resonant tunneling. The completely coherent tunneling process is usually called by resonant quantum tunneling (RQT). However, when electrons tunnel incoherently to and from intermediate state (on the QD), such a process is called by sequential tunneling (ST)~\cite{Krive2010}. The later is exactly the case we are interested in below. Thus, the electron transport through such a double-barrier system can be described within the Landauer-B{\" u}ttiker approach~\cite{Landauer1957,Buttiker1986}. Within this method the average current $I$ determined by tunneling events though the system is related to the transmission coefficient $T(\varepsilon)$,
\begin{equation}\label{IVLandauer}
    I(V)=\frac{2e}{h}\int d\varepsilon T(\varepsilon) \left[ f_L(\varepsilon)-f_R(\varepsilon)\right],
\end{equation}
where $f_{L,R}$ is the Fermi-Dirac distribution function in the Left (source) or Right (drain) electrode, $f_{L,R}(\varepsilon)=1/\left[ \exp{\left\{(\varepsilon-\mu_{L,R})/T\right\}+1}\right]$, $\mu_{L,R}$ is the chemical potential, $T$ stands for temperature and $e$ is the electron charge. For weak tunneling barriers the transmission coefficient is determined by the Breit-Wigner formula,
\begin{equation}\label{TBW}
T_{BW}(\varepsilon)=\frac{\Gamma^2}{(\varepsilon-\varepsilon_i)^2+\Gamma^2},
\end{equation}
where $\varepsilon_i$ corresponds to the resonant level energies of the QD (inside the double-barrier structure), and $\Gamma \propto \vert t_0\vert^2$ is the energy level width witch is associated with the decay rate of the resonant state, $\Gamma/\hbar$, ($t_0$ is the tunneling amplitude). It is useful also to note that for the linear conductance $G=I/V$ one can obtain the following expression from Eq.~(\ref{IVLandauer}) (in the linear response regime, $V\to 0$), which is the well-known Landauer formula for conductance,
\begin{equation}\label{GLandauer}
    G=G_0\int_0^\infty d\varepsilon T(\varepsilon)\left(-\frac{\partial f}{\partial \varepsilon}\right),
\end{equation}
where $G_0=2e^2/h$ is the so-called conductance quantum. From this equation, Eq.~(\ref{GLandauer}), one can obtain the high-temperature ($T\gg \Gamma$) $1/T$ scaling of the linear conductance for sequential electron tunneling, $G\propto \Gamma/T$. Also, Eq.~(\ref{GLandauer}) can serve as an ground of the note "conductance is transmission"~\cite{Houten1996}. 

The partial derivative in the integrand of Eq.~(\ref{GLandauer}) gives the temperature dependence of the conductance, while from transmission coefficient, Eq.~(\ref{TBW}), it is seen that the conductance has the maxima at $\varepsilon=\varepsilon_i$. Here positions of the peaks defined by the quantum dot energy levels (intrinsic effect) and the chemical potential in the electrodes (leads). But it is not only the situation. Modern experimental techniques allow one change parameters of the central part of the transistor (QD). In particular, using an additional electrode (gate) contact which create a capacitor with the island, one can change potential of the quantum dot with respect to leads, so that the energy of the each level in the dot will be shifted on the value determined by the voltage applied to the gate electrode, $eV_G$~\cite{Moskalets}. As a consequence, the linear differential conductance has a resonant dependence on the gate voltage. This, for example, brings a possibility to identify a quantum dot energy spectrum via electric current measurements.

Transport properties of single-electron transistors based on one or several quantum dots, quantum wires are belong to the hot topic in theoretical as well as experimental studies, see, e.g., Refs.~\cite{Park2000,Kate2022,Leturcq2009,Park2021,Wang2018,Braakman2013,Rastelli2019,Wei2016}. 

\subsection{Coulomb blockade regime of electron tunneling.}\label{subsect1_1_1}

In the above consideration we treated the electrons as non-interacting ones. This assumption can work very well for the macroscopic leads (and we will maintain it for the electrodes in what follows). However, it is not so for the quantum dot where one electron can affect its state due to the discreetness of the electron charge. 

The important energy parameter scale is an energy associated with adding one electron to a neutral quantum dot, i.e., the charging energy, which is an electrostatic one caused by own capacity $C$ of the dot,
\begin{equation}\label{Ec}
    \varepsilon_C=\frac{e^2}{2C}.
\end{equation}
It means that at low temperatures, $T\ll\varepsilon_C$, and if the bias voltage applied to the electrodes is small, $V<V_C=e/2C$, the transport of electrons through the system is completely blocked. This corresponds to the Coulomb blockade effect. There are a lot of literature on the Coulomb blockade phenomena, see, e.g., Refs.~\cite{Datta1997,Devoret1992, Nazarov2003,Moskalets,Ryndyk2019}.
It is important to note that the Coulomb blockade can be lifted by adjusting the gate voltage $V_G$ (now in the condition for the resonant tunneling we need to include the charging energy because when an electron come to the island, it shifts the energy levels by the value of the charging energy), in contrast, for example, to the Franck-Condon blockade which we will consider in details in the next subsection,~\ref{subsect1_1_2}. This leads to the Coulomb blockade oscillations, i.e., resonant behaviour of the differential conductance with the oscillation period equals $e/C$ (in gate voltage). Also, we should say that if the size of a QD become very small (the capacity increases), the charging energy start to be negligible in comparison to the energy spacing between quantized energy levels on the island. This results in the fact that the Coulomb blockade oscillations are linear conductance oscillations (as a function of the gate voltage) with the period determined by the energy level spacing. 

The suppression of the electric current through a SET in the Coulomb blockade regime can be understood in the following ostensive picture. The chemical potentials of the source and drain electrodes define the so-called "conducting window", so that when we increase the bias voltage, the width of the "window" became bigger. The resonant transport of electron is possible when (resonant) energy levels of the QD are within the "conducting window". The gate voltage allows one to shift the levels with respect to the Fermi energy of the electrodes and thus bring them to the "conducting window" (or vice versa).

One of the features of the Coulomb blockade phenomenon in single-electron transistors is the Coulomb staircase of current-voltage characteristics which is more pronounced in the case of asymmetric tunneling barriers, see, e.g., Ref.~\cite{Ryndyk2019}. The effect takes place due to the contribution (to the electron transport) of states with the higher energy at higher bias voltages. It may be clear within the above-mentioned picture of the "conducting window". For example, for a given value of the gate voltage, the electric current is suppressed at low bias voltages up to the critical value $V_C$, and electrons start to tunnel after that through the first state on the dot. At higher voltages next electron states became available which results in a jump of the current. As a consequence, in general case for $n$th step for the the current one finds~\cite{Ryndyk2019},
\begin{equation}\label{vn}
    V_C^n=\frac{2(2n-1)\varepsilon_C}{e}.
\end{equation}
The heights of the steps drastically decrease withal and a current-voltage dependence becames linear at big values of the bias voltage. Also, the temperature smooths and eventually smears out the steps, i.e., one can estimate a relevant temperature to observe the effects as $T\approx 1K$~\cite{Leturcq2009}.

A very informative and usually used to present experimental results are stability diagrams. They are contour plots of electric current $I(V,V_G)$ and differential conductance $dI/dV(V,V_G)$ dependencies on the bias and gate voltages. In such plots effects of the Coulomb blockade are clearly seen via the so-called Coulomb diamonds (the regime of Coulomb blockade corresponds to regions of a diamond (rhombus) form). It is worth to mention that additional lines (peaks) can be seen on stability diagrams due to presence of several energy level on a QD ($\Delta\varepsilon\approx\varepsilon_C$, where $\Delta\varepsilon$ stands for an energy level difference) or due to the influence of vibration effects (in molecular transistors, see subsection~\ref{subsect1_1_2})~\cite{Leturcq2009, Rastelli2012}.

As a next point, let us note about a generalization of the Landauer formula for the electric current, Eq.~(\ref{IVLandauer}). 
In case of interacting electrons, one usually uses approaches based on the Green function formalism in order to calculate the transport properties of such a system. The generalized Landauer-like expression for the electric current is the following:
\begin{equation}\label{IVMV1}
    I(V)=\frac{\imath e}{2e}\int d\varepsilon \text{Tr}\{ \left[ f_l(\varepsilon)\Gamma_L-f_R(\varepsilon)\Gamma_R\right]\left(G^r-G^a\right)+\left(\Gamma_L-\Gamma_R\right)G^<,
\end{equation}
which is called by the Meir-Wingreen formula obtained using the Keldysh technique (non-equilibrium Green function approach)~\cite{Meir1992}. In particular, for the case when partial level width are proportional~\cite{Meir1992}, it has a form:
\begin{equation}\label{IVMW2}
    I(V)=-\frac{2e}{h}\int d\varepsilon\left[ f_L(\varepsilon)-f_R(\varepsilon)\right]\text{Tr}\{T\text{Im}G^r\}.
\end{equation}
Moreover, this formula was enlarged to the time-dependent electron transport in Refs.~\cite{Jauho1994,Wingreen1993}.
Here Tr denotes the trace operation, transmission coefficient $T$ now is a matrix, and $G^{r,a,<}$ is a retarded, advanced, or Keldysh Green function, respectively,  which is a correlation function with QD operators. For time-independent case and single-level quantum dot, the retarded Green function has the following form~\cite{Jauho1994, Krive2010} in the energy representation,
\begin{equation}\label{Grsl}
    G^r(\varepsilon)=\left[ \varepsilon-\varepsilon_0-\Lambda(\varepsilon)+\imath\Gamma(\varepsilon) \right]^{-1},
\end{equation}
where $\Lambda(\varepsilon)$ and $\Gamma(\varepsilon)$ are the shift and broadening of the energy level (linewidth) of the quantum dot, $\varepsilon_0$~\cite{Krive2010, Wingreen1993}. For the non-interacting electrons in the wide band approximation~\cite{Wingreen1989,Glazman1988} (the level width is energy-independent) one gets from Eq.~(\ref{IVMV1}) the Landauer formula, Eq.~(\ref{IVLandauer}).

\subsection{Transport properties of molecular transistors.}\label{subsect1_1_2}

A single-molecular transistor (SMT) can be meant as a SET where the central part of the system is movable. It can be a macromolecule (like fullerene one) or a quantum nanowire (particularly, a carbon nanotube) placed between two massive electrodes. Quantitatively new effects arise in transport properties of such devices (as well as in molecular junctions) due to the electron-vibron coupling and new phenomena as polaronic effects and the phonon-assisted electron tunneling, mechanical instability and the electron shuttling come to pass. There are comprehend reviews on the topic of polaronic effects, see, e.g., Refs.~\cite{Krive2010,Lindsay2007, Galperin2007}. Thus, in this subsection we only introduce the basic concepts and review recent achievement in this field, where the vibrons (quants of the QD oscillations) are associated with a mode unrelated to the direction of the electron tunneling~\cite{Krive2010}. However, the electron shuttling will be considered in the section~\ref{sect1_3}. 

Tunneling spectroscopy is a well-known method to study of
electron-phonon interaction in bulk metals (see, e.g.,
Ref.~\cite{Naidyuk2005}). Electron transport spectroscopy can be used
for studying of vibration properties of molecules in
single-molecule-based transistors~\cite{Park2000,Utko2010}.
Current-voltage characteristics of single-electron transistors, where fullerene molecule~\cite{Park2000}, suspended
single-wall carbon nanotube~\cite{Poot2012, Leturcq2009, Babic2003} or
carbon nano-peapod~\cite{Utko2010} are used as a base element,
demonstrate at low temperatures additional sharp features (steps)
at bias voltages $eV_n\simeq n\hbar\omega$ ($\omega$ is the
angular frequency of vibrational degree of freedom). The simplest
models (see, e.g., review Ref.~\cite{Krive2010}) that describe
step-like behavior of $I-V$ curves are based, as a rule, on a
theory where phonon excitations are dispersion-less (vibrons with
a single frequency) and they are assumed to be in equilibrium with
the heat bath at temperature $T$ (bulk metallic electrodes can
play the role of this heat bath). Steps in current-voltage
dependencies (equidistant peaks in differential conductance) are
associated with the opening of inelastic channels of electron
tunneling through vibrating quantum dot. For strong
electron-vibron interaction these models predict: 
\renewcommand{\labelitemi}{}
\begin{itemize}
    \item (i) Franck-Condon blockade~\cite{Koch2005} (exponential suppression)
of conductance at low temperatures $T\ll\hbar\omega$ ,
    \item (ii) non-monotonous temperature dependence of the differential linear conductance.
\end{itemize}
All these
effects were observed in experiments with molecular transistors based on a single fullerene molecule~\cite{Park2000} or carbon nanopeapods~\cite{Utko2010}.

The minimal model that describes the effects consists of a vibrating single-level quantum dot placed between two bulk normal metal electrodes biased by a constant voltage. The position of the QD energy level is tuned by a gate voltage (usually, in a way to obtain the maximal current, $\varepsilon_0(V_G)=\varepsilon_F$). For simplicity let us firstly consider spinless electrons (non-interacting). The Hamiltonian of such a system (tunnel model or Anderson-Holstein Hamiltonian) includes the following parts,
\begin{equation}\label{mH1}
    H=H_l+H_d+H_{v}+H_{int}+H_t.
\end{equation}
Here $H_l$ is the Hamiltonian of non-interacting electrons in the leads $\kappa=L,R$,
\begin{equation}\label{ml}
H_l=\sum_{k\kappa} \varepsilon_{k\kappa} a^\dag_{k\kappa}a_{k\kappa},
\end{equation}
where $a_{k\kappa}^\dag (a_{k\kappa})$ is the creation (annihilation) electron operator with the standard anti-commutation relation. 
In the Hamiltonian of the single-level quantum dot,
\begin{equation}\label{mdot}
    H_d=\varepsilon_0 c^\dag c,
\end{equation}
$c^\dag (c)$ is the creation (annihilation) operator of the electron state in the QD with the energy $\varepsilon_d$.
The vibrational subsystem is descried by the harmonic oscillator Hamiltonian,
\begin{equation}\label{mv}
    H_v=\frac{p^2}{2m}+\frac{m\omega^2x^2}{2},
\end{equation}
with the canonically conjugate operators of coordinate and momentum, $[x,p]=\imath\hbar$.
The electron-vibron interaction Hamiltonian reads as follows
\begin{equation}\label{mint}
    H_{int}=\Delta xc^\dag c.
\end{equation}
The tunnel Hamiltonian is the following,
\begin{equation}\label{mt1}
    H_t=\sum_{k\kappa}t_\kappa a^\dag_{k\kappa} c+\text{H.c.},
\end{equation}
where $t_\kappa$ is the tunneling amplitude which we consider in the symmetric case, $t_L=t_R=t_0$, for simplicity in what follows. The case of position-dependent tunneling amplitude will be taken into account in the section~\ref{sect1_3}.

In order to diagonalize the Hamiltonian, Eq.~(\ref{mH1}), one can perform the unitary Lang-Firsov~\cite{Lang1963} or "small polaron"~\cite{Krive2010} transformation,
\begin{equation}\label{mV}
    \hat{V}=\exp{[\imath\lambda pc^\dag c]},
\end{equation}
with $\lambda=\Delta/\hbar m\omega^2$ is the electron-vibron coupling constant. After this transformation we get the re-normalization of the tunneling amplitude as follows
\begin{equation}\label{mt2}
      H_t\to H_t=\sum_{k\kappa}t_0\text{e}^{-\imath\lambda p} a^\dag_{k\kappa} c+\text{H.c.},
\end{equation}
and shift (decrease) of the electron energy in a vibrating QD is known as the polaronic shift, $\varepsilon_p= \varepsilon_d-\lambda^2\hbar\omega$, see, e.g., Refs.~\cite{Krive2010,Galperin2007,Koch2005,Shkop2021}, and the electron-vibron bound state is called by the \textit{polaron} one.

There are several approaches to calculate transport trough a single-molecular transistor, such as, equation of motion (EOM) method~\cite{Fedorets2003}, master equation~\cite{Krive2010} using the Fermi Golden rule, Keldysh technique~\cite{Koch2005,Meir1992}, or density matrix approximation~\cite{Novotny2003}. However, in all these methods one of the crucial point is how to treat electron-vibron correlations in the averaging procedure. The usual approach which is valid in the perturbation theory over the small parameter of the electron level width (small junction transparency), is to disregard correlations between electrons and vibrons and evaluate the averages with the Hamiltonian of the non-interacting vibrons or with fermion art of the Hamiltonian for the fermionic averages~\cite{Krive2010}. Even so, one needs to be accurate in the considering of regimes of the transport within the validity of the perturbation theory because some of the results can become questionable~\cite{Lundin2002}. Also, it is a usual assumption to take into account the case of the strong coupling of the vibrons to bosonic environment (heat bath with a temperature $T$) so that the process of the equilibration of their distribution function is sufficiently fast (faster than the time corresponding to a tunneling event)~\cite{Krive2010}. The main average one needs to calculate is a correlation function with exponential function of the vibron operators. It can be done via several methods, see, e.g., Ref.~\cite{Mahan}, where the Feynman disentangling of operators technique is well explained. Moreover, it is convenient to use the well-known Campbell-Baker-Hausdorff-Dynkin formula. The result of the calculations of the correlation functions is the following~\cite{Mahan,Koch2005,Skorobagatko2011,Shkop2021,Krive2010},
\begin{equation}\label{mAn1}
    \langle \text{e}^{\mp \imath\lambda\hat{p}(t)}\text{e}^{\pm \imath\lambda\hat{p}(t')}\rangle_0=\sum_{n=-\infty}^{+\infty}A_n\text{e}^{-\imath\omega n (t-t')},
\end{equation}
where $\langle\dots\rangle_0$ indicates averaging with the vibron equilibrium distribution function (density matrix which has the Gibbs form), and
\begin{equation}\label{mAn2}
    A_n=\text{e}^{-\lambda^2(1+n_B)}I_n(z)\text{e}^{-n\omega/(2T)},
\end{equation}
with the normalization condition $\sum_n A_n=1$. In this equation $I_n(z)$ is the modified Bessel function of the first kind~\cite{Grandshtein2014}, $z=2\lambda^2\sqrt{n_B(1+n_B)}$, where $n_B$ is the Bose-Einstein distribution function,
\begin{equation}\label{mnb}
    n_B=1/\left[\text{e}^{\hbar\omega/T}-1\right].
\end{equation}
Then for the electric current, using, for example, the Meir-Wingreen formula, Eq.~(\ref{IVMW2}), we get the following expression~\cite{Lundin2002},
\begin{equation}\label{mIv1}
    I(V)=-\frac{e}{h}\text{e}^{-\lambda^2(1+2n_B)}\sum_{n=-\infty}^{+\infty}I_n\text{e}^{-n\omega/(2T)}\int d\varepsilon T^n_{BW}(\varepsilon)\left[f_L(\varepsilon)-f_R(\varepsilon)\right],
\end{equation}
witch is presented in the form of the sum over the inelastic vibron channels, and now the Breit-Wigner transmission coefficient reads as
\begin{equation}\label{mTBW2}
    T_{BW}^n=\frac{\Gamma^2}{(\varepsilon-\varepsilon_p+n\hbar\omega)^2+\Gamma^2}.
\end{equation}

By analyzing Eq.~(\ref{mIv1}), one can conclude several features in the electron transport due to the electron-vibron coupling. Firstly, the appearance of inelastic resonant channels for electron tunneling with emission ($n<0$) or absorption ($n>0$) of vibrons. This is associated with the so-called \textit{phonon-assisted tunneling}. As a consequence, the current-voltage characteristics of a molecular transistor at low temperatures are step-like functions (see, e.g., Fig.~\ref{fig:fig2_2}). Hence, each step (Franck-Condon step) corresponds to the opening of new inelastic channel for electron tunneling when the bias voltage increase (more inelastic channels enter the "conducting window"). Secondly, for sufficiently ($\lambda\gtrsim 1$) strong electron-vibron interaction, the electric current is extremely suppressed at low temperatures and low bias voltages (see, e.g., Fig.~\ref{fig:fig2_2}). This effect is refereed to the Franck-Condon~\cite{Koch2005} or polaronic blockade~\cite{Krive2010}. And the important point here that this blockade cannot be lifted via tuning the gate voltage as it is for the Coulomb blockade. Also, the electron-vibron coupling results in the non-monotonic temperature dependence of the linear conductance due to the polaronic narrowing of the QD energy level width $\Gamma$,
\begin{equation}\label{mGT}
    G_\lambda(T)\propto G_0(T)\text{e}^{-\lambda^2},
\end{equation}
for the low temperatures, $\Gamma\ll T\ll\hbar\omega$. In Equation~(\ref{mGT}), $G_0=\Gamma/T$ is the high-temperature conductance through a unmovable single-level quantum dot.

As a next step let us take into account the electron spin. It is easy to include in the Hamiltonian, Eq.~(\ref{mH1}), spin $\sigma=\uparrow,\downarrow$, degree of freedom, and add the term corresponding to the Coulomb interaction (with the strength $U$) on the QD,
\begin{equation}\label{mHU}
    H_U=Uc^\dag_\uparrow c_\uparrow c^\dag_\downarrow c_\downarrow .
\end{equation}
Thus, the unitary transformation, Eq.~(\ref{mV}), results in the re-normalization of the Coulomb interaction strength, $U_p=U-2\lambda^2\hbar\omega$. We can see, that this may lead to the Coulomb attraction case for strong electron-vibron interaction, $U<2\lambda^2\hbar\omega$. However, this case needs special consideration~\cite{vonOppen2008}. All the features we have been discussed for spinless electron remain in presence of the electron-electron interaction. 

In Ref.~\cite{Pasupathy2005} in molecular transistors made from $C_{140}$ fullerene molecules, the vibration-assisted tunneling associated with an internal
stretching mode of the molecule was observed, see also Refs.~\cite{Leturcq2009,Park2000}. The strong coupling of this
mode to tunneling electrons, relative to the other molecular
modes, is consistent with molecular modeling. Variations in
the measured strength of vibration-assisted tunneling between
different devices are presented~\cite{Pasupathy2005}.

In a recent paper, Ref.~\cite{Shkop2021}, the transport properties of a single-molecular transistor with the spin-polarized leads was considered in the presence of not only the electron-vibron and Coulomb interaction but the influence of the magnetic field directed perpendicular to the current flow was taken into account. The case of fully spin-polarized leads allows to emphasize the effect emerged due to the interplay between the above-mentioned phenomena. The magnetic field, which induces the spin-flip processes on the quantum dot, leads to the lifting of spin blockade in the spintronic device. The term in the Hamiltonian emerged due to the Zeeman effect (splitting) in the magnetic field $H$ reads as
\begin{equation}\label{mHh}
    H_H=-\frac{g\mu_B H}{2}\left( c^\dag_\uparrow c_\downarrow+c^\dag_\downarrow c_\uparrow\right),
\end{equation}
where $g$ and $\mu_B$ stand for the gyromagnetic ratio and the Bohr magneton, respectively. This non-diagonal term in the Hamiltonian can be vanished out by the performing the canonical transformation of the dot fermionic operators~\cite{Zubov2018, Shkop2021,BahrovaC2018,ShkopC2018,Shkop2018,ShkopC2019} which results in the re-normalization of the dot energy level (splitting) $\varepsilon_{1,2}=\varepsilon_d\pm g\mu_B H/2$, and tunneling amplitude, 
\begin{equation}\label{mHt2}
    H_t\to H_t=\frac{t_0}{\sqrt{2}}\text{e}^{-\imath\lambda\hat{p}}\sum_{k\kappa}a^\dag_{k\kappa}\left( j_\kappa d_1+d_2\right)+\text{H.c.},
\end{equation}
where $j_{L,R}=\pm 1$ and $d_{1,2}$ is the new dot operator. The reduced density matrix technique and the perturbation theory over the QD level width was used in Ref.~\cite{Shkop2021}, see also Ref.~\cite{ShkopC2020}, in order to calculate the transport properties of the system. It has been obtained that the current-voltage dependencies have doubled
number of Franck-Condon steps compared to a conventional molecular transistor. 
Every voltage interval $eV=2\hbar\omega$ has two steps. The doubling
is explained by the fact that the system with the Zeeman splitting has doubled number of
elastic channels, with the inelastic channels associated with each of them~\cite{Shkop2021}. The doubling of the steps can also be observed in the presence of the Coulomb interaction. Moreover, it is found that the lifting of the Coulomb
blockade by the bias voltage proceeds in stages, so that there are two elastic channels for
tunneling of the second electron to the quantum dot, and one of these channels opens
earlier than the other in energy. The steps are separated by the voltage interval equal to
energy splitting in the external magnetic field~\cite{Shkop2021}. In addition, it is obtained that for strong electron-vibron interaction, $\lambda\gtrsim 1$, the temperature dependence of the linear conductance is non-monotonic and anomalous growth of a conductance maximum depends on the Coulomb interaction strength as well as on the external magnetic field~\cite{Shkop2021}. Also, thermoelectric properties of the device under consideration in Ref.~\cite{Shkop2021} and with the temperature drop through the system are studied in Ref.~\cite{Shkop2022}.

In addition, transport properties of the following spintronic device are investigated in Ref.~\cite{Zubov2018}. (Let us omit the electron-vibron interaction for a moment.) Within the model under consideration, a unmovable QD is placed between magnetic fully spin-polarized (for simplicity) leads (half-metals) witch are held at different temperatures and chemical potentials (tuned by a bias voltage). In such a setup the effect of spin blockade occurs: an electron with the spin $\uparrow$ cannot tunnel to the lead with the spin polarization $\downarrow$ and vice versa. However, an external magnetic filed applied perpendicular to the lead magnetization, induces spin-flip processes of an electron in the quantum dot. An arbitrary direction of the magnetic field was considered in Ref.~\cite{Pedersen2005}, where the case of non-interacting as well as interacting electron was taken into account and the dependence of the conductance was found using the Green function approach. The equation of motion method was used in Ref.~\cite{Zubov2018} to calculate the electric and heat currents in case of non-interacting electrons explicitly. It was shown that in an optimal regime the figure of merit ($ZT$) of the proposed spintronic device is essentially enhanced~\cite{Zubov2018} in comparison with the analogous device with unpolarized electrons~\cite{Kennes2013}. In particular, it was shown~\cite{Zubov2018} that (in the simplest case) the electric current (transmission coefficient) has the following dependence of the magnetic field, see also Refs.~\cite{BahrovaC2018},
\begin{equation}\label{mIh}
    I=I_0 \frac{h^2/2}{h^2+\Gamma^2}\left[f_L^+-f_R^+\right],
\end{equation}
where $I_0=e\Gamma$ is the maximal current through the SET with an unmovable dot and $h=g\mu_B H$ and $2f_{L,R}^+=f_{L,R}(\varepsilon_1)+f_{L,R}(\varepsilon_2)$.
Also, the influence of the Coulomb interaction on thermoelectric properties was calculated using the density matrix approximation. It has been obtained that in the Coulomb blockade regime the figure of merit is not suppressed due to electron-electron interaction~\cite{Zubov2018}.

Nevertheless, when coupling of vibron subsystem to the heat bath is weak and
vibrons are not in equilibrium during the time of electron
tunneling through the system, their density matrix can not be in
the Gibbs form and it has to be evaluated from the solution of
kinetic equations. This problem can be solved only numerically
(see, e.g., Ref.~\cite{Kim2003}). There are only few papers~\cite{Mitra2004,Liu2019,Liu2019a,Boese2001,McCarthy2003}, where vibrons in electron transport through a SET were
considered as non-equilibrated. In Ref.~\cite{Liu2019a} it was assumed
that vibron subsystem is in a coherent state. In the approach used
in the cited paper, the density matrix of coherent state was
time-independent, that contradicts Liuville-von Neumann equation
for density matrix of {\it non-interacting} vibrons. Therefore the
results of this approach are questionable and the problem of
electron transport through a vibrating quantum dot with coherent
vibrons has to be re-examined.

It is worth to mention about one case of non-interacting vibrons, particularly, when the vibron subsystem is in a Fock state with the vibron number $n$ or in a superposition of ones. In Ref.~\cite{Chu2018} it is reported about an experimental generation of the multi-phonon Fock states in a bulk acoustic-wave resonator with a sufficient fidelity (up to $n=8$). A Wigner tomography and state reconstruction to highlight the quantum nature of the prepared states was also performed~\cite{Chu2018}. Thus, in the case when the density matrix of mechanical subsystem describes the one being in a Fock state, the correlation function, Eq.~(\ref{mAn1}), has the following form~\cite{Liu2019,Mahan},
\begin{eqnarray}\label{mAn3}
       && \langle n\vert\text{e}^{\mp \imath\lambda\hat{p}(t)}\text{e}^{\pm \imath\lambda\hat{p}(t')}\vert n\rangle=\text{e}^{-\lambda^2[1-\text{e}^{\imath\omega (t-t')}]}L_n\left[2\lambda^2 \left(1-\cos{\{\omega (t-t')\}}\right)\right]=\nonumber\\
     &&\hspace{4.5cm}=\sum_{m=0}^n\sum_{k=0}^\infty A_{mk}^n\text{e}^{\imath\omega [m-k](t-t')},
\end{eqnarray}
with 
\begin{equation}\label{mAn4}
    A^n_{mk}=\text{e}^{-\lambda^2}\frac{(-1)^{m+k} n!}{(m!)^2 (n-m)!}(2\lambda)^{2m}L_k^{2m-k}(\lambda^2),
\end{equation}
where $L_i^j(z)$ is a generalized (associated) Laguerre polynomial~\cite{Grandshtein2014}. Note that Eq.~(\ref{mAn1}) can be obtained from Eq.~(\ref{mAn3}) by the summation oven $n$ with corresponding coefficients~\cite{Mahan,Beitmen1966}.

In contrast, in the next chapter~\ref{chapt2} we will consider single-electron transistor with vibrating
quantum dot, where vibronic subsystem is described by \textit{time-dependent} density matrix. Physically this approach corresponds to
coherent oscillations of quantum dot treated as harmonic quantum
oscillator. Coherent states of harmonic oscillators are well known
in physics (see, e.g., Refs.~\cite{Gazeau2009,Baz}). In tunnel electron transport
they appear, for instance, in weak superconductivity
(Josephson current through a vibrating quantum dot, see
Ref.~\cite{Zazunov2006} and referencies therein). Last years coherent
states of photons ("Schroedinger-cat" states) coupled to qubits
and qubits formed by the coherent photon states became a hot topic
of studies in quantum computing science, see, e.g., review
Ref.~\cite{Girvin2017}.

\section{Coherence effects in electron transport through a nanoelectromechanical system.}\label{sect1_2}

Electro-mechanical phenomena on the nanometer scale attract
significant attention during the last two decades~\cite{Ekincia2005}.
Recent advantages in nanotechnologies acquire a promising platform 
for studying the fundamental phenomena generated by the interplay between quasi-classical and
pure quantum subsystems. A charge qubit formed by a tiny
superconducting island (Cooper-pair box (CPB)) whose basis states
are charge states (e.g. states which represent the presence or
absence of excess Cooper pairs on the island), is one of a large
group of pure quantum systems~\cite{Nakamura1999}. 
There are many types of solid-state systems which qubit based on, such as quantum superconducting circuits (including biased Josephson junctions, SQUIDs and CPBs), see, e.g., review Refs.~\cite{Wendin2007,Pashkin2009}; quantum dots~\cite{Ranni2021,Hensen2019} and atomic ones~\cite{Dupont2013}.

In general, a qubit is one of the physical realizations of a two-level system~\cite{Ivakhnenko2022}, including ultracold atoms, classical nanomechanical resonators and semiconductor microcavities, where extremely controllable qubits can be realized on the exciton-polariton condensates~\cite{Xue2021,Ghosh2020}. One of the main features related to a two-level system is the fact that it usually exhibits an avoided level crossing (anticrossing) of its energy levels as an external parameter is varied~\cite{Shevchenko2010}. A driven two-level system is described by a standard Hamiltonian, 
\begin{equation}\label{HTLS}
    H=-\frac{\Delta}{2}\sigma_x-\frac{\varepsilon(t)}{2}\sigma_z,
\end{equation}
where $\varepsilon(t)$ is a bias energy and $\Delta$ stands for an energy gap~\cite{Shevchenko2019}. By solving the time-dependent Schr{\"o}dinger equation for linearly driven system, $\varepsilon (t)=v t$, one gets the following expression for the transition probability that is the probability to find the system in the excited state known as Landau-Zener-St{\"u}ckelberg-Majorana (LZSM) formula,
\begin{equation}\label{PLZSM}
    P_{LZSM}=\text{e}^{-2\pi\delta},
\end{equation}
with $\delta=\Delta^2/(2\hbar v)$ being the adiabaticity parameter, see review Ref.~\cite{Ivakhnenko2022}. The non-linear driving is considered in Ref.~\cite{Ashhab2022}.
Additionally, the LZSM transitions in a periodically driven  Cooper-pair box system was investigated~\cite{Parafilo2018}.
At the same time
modern nanomechanical resonators which dynamics according to
Ehrenfest theorem to great extent is described by classical
equations, are ideal representatives of quasiclassical subsystem~\cite{Schmid2016}. Systems, which dynamics is determined by the mutual influence between a superconducting qubit and a nanomechanical resonator, are a subject of cutting-edge research in quantum physics, especially, in quantum communication, see, e.g., Refs.~\cite{Satzinger2018,Bienfait2019,Chu2017,Chou2020,LaHaye2009,Tian2005}

There are two main questions that arise related to an
interplay between quasi-classical dynamics of the mechanical
resonator and quantum dynamics of the charge qubit. The first one
is: how quasi-classical motion may affect pure quantum
phenomena? Considering this question, it was shown that the
superconducting current between two remote superconductors can be
established by mechanical transportation of Cooper pairs performed
by an oscillating CPB~\cite{Gorelik2001}. Even more, it was
demonstrated that such transportation could generate correlations
between the phases of space-separated superconductors~\cite{Isacsson2002}. Another question is how coherent Josephson dynamics
of a charge qubit will affect the dynamics of the quasi-classical
resonator, in particular, whether or not the quantum
entanglement between a superconducting qubit and mechanical vibrations can be
achieved?  Recently it was demonstrated that individual phonons can
be controlled and detected by a superconducting qubit enabling
coherent generation and registration of quantum superposition of
zero and one-phonon Fock states~\cite{Satzinger2018, Bienfait2019}.  At the
same time nanomechanical resonators provide the possibility to store
quantum information in the complex multi-phonon coherent states.
Such states, in contrast to single-phonon states, where mechanical
losses irreversibly delete the quantum information, allow their
detection and correction~\cite{Hann2019, Girvin2017}. 

A huge challenge in the realisation of full-scale quantum computer systems is controlling qubits in an error-free way. Quantum error correction (QEC) protocols offer a solution to this
problem, in principle allowing for arbitrary suppression of the logical error rate provided certain
threshold conditions on the physical qubits are met~\cite{Roffe2019}. For now several quantum error correction methods are proposed. QEC codes~\cite{Devoret2013} based on cat states are widely used~\cite{Girvin2017}. Because of phases are more robust against photon loss errors, information is typically encoded in the phase of a coherent state. In analogy to classical phase-shift keying, quantum information can also be encoded to the phase of a coherent state. The simplest code (two-component cat code) is thus to use two coherent states with opposite phases that is, the cat states~\cite{Cai2021}. Another QEC protocols are GKP codes. They are quantum error-correcting codes that protect a state of a finite-dimensional quantum system (qudit) that is encoded in
an infinite-dimensional system (harmonic oscillator)~\cite{Gottesman2001}. For a typical two-level logical qubit, the GKP code is defined as coherent superposition of infinitely squeezed states or the eigenstates of the position operator $\hat{x}$  with a spacing of $2/\sqrt{\pi}$~\cite{Cai2021}. Besides, recently binomial codes for QEC were proposed~\cite{Michael2016,Hu2019}. These “binomial quantum codes” are formed from a finite superposition of Fock states weighted with binomial coefficients. It was shown that the binomial codes are protected to given order in the time step against continuous dissipative evolution under loss, gain, and dephasing errors~\cite{Michael2016}.

Motivated by such a
challenge, in the chapter~\ref{chapt3} we will discuss the possibility to generate
quantum entanglement between the charge qubit states and mechanical coherent ones in
a particular nanoelectromechanical system (NEMS) where mechanical
vibrations are highly affected due to the weak coupling with movable a
Cooper-pair box.
Moreover, a protocol of bias
voltage manipulation which results in the formation of entangled
states incorporating so-called cat-states (the quantum superposition
of the coherent states) which are robust in manipulation, is proposed.

\section{Mechanical instability in nanoelectromechanical devices.} \label{sect1_3}

In contrast to the previous sections~\ref{sect1_1},~\ref{sect1_2}, where we have considered the influence of electron-vibron interaction on transport properties of single-molecule transistors within the approach when this coupling is associated with mechanical modes unrelated to the direction of the electron flow, in this section we pay attention to the case when the position of a QD between the toward the leads exponentially modifies the tunneling probability~\cite{Krive2010}. This can result in the fact that an equilibrium position of the QD is no more stable, that is, mechanical instability and electron shuttling regime~\cite{Gorelik1998} can take place. This is usually the case of weak electromechanical coupling, the influence of polaronic effect on electron shuttling phenomenon was studied in Ref.~\cite{Skorobagatko2009,Parafilo2012}. There are several comprehend reviews on the topic, see Refs.~\cite{Shekhter2014,Shekhter2003,Shekhter2007,Shekhter2009, Parafilo2012}. Thus, we will briefly introduce effects of mechanical instability in such a system and review some recent results.

The simplest model that can catch the electric shuttle and mechanical instability effects, is described by the Hamiltonian Eq.~(\ref{mH1}), where now the tunneling amplitude is position-dependent and the term Eq.~(\ref{mint}) can be presented as $H_{int}=(\varepsilon_d-e\mathcal{E}x)c^\dag c$, where $\mathcal{E}$ is an electric field due to presence an electron on the quantum dot,
\begin{equation}\label{mHt3}
    H_t=\sum_{k\kappa}t_\kappa (\hat{x})a^\dag_{k\kappa}c+\text{H.c.},
\end{equation}
with $t_\kappa=t_0\text{e}^{\pm \hat{x}/\lambda}$, where $\lambda$ is the tunneling length. 
In order to solve the problem, the Liouville-von Neumann equation (or, the Lindblad equation, more generally) is used as well as Green function approach in the perturbation theory over the parameter of the electromechanical coupling. In this case if we neglect the effects of zero-point fluctuations of a quantum dot, we can use the semi-classical treatment within which $\langle \hat{x}\rangle=x$ and we are interested in the big values of the dot oscillation amplitude. Hence, the dot coordinate is governed by the Newton equation,
\begin{equation}\label{mNew}
    \ddot{x}+\omega^2 x=\mathcal{F}(t)/m,
\end{equation}
where the average force has a form:
\begin{equation}
    \mathcal{F}(t)=-\text{Tr}\left[ \hat{\rho}\frac{\partial H}{\partial x}\right].
\end{equation} 
It was shown analytically~\cite{Fedorets2002} that the mechanical (or shuttle) instability can occur, that is, the amplitude of the QD oscillations which being small after an initial fluctuation from an equilibrium position, start to grow exponentially with the increment $r_s\sim \lambda\Gamma$ if the bias voltage is bigger than the threshold one, $eV>2(\varepsilon_d+\hbar\omega)$. In this case it can develop into a limit cycle in presence of small but finite mechanical friction (term $\gamma\dot{x}$ in the l.h.s. of Eq.~(\ref{mNew}))~\cite{Fedorets2002,Fedorets2003}. In addition, the fully quantum-mechanical approach (treatment of a QD coordinate quantum-mechanically with the help of the Wigner function representation) can be used in order to investigate the mechanical instability. It was obtained that the Wigner function in the regime of developed self-oscillations has a circle-like form (not Gaussian shape), see Refs.~\cite{Novotny2003,Fedorets2004}. A sharp increase in current with the transition to the stationary regime was also obtained. Hence, in the limit cycle regime $I\sim e\omega$~\cite{Fedorets2004}.

The internal friction caused by the temperature drop $T$ (not a bias voltage) across the system in case of spinless electrons is considered in Ref.~\cite{Ilinskaya2018a}. The following temperature dependence of it was found,
\begin{equation}\label{mrT}
    \gamma (T)\sim T^{-1}\left[\cosh^2{\left\{\varepsilon_d/(2T)\right\}}\right]^{-1}.
\end{equation}

The next step in generalization is to take into account the electron spin. A spintronic nanoelectromechanical single-electron transistor with spin-polarized leads is considered in Ref.~\cite{Fedorets2004}. The density matrix approximation and the high bias voltage limit was used to find a steady-state solution. It was found that there are two types of transitions between steady states when the electric or magnetic field is varied~\cite{Fedorets2004}. In addition, the hysteresis behavior of a steady-state amplitude and electric current in the hard transition regime was obtained.
The so-called spin-mechanical coupling was considered in Refs.~\cite{Parafilo2016, Atalaya2012}. The semi-classical approach was used to derive the increment of the dot oscillation growth in the mechanically unstable regime in the system under periodic magnetic field~\cite{Parafilo2016}. An opposite regime of the mechanical ground-state cooling was proposed in Ref.~\cite{Atalaya2012}. 

Another type of electromechanical coupling in the magnetic shuttle structures is based on the ferromagnetic exchange coupling between a QD and magnetic leads. The region of the mechanical instability for such a nanoelectromechanical device with the spin-polarized leads and for interacting electrons was obtained in Ref.~\cite{Kulinich2014}, see also Refs.~\cite{Ilinskaya2015,Ilinskaya2015a, Ilinskaya2023}. It ha           
          s been found that a shuttle regime of the electron transport occurs at sufficiently low magnetic field strength ($h\ll \Gamma$) in contrast to the electric one. 
This setup under the temperature drop across the system was investigated in Refs.~\cite{Ilinskaya2020,Ilinskaya2019,Ilinskaya2019a,Ilinskaya2018}. The region of the mechanical instability was obtained analytically in the adiabatic limit within the semi-classical approach. It is shown that the shuttle instability occurs in the region of external magnetic fields between a lower, which depends only on the phenomenological friction, and upper (which is temperature-dependent and saturates at the value $h_{c2}/\hbar\omega=\sqrt{7/2}$ at high temperatures) critical values~\cite{Ilinskaya2018}. The regime of the instability does not emerge at high values of the magnetic field strength because in this case the spin-flip time exceeds the characteristic time scales determined by mechanical ($\omega^{-1}$) and electronic ($\hbar/\Gamma$) time scales~\cite{Ilinskaya2018}. Moreover, temperature dependence of the friction coefficient remains the same as for the case of spinless electrons, Eq.~(\ref{mrT}). The effects of Coulomb interaction on the mechanical instability of the magnetic shuttle device are described in Refs.~\cite{Ilinskaya2019, Ilinskaya2019a}. It was shown that such a spintro-mechanical shuttle instability can be triggered by the electron-electron repulsion. The critical value of the strength of this interaction crucially depends on the temperature and the strength of the magnetic field~\cite{Ilinskaya2019a}. Also, the self-saturation effect was predicted for a magnetic shuttle device~\cite{Ilinskaya2019a,Kulinich2014}. This effect come out in the fact of the presence of the stationary regime of mechanical self-oscillations even without the influence of an external friction determined by the quality factor of a nanomechanical system bath as it is for an electrical one. In addition, Coulomb correlation effects in a thermally driven and voltage-biased magnetic devices were investigated numerically in Ref.~\cite{Ilinskaya2019}. It has been obtained that thermally induced magnetic shuttling of spin-polarized electrons is a threshold phenomenon~\cite{Ilinskaya2019} as it is for electric shuttle device. Eventually, in Ref.~\cite{Ilinskaya2020} electric and magnetic exchange forces was taken into account at the same time. It leads to the several non-trivial effects which can be seen in an experiments with electric current measurements as such a possibility has been demonstrated in Ref.~\cite{Ilinskaya2020} by obtaining numerically \textit{I-V} curves of the considered spintro-mechanical transistor. The non-monotonic dependence of the differential conductance in the stable (vibronic) regime was obtained, on the one hand. On the other hand, the effect of negative differential conductance in the stationary regime of mechanical self-oscillation was shown~\cite{Ilinskaya2020}.

There are a number of the experiments where the regime of the mechanical instability in nanoelectromechanical systems was observed, see, e.g., Refs.~\cite{Vincent2007,Scheible2004, Kim2010,Kim2012,Zhao2016,Chan2011,Moskalenko2009,Steele2009,Etaki2013}. A coherent spin shuttle processes in a GaAs/AlGaAs quantum dot array was considered in Ref.~\cite{Fujita2017}.

The effect of self-sustained oscillations is itself an interesting problem from a fundamental point of view, opening new
possibilities for mass and force sensing~\cite{Jensen2008,Braakman2019}, while its underlying physical processes show potential
applications for mechanical cooling~\cite{Urgell2019}. Self-sustained mechanical oscillations were first
observed in a carbon nanotube (CNT)-based transistor~\cite{Steele2009}, with further studies later verifying
their transport signatures~\cite{Schmid2012,Schmid2015,Weldon2010}. Recently, the experimental observation of self-driven
oscillations of a CNT-based quantum dot in the Coulomb blockade regime has been reported~\cite{Willick2020}.

Nevertheless, superconducting (SC) elements incorporated into nanoelectromechanical systems extend the horizon of this phenomenon, namely through the effects of superconducting phase coherence; see, e.g., the following reviews~\cite{Parafilo2012, Meden2019}.  A SC electrode located near a quantum dot can affect its electronic state via the tunneling exchange of Cooper pairs due to SC proximity effect. In Ref.~\cite{Gorelik2001} (see also~\cite{Isacsson2002}) it was demonstrated that a movable Cooper-pair box oscillating periodically between two remote superconducting electrodes can serve as a mediator of Josephson coupling leading to coherent transfer of Cooper pairs between the SC leads. The polaronic effects influence on the Josephson current in a S-QD-S system was considered in Ref.~\cite{Parafilo2013}. Also, analogously to a system in a normal state where the lifting of the Franck-Condon blockade is leads to the non-monotonic temperature dependence of the differential conductance~\cite{Utko2010,Krive2008}, for the SC one is accompanied by non-monotonic temperature dependence of the critical Josephson current~\cite{Parafilo2014, Parafilo2013, Shekhter2013}. The polaronic narrowing of the Josephson critical current was considered in Refs.~\cite{Novotny2005,Skoldberg2008,Zazunov2006}.

Furthermore, if the tunneling amplitude depends on the distance between the QD and the SC leads, such exchange also provides a connection between the electronic and mechanical degrees of freedom.
 Additional injection of electrons from a biased normal metal electrode into the QD generates peculiar dynamics of Cooper pairs on it. Interplay between electromechanical effects and phase coherence gives new and
unusual properties to a number of normal metal/superconducting hybrid junctions~\cite{Parafilo2020,Baranski2015, Baran2019, Moghaddam2012}.
In particular, it has recently been shown that in a normal metal--suspended CNT--superconductor
transistor, Andreev reflection~\cite{Andreev1964,Kulik1970} may give rise to a cooling of the mechanical subsystem~\cite{Stadler2016,Stadler2017,Rastelli2019} or
generate a single-atom lasing effect~\cite{Rastelli2019} if certain conditions are fulfilled. The resonant Andreev tunneling in a N-QD-S system was observed in Ref.~\cite{Gramich2015}.

The mechanical functionality of NEMS is to a large extent determined by the physical principles
underlying the interaction between the electronic and mechanical subsystems. In all studies
mentioned above, this interaction was due to the \textit{localization} of the charge~\cite{Fedorets2002,Parafilo2020} or spin~\cite{Kulinich2014, Ilinskaya2019a} carried by
electrons in the movable part of the system. In the chapter~\ref{chapt4}, we will consider a fundamentally new type
of electromechanical coupling based on the \textit{quantum delocalization} of Cooper pairs (see also Ref.~\cite{Parafilo2022}). We
demonstrate that such coupling can promote a self-saturated mechanical instability resulting in
the generation of \textit{self-sustained} mechanical oscillations. The effect of the \textit{ground-state cooling} of nanomechanical vibrations in the considered system is also proposed. It is also shown that regime of pumping or either cooling significantly
affect the average current through the system, making it possible to carry out direct
experimental detection.

\chapter{POLARONIC EFFECTS INDUCED BY NON-EQUILIBRIUM VIBRONS IN A SINGLE-MOLECULE TRANSISTOR} \label{chapt2}

In this chapter electron transport in a molecule transistor is considered in the assumption that the mechanical subsystem is in a non-equilibrium, namely, coherent state. The current-voltage characteristic of such a transistor based on a vibrating quantum dot are calculated. Also, the obtained electric current dependencies on the oscillation amplitude of the QD are analyzed. 

\section{Model of a single-electron transistor.} \label{sect2_1}

\begin{figure}
\centering
\includegraphics[width=0.45\columnwidth]{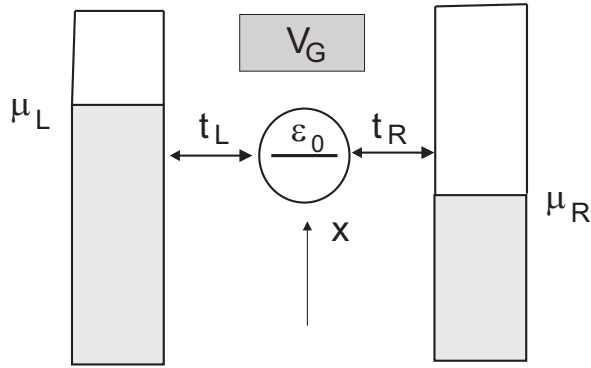}
\caption{\textit{Sketch of the single-electron transistor. A vibrating
one-level ($\varepsilon_0$ is the level energy) quantum dot
(macromolecule) is placed between two bulk electrodes biased by
the voltage $V$. The dot tunnel couples ($t_L=t_R=t_0$ is the
tunneling amplitude) to the leads with the chemical potentials
$\mu_{L,R},\;\mu_L-\mu_R=eV$ and the temperature $T$. The gate
voltage $V_G$ is set $\varepsilon_0(V_G)=\varepsilon_F$, where
$\varepsilon_F$ is the Fermi energy, to get maximal current. The
dot oscillates in $x$ direction perpendicular to the electric
current flow. QD oscillations are modelled  by the coherent state
of one-dimensional harmonic oscillator.}} \label{fig:fig2_1}
\end{figure}

The model device we are interesting in is depicted in Fig.~\ref{fig:fig2_1}. It
consist of two bulk electrodes, source (Left) and drain (Right)
leads, with chemical potential biased by voltage, $\mu_L-\mu_R=eV$,
and a single-level quantum dot, which oscillates in the
direction $x$ perpendicular to the direction of electron
current flow. The gate voltage $V_G$ is adjusted to obtain maximal tunnel
current, $\varepsilon_0(V_G)=\varepsilon_F$, where
$\varepsilon_0(V_G)$ is the dot level energy and $\varepsilon_F$
is the Fermi energy of the leads. For simplicity we consider
tunneling of spinless electrons in a symmetric junction and it is
assumed that the vibration of QD does not change tunneling matrix
elements $t_L=t_R=t_0$. Here we consider the process of
sequential electron tunneling, when $\max(eV, T)\gg\Gamma$, where
$\Gamma\propto |t_0|^2$ is the QD level width which is a characteristic energy
of dot-leads tunnel coupling. This model device can simulate, for
instance, a single-electron transistor based on a suspended single-wall carbon nanotube. 

\section{Hamiltonian of the system and equations for density matrix.} \label{sect2_2}

The Hamiltonian of the system, which is schematically illustrated in Fig.~\ref{fig:fig2_1}, consists of four terms,
\begin{equation}\label{1}
H= H_l+ H_{dot}+H_{v-d}+ H_{tun},
\end{equation}
where $H_l, H_{dot}$ are the Hamiltonians of the non-interacting
electrons in the leads and the dot, respectively,
\begin{equation}\label{2}
H_l=\sum_{k,\kappa}\varepsilon_{k,\kappa}
a^\dag_{k,\kappa}a_{k,\kappa},\, H_{dot}=\varepsilon_0 c^\dag c,
\end{equation}
$a_{k,\kappa}^\dag (a_{k,\kappa})$ is the creation (annihilation)
operator of an electron in the lead $\kappa=L,R$ with momentum $k$ and energy
$\varepsilon_{k,\kappa}$, $c^\dag(c)$ is the creation
(annihilation) operator of electron state in the dot with the
energy $\varepsilon_0$. These operators in the occupation-number representation (second quantization) obey the standard anti-commutation relations $\{a^\dag_{k,\kappa},a_{k',\kappa '}\}=\delta_{kk',\kappa\kappa '}$, where $\delta_{k,\kappa}$ is the Kronecker delta.

Hamiltonian $H_{v-d}$ describes the vibronic (mechanical) subsystem and the
interaction between electrons and vibrons,
\begin{equation}\label{4}
\hat H_{v-d}=\frac{p^2}{2m}+\frac{m\omega^2 x^2}{2}+\Delta
xc^\dag c.
\end{equation}
In Equation~(\ref{4}) $x,p$ are the canonical conjugating operators of
coordinate and momentum, $[x,p]=\imath \hbar$. Here $\omega, m$ are
the frequency of dot oscillations and the mass of the dot,
$\Delta$ is the electron-vibron coupling constant.

Hamiltonian $H_{tun}$ describes the tunnelling of electrons
between the dot and the leads and it takes the standard
form,
\begin{equation}\label{5}
 H_{tun}=\sum _{k,\kappa} t_\kappa a^\dag_{k,\kappa}c+\text{H.c.},
\end{equation}
where $t_\kappa$ is the tunnelling amplitude. In what follows we
restrict ourselves to the symmetric case, $t_L=t_R=t_0$, which does not affect the obtained results qualitatively. 

In order to diagonalize the Hamiltonian~(\ref{1}), it is convenient to perform the unitary transformation,
$UHU^\dag\rightarrow H$, where $U=\exp[i\lambda p c^\dag c]$, so-called Lang-Firsov canonical transformation or polaronic transformation~\cite{Mahan}. By equating coefficients of bosonic operators of the same power, one can obtain the re-normalized constant of the electron-vibron interaction, $\lambda=\Delta/\hbar m \omega^2$. Then Eq.~(\ref{4}) takes the diagonal form,
\begin{equation}\label{7}
H_{v-d}\rightarrow H_v=\frac{p^2}{2m}+\frac{m\omega^2 x^2}{2},
\end{equation}
while the tunnelling Hamiltonian $H_{tun}$ is transformed to the following one,
\begin{equation} \label{8}
H_{tun}\rightarrow H_{tun}=t_0\sum_{k,\kappa}\text e^{-\imath\lambda p}
a^\dag_{k,\kappa}c+\text{H.c.}.
\end{equation}

The quantum consideration of electron-vibron interacting system is
based in what follows on the approximation that the density matrix
of the system is factorized~\cite{Novotny2003} to direct product of the leads
equilibrium density matrix, the vibron  density matrix and the
density matrix of the dot,
\begin{equation}\label{9}
\rho \approx\rho_l\otimes\rho_v\otimes\rho_{dot}.
\end{equation}

This approximation corresponds to the case of sequential electron
tunneling, which holds when $\text{max}\{eV,T\}\gg \Gamma$, where
$\Gamma$ is the electron level width, $T$ is the temperature and
$V$ is the bias voltage. Here we assume that vibrons are described
by a time-dependent coherent state $\vert z(t) \rangle$. Note,
that in Ref.~\cite{Liu2019} current-voltage characteristics of a
single-electron transistor were calculated for time-independent
coherent state of vibrons. This assumption contradicts the equation
of motion of non-interacting vibrons in our model,
\begin{equation}
   \vert
z(t)\rangle=\exp\left(-\imath H_{v}t\right)\vert z\rangle, 
\end{equation}
$(\hbar=1)$. Here $\vert z \rangle$ is the eigenfunction of vibron
annihilation operator $b, b\vert z\rangle =z \vert z \rangle$ ($z$
is the complex number). The corresponding density matrix takes the
standard form,
\begin{equation}\label{93}
\rho_v(t)=\vert z(t)\rangle \langle z(t)\vert.
\end{equation}
Time evolution of the system is described by the Liouville-von Neumann equation for the density matrix,
\begin{equation}\label{10}
\frac{\partial \rho}{\partial t}+\imath[H_0+H_{tun},\rho]=0,
\end{equation}
where $H_0=H_l+H_v+H_{dot}$. It has the following formal solution,
\begin{equation}\label{11}
\rho(t)=\rho(-\infty)-\imath\int_{-\infty}^t dt'\text e^{-\imath
H_0(t-t')}[H_{tun},\rho(t')]\text e^{\imath H_0(t-t')}.
\end{equation}
Then by substituting Eqs.~(\ref{9}), (\ref{11}) into Eq.~(\ref{10}) and tracing out both the electronic degrees of
freedom of the leads and vibronic degrees of freedom of the dot, one gets the following equation for the reduced density matrix of the QD, $\rho_{dot}=\text{Tr}_{leads, v}\rho$,
\begin{eqnarray}\label{12}
\frac{\partial \rho_{dot}}{\partial
t}+\imath[H_{dot},\rho_{dot}]=-\text{Tr}\int_{-\infty}^t dt'[H_{tun}, \text e^{-\imath
H_0(t-t')}[H_{tun},\rho(t')]\text e^{\imath H_0(t-t')}].
\end{eqnarray}
Now we can explicitly calculate averages of electronic and
vibronic operators in the approximation of the factorized density
matrix, Eq.~(\ref{9}). For equilibrium density matrix of electrons
in the leads we use the standard expression,
\begin{equation}\label{13}
\langle a^\dag_{k,\kappa}a_{k',\kappa'}\rangle=
f_\kappa(\varepsilon_{k,\kappa})
\delta_{k,k'}\delta_{\kappa,\kappa'},
\end{equation}
where
$f_\kappa(\varepsilon)=(\exp((\varepsilon-\mu_\kappa)/T)+1)^{-1}$ is the Fermi-Dirac distribution function, $\mu_{L,R}= \mu_0\pm
(eV/2)$ is the electrochemical potential in the lead $\kappa$. Calculations of the vibronic correlation function
\begin{equation}
  F(t,t_1;\lambda)=\langle\exp[-\imath \lambda p(t)] \exp[\imath
\lambda p(t_1)]\rangle ,  
\end{equation}
in coherent state representation result in the equation,
\begin{eqnarray}\label{130}
&& F(t,t_1;\lambda)=\text{Tr[ e}^{-\imath \lambda p(t)} \vert
z\rangle\langle z\vert
\text e^{\imath \lambda p(t_1)}]= \nonumber \\
&&\exp\left\{-\lambda^2\left[1-\text e^{\imath
\omega(t-t_1)}\right]-\lambda z
\left[\text e^{-\imath \omega t}- \text e^{-\imath \omega
t_1}\right]+\lambda z^\ast\left[\text e^{\imath \omega t}- \text
e^{\imath \omega t_1}\right]\right\}.
\end{eqnarray}
Here we introduce the dimensionless constant of the electron-vibron interaction $\lambda\hbar\sqrt{2}/ l_0\rightarrow
\lambda$, where $l_0=\sqrt{\hbar/m\omega}$ is the amplitude of zero-point oscillations. The parameter $\lambda$ can be rewritten in the form $\lambda=\sqrt{2}l/l_0$, where $l=\Delta/m\omega^2$ is the characteristic displacement length of classical oscillator.
Note that in the case of averaging with a nonequilibrium vibronic density matrix, the using of a well-known formula, which still  $\langle\text{e}^{\hat{A}}\rangle=\text{e}^{\frac{1}{2}\langle \hat{A}^2\rangle}$~\cite{Landau11} leads to a wrong result.

With the help of Eqs.~(\ref{13}), (\ref{130}), Eq.~(\ref{12}) for the reduced density matrix of the QD can be represented as follows,
\begin{eqnarray}\label{15}
&&\frac{\partial \rho_{dot}}{\partial
t}+\imath[H_{dot},\rho_{dot}]=\frac{\Gamma}{4\pi}\sum_\kappa \int
d\tau\int d\varepsilon  \times\nonumber\\
&&\left\{F(t,t-\tau;\lambda)\text e^{\imath \varepsilon \tau}
\left[1-f_\kappa (\varepsilon)\right]c \text e^{-\imath
H_{dot}\tau}\rho_{dot}(t-\tau)c^\dag \text e^{\imath
H_{dot}\tau}+\right.
\nonumber\\
&&+F(t,t-\tau;-\lambda)\text e^{-\imath \varepsilon \tau} f_\kappa
(\varepsilon)c^\dag \text e^{-\imath H_{dot}\tau}\rho_{dot}(t-\tau)c
\text e^{\imath H_{dot}\tau}-\nonumber\\
&&-F^\ast(t,t-\tau;-\lambda)\text e^{\imath \varepsilon
\tau}f_\kappa(\varepsilon) c \text e^{-\imath
H_{dot}\tau}c^\dag\rho_{dot}(t-\tau)\text e^{\imath
H_{dot}\tau}-\nonumber\\
&&\left.-F^\ast(t,t-\tau;\lambda)\text e^{-\imath \varepsilon
\tau}[1-f_\kappa(\varepsilon)]c^\dag \text e^{-\imath H_{dot}\tau} c
\rho_{dot}(t-\tau)\text e^{\imath H_{dot}\tau}+\text{H.c.}\right\},
\end{eqnarray}

where $\Gamma=2\pi\nu t_0^2$ is the level width of electron state
in the dot, $\nu$ is the density of states of the leads, which we
assume to be energy independent (wide-band approximation, see, e.g.,
Ref.~\cite{Wingreen1989}). In should be noted that unlike the case of
equilibrated vibrons (see, e.g., Ref.~\cite{Shkop2021}), the vibronic correlation function, Eq.~(\ref{130}), depends on two times
independently. This means that time-invariance in our system is
explicitly broken. The vibrons in coherent state $\vert z(t)
\rangle$, (which physically describes oscillations of a quantum
pendulum) violates time-invariance.

The reduced density matrix (operator) $\rho_{dot}$ acts in Fock space which in our
case is a two dimensional space of a spinless electron level in
a dot. The matrix elements of the density operator are: $\rho_0(t)=\langle 0\vert\rho_{dot}(t)\vert
0\rangle,\rho_1(t)=1-\rho_0(t)=\langle 1\vert\rho_{dot}(t)\vert
1\rangle$, where $\vert 1\rangle=c^\dag \vert 0\rangle$, а $\vert
0\rangle$ is a vacuum (ground) state. From Eq.~(\ref{15}) it follows that
the probability $\rho_0(t)$ satisfies the integro-differential equation,
\begin{eqnarray}\label{16}
&&\frac{\partial \rho_0}{\partial
t}=\frac{\Gamma}{4\pi}\sum_\kappa\int d\tau \int
d\varepsilon\left\{ F(t,t-\tau;\lambda)\text
e^{\imath(\varepsilon-\varepsilon_0)\tau}
\left[1-f_\kappa(\varepsilon))\right]\left[1-\rho_0(t-\tau)
\right]-\right.\nonumber\\
&&\hspace{2cm}\left.-F^\ast(t,t-\tau;-\lambda)\text
e^{\imath(\varepsilon-\varepsilon_0)\tau}
f_\kappa(\varepsilon))\rho_0(t-\tau)\right\}.
\end{eqnarray}

This equation is strongly simplified after integration over
$\varepsilon$. This integration can be done by using the Sokhotski–Plemelj theorem, see, e.g.,~\cite{Rashba},
\begin{equation}\label{17}
\int d\varepsilon \text e^{-\imath \varepsilon \tau} f_\kappa
(\varepsilon)=-\imath \pi \delta (\tau) +
\text{p.v.}\frac{\imath\pi T\text e^{-\imath \mu_\kappa
\tau}}{\sinh \pi T\tau},
\end{equation}
where the symbol p.v. denotes the principal value of an integral (Cauchy principal value).
In the limit $T\gg\Gamma$ one can neglect the retardation effects and Eq.~(\ref{17}) takes a simple local form,
\begin{equation}\label{18}
-\frac{\partial \rho_0}{\partial t}=M_1(t)\rho_0-M_2(t),
\end{equation}
where
\begin{equation}\label{19}
M_i(t)=1-\frac{1}{2}\sum_n A_n^{(i)}(t)[f_L(\varepsilon_0-n\omega)+
f_R(\varepsilon_0-n\omega)].
\end{equation}
The coefficients $A_n^{(i)}(t)$ are periodic
functions of time (with the period $2\pi/\omega$) and they can
be presented as the Fourier series,
\begin{eqnarray}\label{191}
&&\hspace{1cm}A_n^{(i)}(t)=\sum_p a_{n,p}^{(i)}e^{\imath \omega p t},\\
&&a_{n,p}^{(1)}=\frac{1}{\pi}\int_{-\pi}^\pi d\vartheta \text
e^{-\lambda^2(1-\cos \vartheta)}\sin\left( n\vartheta-\frac{\pi
p}{2}\right)\times\nonumber\\&& \times \sin\left(\lambda^2
\sin\vartheta\right)\cos\left(\frac{p\vartheta}{2}\right)J_p\left(4\lambda
\vert z\vert \sin\frac{\vartheta}{2}\right),\label{192}\\
&&a_{n,p}^{(2)}=\frac{1}{2\pi}\int_{-\pi }^\pi d\vartheta \text
e^{-\lambda^2(1-\cos
\vartheta)}\cos\left(\frac{p\vartheta}{2}\right)\times\nonumber\\&&
\times\cos\left(\frac{\pi
p}{2}-n\vartheta+\lambda^2\sin\vartheta\right) J_p\left(4\lambda
\vert z\vert \sin\frac{\vartheta}{2}\right).\label{193}
\end{eqnarray}
In Equations~(\ref{192}), (\ref{193}) $J_p(x)$ is the Bessel function of
the first kind and we parameterized the coherent state eigenvalue
$z$ in the form $z=\vert z\vert \exp(\imath\varphi)$, where the parameter $\vert z\vert$ determines the amplitude of the dot
oscillations.

In the asymptotic ($t\gg 1/\Gamma$) steady state regime of
oscillations the probability $\rho_0(t)$ is a periodic function of
time, $\rho_0(t+T_0)=\rho_0(t)$, and therefore it can be presented
as the Fourier series,
\begin{equation}\label{24}
\rho_0(t)=\sum_n\rho_n e^{\imath\omega n t},\,\rho_{-n}=\rho_n^\ast.
\end{equation}
Then the equation for the Fourier harmonics takes the following form,
\begin{eqnarray}\label{26a}
&&\imath p \rho_{p}=\delta_{p,0}-\rho_p-\frac{1}{2}\sum_n
\left[a_{n,p}^{(2)}-\sum_{k}a_{n,p+k}^{(1)}\rho_{k}\right]\times\nonumber\\
&&\hspace{1cm}\left[f_L(\varepsilon_0-n\omega)+f_R(\varepsilon_0-n\omega)\right],
\end{eqnarray}
and is a basic equation of this chapter. Its solutions are discussed in the subsection~\ref{sect2_4}.

\section{Electric current.} \label{sect2_3}

We are interested in current-voltage ($I-V$) characteristics of the single-electron
transistor. Therefore we need to calculate time-averaged current
through the system in the stationary regime,
\begin{equation}\label{20}
I=\frac{1}{T_0}\int_{T_0}^{}J(t)dt,
\end{equation}
where $J(t)=(J_L+J_R)/2$, with left (L) and right (R) electric currents are defined as change of electrons in the corresponded lead,
\begin{equation}\label{21}
J_\kappa= \eta_\kappa e \text{Tr}\left(\rho\frac{\partial
N_\kappa}{\partial t}\right), \; \; \; N_\kappa=\sum_{k}a_{k,\kappa}^\dag
a_{k,\kappa},
\end{equation}
where $\eta_{L/R}=\pm 1$, а $N_\kappa$ is the electron number operator in the lead $\kappa$. With the help of Eq.~(\ref{11}) the averaged current takes the form,
\begin{eqnarray}\label{22}
&&J_\kappa=\eta_\kappa\text{Tr}\int_{-\infty}^tdt'\text e^{\imath
H_0(t-t')}I_\kappa \text e^{-\imath H_0(t-t')}[H_{tun},
\rho]+\text{c.c.}, \nonumber\\&&I_\kappa= et_0 \text
e^{-\imath\lambda p}\sum_k ca_{k,\kappa}^\dag.
\end{eqnarray}
The straightforward calculation of Eq.(\ref{22}) yields the
following equation analogous to Eq.(\ref{18}),
\begin{equation}\label{23}
\frac{J(t)}{I_0}=-\rho_0(t)P_1(t)+P_2(t),
\end{equation}
where $I_0= e\Gamma/2$ is the saturation current through a single-level symmetric junction,
\begin{equation}\label{23a}
P_i(t)=\sum_n A_n^{(i)}(t)\left[f_L(\varepsilon_0-n\omega)-
f_R(\varepsilon_0-n\omega)\right],
\end{equation}
and coefficients $A_n^{(i)}$ are determined by Eqs.~(\ref{191})-(\ref{193}). From Equations~(\ref{24}),(\ref{20}) and (\ref{23}) one gets the following expression for the averaged current,
\begin{equation}\label{25}
I=I_0\sum_{n,k}\left[a^{(2)}_{n,k}\delta_{k,0}-a^{(1)}_{n,k}
\rho_k\right]\left[
f_L(\varepsilon_0-n\omega)-f_R(\varepsilon_0-n\omega)\right].
\end{equation}
Note that the average current does not depend on the phase $\varphi$ of the coherent state.


\section{Results of numerical calculations of \textit{I-V} characteristics.} \label{sect2_4}

\begin{figure}
\centering
\includegraphics[width=0.5\columnwidth]{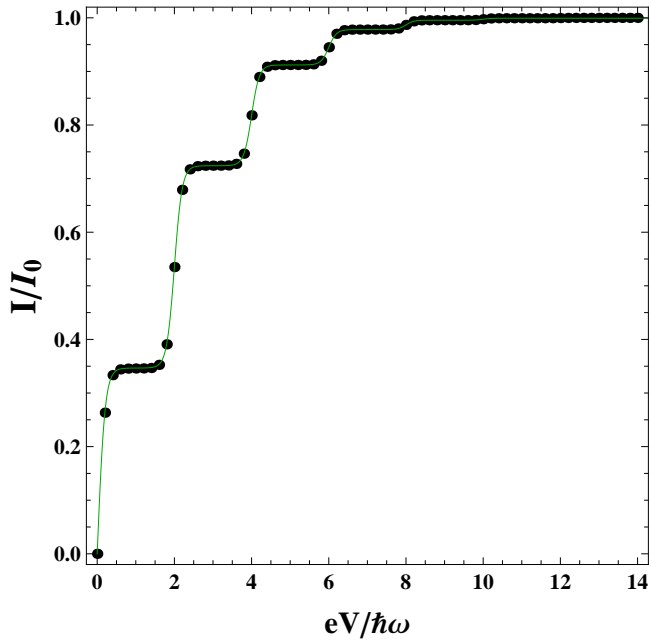}
\caption{\textit{The current-voltage dependencies for small value of
coherent state parameter of vibrons $\vert z\vert=0.25$ and for
strong electron-vibron interaction $\lambda=1$. The black dotted
curve corresponds to numerical calculation of current when the
vibrons are in the coherent state. The thin green curve represents
$I-V$ characteristics when the vibrons are in equilibrium and
characterized by the effective temperature $T^*$ determined by
Eq.~(\ref{26}). In calculations the values $T/\hbar \omega =0.05,
\Gamma / \hbar \omega=0.001$ was used. }} \label{fig:fig2_2}
\end{figure}

\begin{figure}
\centering
\includegraphics[width=0.5\columnwidth]{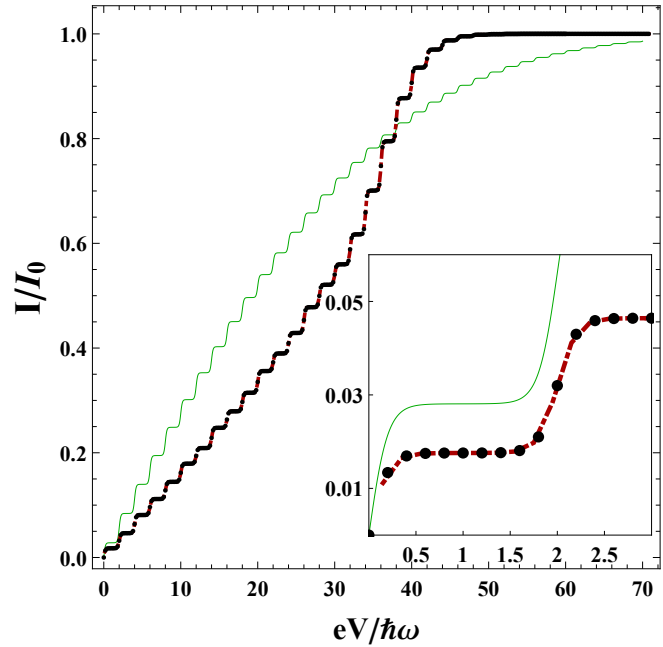}
\caption{\textit{$I-V$ plots for the large value of the parameter $\vert
z\vert=10$. All other parameters are the same as in Fig.~\ref{fig:fig2_2}. The
thin green curve corresponds to the case of equilibrated vibrons
with the effective temperature determined by the parameter $\vert
z\vert=10$. The red dash-dotted curve represents calculation of
current in the approximation when $\rho_0=0.5$ (see subsection~\ref{sect2_5}). Inset shows the region of low voltages.}}  \label{fig:fig2_3}
\end{figure}

Equation~(\ref{26a}) is the system of infinite number of equations for the Fourier harmonics. However, in this case the series for the coefficients of the harmonics converge quickly enough to allow one to consider only first several terms. Meanwhile, the sufficient number of the terms depends on parameters of the system, especially on the coherent state parameter $|z|$. Results of numerical calculations of Eq.~(\ref{25}) and~(\ref{26a}) are presented in Figs.~\ref{fig:fig2_2},\ref{fig:fig2_3}. 

As one can see, the plots for coherent vibrons (black dotted
curves) demonstrate step-like behavior of current versus bias
voltage at low temperatures, $T\ll\hbar\omega$. This behavior is
similar (however, in general case not identical) to Franck-Condon
steps in $I-V$ curves known for equilibrated vibrons (see e.g.
review paper Ref.~\cite{Krive2010} and references therein).
The curves for equilibrated and coherent vibrons coincide (see
Fig.~\ref{fig:fig2_2}) when the amplitude of oscillations of QD is less or of the
order of the amplitude of zero-point oscillations $l_0$, $|z|\leq
1$.

It is physically clear that in this case both systems are close to
their ground state (the average number of vibrons $<n>\ll 1$) and there is no difference in the behavior of
coherent and non-coherent vibrons. The strong differences appear
for large amplitudes of oscillations when $|z|\gg 1$ (see Fig.~\ref{fig:fig2_3}
where the dotted curve corresponds to vibrons in the coherent
state with parameter $|z|=10$). It is useful to introduce
effective temperature of vibrons $T^\ast$ by equating the average
number of vibrons in coherent and equilibrium state,
\begin{equation}\label{26}
|z|^2=(\exp(\hbar\omega/T^\ast))-1)^{-1}.
\end{equation}
Then for large amplitudes of oscillations ($|z|\gg 1$) and
moderately strong electron-vibron interaction ($\lambda\sim 1$)
$T^\ast\simeq |z|^2\hbar\omega\gg \lambda^2\hbar\omega$. It is
clear that at these high temperatures of the leads the Franck-Condon
steps in $I-V$ characteristics will be smeared out. It means that
coherent vibrons for large amplitudes of QD oscillations lead to
strong suppression of current at low biases and to pronounced
step-like behavior of $I-V$ curves. It is interesting to compare
this behavior with the Franck-Condon theory by assuming that the
vibronic subsystem is hot (it is described by Bose-Einstein
distribution with the temperature $T^\ast$), while the leads are
kept at low temperatures $T\ll\hbar\omega$. The thin green curve in Fig.~\ref{fig:fig2_3} demonstrates this case.
We see rather strong differences in current-voltage dependencies:
(i) the height of the steps for coherent vibrons are not regular,
and (ii) the current in the case of coherent vibrons saturates at
lower voltages ($eV_s\simeq |z|\hbar\omega$) than for equilibrated
vibrons.

\section{Estimation of the probability and current in steady-state regime}\label{sect2_5}
\begin{figure}
\centering
\includegraphics[width=0.5\columnwidth]{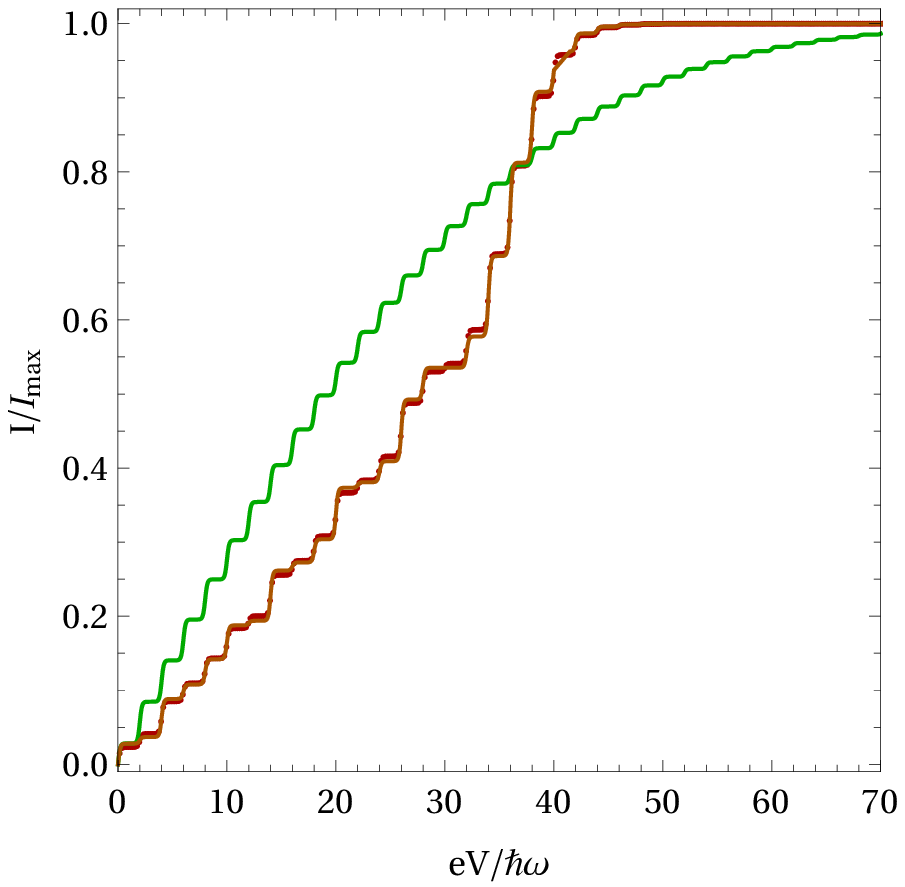}
\caption{\textit{$I-V$ plots for the large value of the parameter $\vert
z\vert=20$ and $\lambda=0.5$. All other parameters are the same as in Fig.~\ref{fig:fig2_2}. The
thin green curve corresponds to the case of equilibrated vibrons
with the effective temperature determined by the parameter $\vert
z\vert=20$. The red dash-dotted curve represents calculation of
current in the approximation when $\rho_0=0.5$ (using Eqs.~(\ref{25}) and~(\ref{26a})) and almost matches the solid orange curve obtained with the help of Eq.~(\ref{27b}).}}  \label{fig:fig2_4}
\end{figure}
While computing of Eqs.~(\ref{25}) and~(\ref{26a}), one can note that the coefficient $\rho_0$ (zeroth harmonic) of the Fourier series~(\ref{24}) in the stationary regime is equal to $\rho_0=0.5$ with with very
high accuracy, $\sim 10^{-5}$. It means that the probability (matrix elements of the reduced density matrix) does not depend on time, $\rho_0=\rho_1=1/2$. Then, by substituting $\rho_0=1/2$ and $\rho_p=0$ for $ p\ge 1$ in Eq.~(\ref{25}), we obtain a simple \textit{analytic} formula for the time-averaged electric current,
\begin{equation}\label{Ias}
I=I_0\sum_{n}a_{n}\left[ f_L(\varepsilon_0-n\omega)
-f_R(\varepsilon_0-n\omega)\right],
\end{equation}
where
\begin{eqnarray}\label{27}
&&a_{n}=\frac{1}{\pi} \int_{0}^{\pi}d\vartheta \text{e}^{-\lambda
^2 (1-\cos{\vartheta})}
\times\nonumber\\&&\hspace{1cm}\times\cos{n\vartheta}
\cos{(\lambda ^2\sin{\vartheta})} J_0\left( 4\lambda \vert z\vert
\sin{\frac{\vartheta}{2}}\right).
\end{eqnarray}
Furthermore, for the case $\lambda\leq 1$ we can proceed with estimation of the integral in Eq.~(\ref{27}),
\begin{equation}\label{27b}
   a_n\simeq J_n^2\left( 2\lambda \vert
z\vert\right). 
\end{equation}
This allows us to strongly simplify numerical calculations. Figures~\ref{fig:fig2_3} and \ref{fig:fig2_4} represent obtained current-voltage characteristics. The red dash-dotted curve corresponds to calculations using Eqs.~(\ref{Ias}) та (\ref{27}). This approximate analytical calculations coincide with the numerical ones with a sufficiently high accuracy ($\sim 10^{-5}$).

It needs to be noted that Eq.~(\ref{Ias}) has the same form as a
well-known equation (see, e.g., Ref.\cite{Krive2010}) for the
current of spinless electrons through a vibrating QD with
equilibrated vibrons (when the mechanical subsystem is described by equilibrium density matrix $\rho_{eq}$),
\begin{equation}\label{Ieq}
I_{eq}=I_0\sum_{n}A_{n}\left[ f_L(\varepsilon_0-n\omega)-f_R(\varepsilon_0-n\omega)\right],
\end{equation}
where now spectral densities $A_n$ are defined by the expression
$\text{Tr [e}^{-\imath \lambda p(t)}\text e^{\imath \lambda
p(0)}\rho_{eq}]=\sum_{n}A_ne^{\imath \omega n t}$, see~subsection~\ref{subsect1_1_2}.

\section*{Conclusions} \label{sect2_6}
\addcontentsline{toc}{section}{Conclusions}

In this chapter the electron transport in a molecular transistor has been considered, assuming vibrations of quantum dot oscillations to be in a coherent state. It was shown that $I-V$
curves at low temperatures have a step-like form which is similar to the
steps that accompany the lifting of the Franck-Condon blockade by bias
voltage. However, for large amplitudes of oscillations there are
strong differences in the predictions of the Franck-Condon theory
and the model considered in this chapter. By using numerical calculations we found strong
suppression of conductance even for a weak or moderately strong
electron-vibron coupling. The lifting of this coherent
oscillations-induced blockade by a bias voltage occurs at voltages
much lower then the ones predicted by the Franck-Condon theory. In addition, $I-V$ characteristics of a single-electron transistor with coherent vibrons do not depend on the phase of coherent state parameter. 

The main statements of this chapter are based on the publications~\cite{Bahrova2020, BahrovaC2020, BahrovaC2018}.           
\chapter{ENTANGLEMENT BETWEEN CHARGE QUBIT STATES AND COHERENT STATES OF NANOMECHANICAL RESONATOR GENERATED BY AC JOSEPHSON EFFECT} \label{chapt3}

In this chapter a superconducting nanoelectromechanical system based on a nanowire is considered. An experimentally simple protocol for bias voltage manipulation is discussed. This protocol results in the formation of the entanglement, which can be controlled by parameters of the device, between the charge qubit and the nanomechanical resonator. An experimentally feasible detection of the effects by measuring average current is also considered.

\section{Model and Hamiltonian of the nanoelectromechanical device.} \label{sect3_1}

\begin{figure}
\centering
\includegraphics[width=0.75\columnwidth]{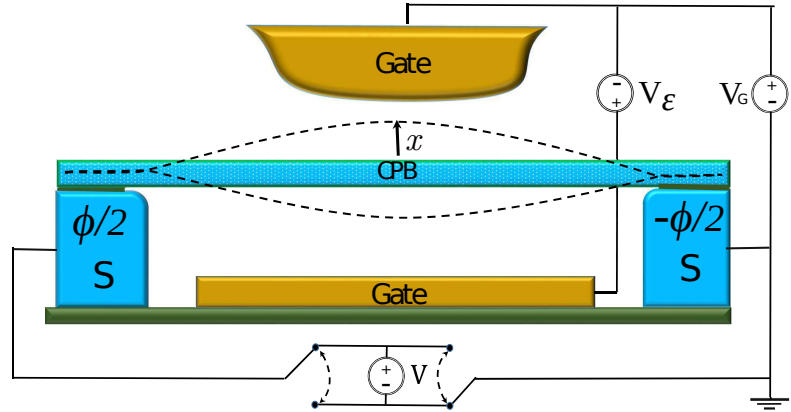}
\caption{\textit{Schematic illustration of the NEMS under consideration.
The superconducting nanowire, treated as a charge qubit, is tunnel
coupled to two bulk superconductors (S) with the superconducting phase
difference $\phi$ and capacitively coupled to the two gate electrodes.
The bending oscillations in the $x$ direction are described by the
harmonic oscillator.}}\label{fig:fig3_1}
\end{figure}

A schematic representation of the nanoelectromechanical system (NEMS) prototype, which is under the consideration, is presented in Fig.~\ref{fig:fig3_1}. It consists of the
superconducting nanowire (SCNW)~\cite{Bezryadin2010,Masuda2016}, which is suspended
between two bulk superconductors and is capacitively coupled to
two side gate electrodes. In what follows we consider the case
when SCNW represents a superconducting island that can be treated as
a charge qubit (Cooper-pair box) whose basis states are charge
states --- states which represent the presence or absence of excess
Cooper pairs on the island. Usually these states are refereed to as
charge and neutral states correspondingly. As this takes place, the
gate voltage $V_{G}$ and the voltage applied between the gates
$V_{\mathcal{E}}$ are chosen in the way that the difference in the
electrostatic energies of the charged and neutral states equals to
zero at the straight configuration of the nanowire, while nanowire
bending removes this degeneracy. We also reduce the bending dynamics
of the SCNW to the dynamics of the fundamental flexural mode
described by the harmonic oscillator.

Joint Cooper pairs dynamics and mechanical one of this system
is described by the Hamiltonian which can be presented in the form,
\begin{equation}\label{1ham}
    H=H_q+H_m+H_{int}.
\end{equation}
Here
\begin{equation}\label{f1}
    H_{q,int}=\frac{[\hat{Q}+Q_G(\hat{x})]^2}{2C(\hat{x})}-\sum_{\sigma} [E_{J,\sigma} \cos{(\phi_\sigma-\hat{\phi})}],
\end{equation}
where $\hat{Q}=2e\hat{n}$ is discrete charge operator on the dot ($\hat{n}$ is Cooper pair number operator on the dot), $Q_G(\hat{x})=V_G C_G(\hat{x})$ is continuous charge generated by the gate voltage $V_G$, $C(\hat{x})=2C_J+C_G(\hat{x})$ is the mutual capacitance that includes gate and Josephson junctions ones. The constant $E_J=E_{J,L}=E_{J,R}$ is the Josephson coupling
energy (we consider only the case of symmetric coupling), $\hat{\phi}$ is the phase operator on the dot that satisfies commutation relation $[\hat{\phi},\hat{n}]=\imath$. Thus, Eq.~(\ref{f1}) can be rewritten as:
\begin{equation}\label{f2}
    H_{q,int}=E_c(\hat{x})\left(\hat{n}+\frac{C_G(\hat{x})V_G}{2e}\right)^2-
    2E_J\cos{\phi}\cos{\hat{\phi}}.
\end{equation}
 Here $E_c(\hat{x})=(2e)^2/(2C(\hat{x}))$ is the charging energy. The coordinate-dependent gate capacitance $C_G(\hat{x})$ is tuned in such a way that the difference of the electrostatic energies between the charge (with one exceed Cooper pair) and neutral state is proportional to the CPB dimensionless displacement $\hat{x}$ and is equal to zero at an equilibrium point of oscillations.
Then in Coulomb blockade regime ($E_C(0)\gg E_J$) in a charge basis with determined number of Cooper pairs on the island the operator function $\text{exp}[\imath \hat{\phi}]$ is:
\begin{equation}\label{f3}
    \text{e}^{\imath \hat{\phi}}\vert n\rangle =\frac{1}{\sqrt{2\pi}}\int_0^{2\pi}d\phi \text{e}^{\imath (n+1)\phi}\vert \phi\rangle =\vert n+1\rangle.
\end{equation}
Therefore, in the basis netted on the neutral $\vert 0 \rangle=(0;1)^T$ and charged $\vert 1\rangle=(1;0)^T$ state and in the particle number representation for mechanical variables one gets:
\begin{eqnarray}\label{1b}
&&H_q=-E_J\sigma_1\cos\phi ,\nonumber\\
&&H_m=\frac{\hbar\omega}{2}\left(\hat x^2+\hat p^2\right),\nonumber\\
&&H_{int}= \varepsilon\hat x\sigma_3.\nonumber
\end{eqnarray}

where $\sigma_i,i=1,2,3,$ are the Pauli matrices, $E_Q$ is actually the energy associated with the electrostatic field generated by gate voltage $V_G$. It implies that only the first order of the gate capacitance matters and the case of toward the degeneracy point in an equilibrium point is considered.
Note that one can diagonalize the non-perturbation part of the Hamiltonian, Eq.~(\ref{1b}), by the unitary transformation:
\begin{equation}
    \hat{U}=\left( I+\imath\sigma_2\right)/\sqrt{2}, 
\end{equation}
which turns $\sigma_1\rightarrow \sigma_3, \sigma_3\rightarrow -\sigma_1$. Then,
\begin{equation}\label{f5}
    H=\varepsilon\hat{x}\sigma_1+E_J\cos\phi (t)\sigma_3+\frac{\hbar\omega}{2}\left(\hat x^2+\hat p^2\right).
\end{equation}

Here Hamiltonian $H_q$ represents Josephson coupling between the Cooper Pair Box (CPB) and
bulk superconductors with $\phi=\phi (t)$ is the superconducting phase difference
between electrodes, $\sigma_i (i=1,2,3)$ are the Pauli matrices
acting in the qubit Hilbert space in a basis where vectors $(1,0)^T$
and $(0,1)^T$ represent charged and neutral states, respectively.
Hamiltonian $H_m$ in Eq.~(\ref{1ham}) represents dynamics of the
fundamental bending mode described by the harmonic oscillator with
frequency $\omega$ (here momentum and coordinate operators, $\hat p$
and $\hat x$, are normalized on the amplitude of zero-point
oscillations $x_0=\sqrt{\hbar/M\omega}$, $M$ is an effective mass of the island,
$[\hat x,\hat p]=\imath$). The third term, $H_{int}$, describes the
electromechanical coupling between the charge qubit and the mechanical
oscillator induced by the electrostatic force acting on the charged
state of the qubit, $\varepsilon =e\mathcal E x_0$. In the last
equality, $\mathcal E$ is an effective electrostatic field that is
controlled by the difference of the applied voltages $V_G$ and
$V_\varepsilon$. Below we assume $\varepsilon\ll\hbar \omega,
E_J$ that corresponds to a typical experimental situation~\cite{Satzinger2018,Tian2005,Arrangoiz2019}.

The states of the system described by the Hamiltonian, Eq.~(\ref{1ham}),
are a superposition of direct products  of qubit states, $\mathbf
e^\pm_i$, and  eigenstates  of the oscillator $\vert n\rangle$. Here
and below $\mathbf e^\nu_i$ denotes the eigenvectors of the Pauli
matrices $\sigma_i$ with eigenvalues $\nu=\pm 1$.

If $\varepsilon=0$, the interaction between the qubit and the
mechanical subsystem is switched off and stationary states of the
Hamiltonian, Eq.~(\ref{1ham}), are pure states. The entropy of
entanglement is an integral of motion, i.e. if the system is
initially in a pure state, it will be in a pure state at any moment
of time. If we apply a constant bias voltage between superconducting leads, an oscillatory ($\propto\sin{\phi(t)}$) current emerges due to ac Josephson effect, $\dot \phi(t)=
2eV/\hbar$, where $V$ is a bias voltage. The synchronous switching on the electrical field $\mathcal E$ and
the bias voltage between the superconducting leads results in the evolution of such pure states in the
states represented by entanglement between the qubit and
oscillator states.

\section{Time evolution of the system.}\label{sect3_2}

To carry out an analysis of time evolution of the system, we introduce the
dimensionless time and energies, $\omega t \rightarrow t,
E_J/\hbar\omega \rightarrow E_J, \varepsilon/\hbar\omega
\rightarrow\varepsilon$ and assume that at the moment of switching
on the interaction between the subsystems ($ t = 0 $), the
difference between the superconducting phases is $\phi = \phi_0$ and
the system has been in a pure state,
\begin{equation}\label{in}
\vert \Psi(0)\rangle = \mathbf e_{in} \otimes \vert
0\rangle.
\end{equation}
At $t>0$, according to the (second) Josephson relation,
\begin{equation}
  \phi(t)=2eVt/\hbar\omega+\phi_0.   
\end{equation}
The Hamiltonian, Eq.~(\ref{1ham}),
and, as a consequence, the time evolution operator $\hat U(t,t')$,
which is defining evolution of the arbitrarily initial state, have the following
properties,
\begin{equation}\label{1a}
\hat{H}(t+T_V)=\hat H(t), \quad \hat U(t,t')=\hat U(t+T_V,t'+T_V),
\end{equation}
that is the are periodic in time with the period
\begin{equation}
    T_V=2\pi/\Omega_V=\pi\hbar\omega/e|V|. 
\end{equation}
To analyze the
evolution operator, one can use the interaction picture (with respect to the interaction Hamiltonian $H_{int}$) taking,
\begin{equation}\label{6}
\hat U(t,t')= \hat{\mathcal U}_\kappa (t) \hat{\mathcal U}_\kappa
(t,t')\hat{\mathcal U}^\dag_\kappa(t'),
\end{equation}
where
\begin{equation}\label{7u}
\hat{\mathcal U}_\kappa(t)=\exp\left[ \frac{\imath
E_J}{\Omega_V}\sigma_1 \sin\left(\Omega_V
t+\kappa\phi_0\right)-\imath a^\dag a t\right],
\end{equation}
is the unitary evolution operator corresponded to the non-perturbed Hamiltonian and describes free evolution of mechanical and electronic subsystem independently.
The parameter $\kappa=\text{sgn}\left(V/|V|\right)=\pm$
characterizes the direction of the bias voltage drop. The operator
$\hat{\mathcal U}_\kappa (t,t')$ obeys the following equations,
\begin{eqnarray}\label{9u}
\imath \frac{\partial \hat{\mathcal U}_\kappa(t,t')}{\partial t}=
\hat{\mathcal H}_\kappa(t)\hat{\mathcal U}_\kappa(t,t'),\nonumber \\
\hat{\mathcal H}_\kappa(t)= \varepsilon \hat x(t)\sigma_3(t),\quad
\hat{\mathcal U}_\kappa(t,t)=\hat I.
\end{eqnarray}
Here
\begin{eqnarray}
&&\hat x(t) = \frac{1}{\sqrt 2}(\hat a \text{e}^{-\imath t}+\hat a^\dag
\mathrm{e}^{\imath t}),\nonumber\\
&&\sigma_3(t)=\sigma_3\cos\left(\frac{E_J}{\Omega_V}
\sin(\Omega_V t+\kappa\phi_0)\right)- \nonumber\\
&&\hspace{1cm}-\sigma_2\sin\left(\frac{E_J}{\Omega_V}
\sin(\Omega_Vt+\kappa\phi_0)\right).
\end{eqnarray}

If the frequencies $\omega$ and $\Omega_{V}$ are incommensurable, the
operator $\hat{\mathcal H}_\kappa(t)$ is a quasiperiodic function of
time. In such a case one can expect that the mechanical subsystem,
being initially in the ground state, does not significantly deviate
from this state in the process of evolution. A rigorous consideration of this case is done numerically~\cite{DankoC2020}. Here let us consider the resonant case
when $\Omega_V=\omega$ and assume that $\varepsilon \ll 1$. The
first condition stipulates the following properties of the evolution
operator,

\begin{equation}\label{N}
\hat {\mathcal U}_\kappa(2\pi N, 2\pi N') =\left(\hat{\mathcal
U}_\kappa(2\pi,0)\right)^{N-N'},
\end{equation}
where $N, N'$ are natural numbers. The second assumption
allows us to make the following substitution in a leading
approximation regarding small $\varepsilon$,
\begin{equation}
\hat{\mathcal U}_\kappa(t,t')
= \hat{\mathcal U}_\kappa (2\pi N,2\pi N'),
\end{equation}
where $N (N')=[t(t')/2\pi] ([x]$ is an integer part of $x$), and
obtain an expression for $\hat{\mathcal U}_\kappa(2\pi,0)$ which can
be written as,
\begin{eqnarray}\label{Evop}
&&\hat{\mathcal U}_\kappa(2\pi,0)= \exp\left[\imath\tilde\varepsilon
\sigma_2\hat p(\kappa\phi_0)+\varepsilon^2{\cal O}(\hat I)\right], \nonumber\\
&&\hat p(\phi)= \hat p\cos\phi+\hat x\sin \phi.
\end{eqnarray}
Here $\tilde\varepsilon= 2\pi \varepsilon J_1(2E_J)$
and $J_1(x)$ is 
the Bessel function of the first kind. Using the above relations one
can obtain an expression for the evolution operator $\hat U(t,t')$,
which in the main approximation regarding $\varepsilon $ has a form,
\begin{equation}\label{U}
\hat U(t,t')=\hat{\mathcal U}_\kappa(t)
\exp\left[\imath\tilde\varepsilon \sigma_2\hat
p(\kappa\phi_0)(t-t')\right]\hat{\mathcal U}^\dag_\kappa(t').
\end{equation}

Using Eqs.~(\ref{in}),(\ref{U}), one gets that at the time $t$, with
the accuracy to small parameter $\tilde\varepsilon\ll 1$, the state
of the system $\vert \Psi(t)\rangle$ is given by an expression,
\begin{equation}\label{WF}
\vert \Psi(t)\rangle=\sum\limits_\nu
A_\nu^\kappa\mathbf{e}^\nu_2(t,\kappa\phi_0)\otimes\vert -\nu
\mathrm{z}(t,\kappa)/\sqrt 2 \rangle.
\end{equation}
Here
$$\mathbf{e}^\nu_2(t,\kappa\phi_0)=\mathbf{e}^\nu_2
\exp\left[\imath E_J\sigma_1\sin(t+\kappa\phi_0)\right],$$
and $\mathbf{e}_2^\nu=\sigma_1\mathbf{e}_2^{-\nu}$ are the
eigenvectors of the Pauli matrix $\sigma_2$ with eigenvalues $\nu
=\pm 1$,
\begin{equation}\label{Anu}
A_\nu^\kappa\equiv\left(\mathbf{e}_2^\nu
(0,\kappa\phi_0),\mathbf{e}_{in}\right)=\cos{\left(\kappa E_J  \sin{\phi_0}\right)}c_2^\nu-\imath \sin{\left(\kappa E_J \sin{\phi_0} \right)}c_2^{-\nu},
\end{equation}
where an initial state can be presented in a basis of eigenvectors of the matrix $\sigma_2$ as:
\begin{equation}\label{einitial}
    \mathbf{e}_{in}=\sum_{\nu =\pm 1} c_2^\nu \mathbf{e}_2^\nu .
\end{equation}
The symbol
$|\alpha\rangle$ (where $\alpha$ is a complex number) denotes the
coherent states of the harmonic oscillator, $\hat a|\alpha\rangle=
\alpha|\alpha\rangle$, while a complex function
$\mathrm{z}(t,\kappa)$ is defined as:
\begin{equation}\label{670}
\mathrm{z}(t,\kappa)=\tilde \varepsilon
t\exp\left[-\imath(t+\kappa\phi_0)\right].
\end{equation}

It should be stressed that Eq.~(\ref{WF}) is valid only for
restricted time interval $t\leq \tilde\varepsilon^{-2}$. Time $t$ should be
also shorter than any dephasing and relaxation times. From
Eq.~(\ref{WF}) one can see that initially pure state
$\vert \Psi(t=0)\rangle=\mathbf{e}_{in}\otimes |0\rangle$
evolves into the state represented by the entanglement between the
two qubit states and two coherent states of the mechanical
resonator. Moreover, the details of this entanglement depend on
switching time (parameter $\phi_0$) and direction of the bias
voltage (parameter $\kappa$). These circumstances allow one to
manipulate the described above entanglement by changing the bias
voltage direction.


\section{Generation of "Schr{\"o}dinger-cat states".} \label{sect3_3}

\begin{figure}
\centering
\includegraphics[width=.55\columnwidth]{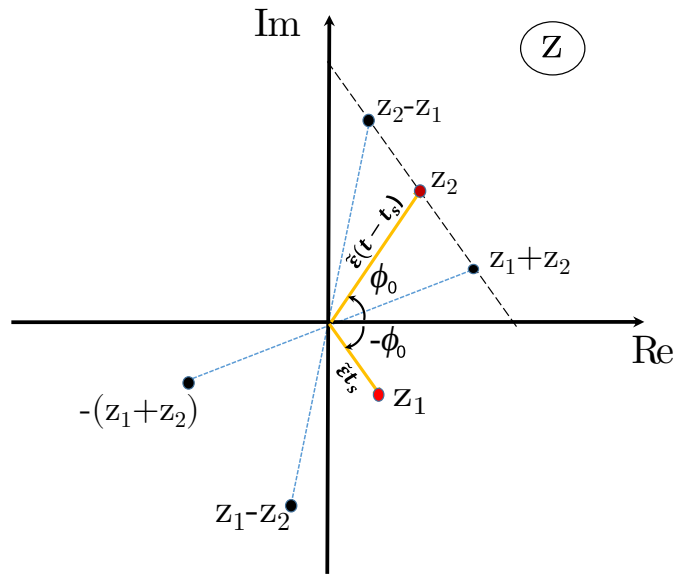}
\caption{\textit{Schematic illustration of the positions of the coherent
states described by the complex numbers $\mathrm{z}_{1,2}$ and their
combinations $\mathrm{z}_\pm$ in the complex plane. It denotes the time evolution of the coherent states, on the one hand, and the dependence on the initial phase difference $\phi_0$, on the other one.}
}\label{fig:fig3_2}
\end{figure}

To demonstrate the effect of the entanglement between the charge qubit and mechanical vibrations that comprehends the formation of so-called Schr\"{o}dinger-cat states of nanomechanical resonator, we consider the following time protocol
for $V(t)$:
$$2eV(t)=-\hbar\omega\theta(t)\left[1-2\theta(t-t_s)\right].$$
Namely, during the time interval $0<t<t_s$ the bias voltage
$V(t)=-\hbar\omega/2e$ and then it switches its sign. Using
Eqs.~(\ref{6}), (\ref{N}), (\ref{Evop}), one gets that at $t>t_s$ the evolution operator has the form:
\begin{equation}\label{Sw1a}
 \hat U(t,0)=\hat{\mathcal {U}}_+(t)
\text{e}^{\imath\sigma_2\tilde\varepsilon (t-t_s)\hat
p(\phi_0)}\hat S \text{e}^{\imath\sigma_2\tilde\varepsilon t_s
\hat p(-\phi_0)}\hat{\mathcal{U}}_-(0), 
\end{equation}
where
\begin{eqnarray}\label{Sw1}
&& \hat S=\hat{\mathcal{U}}_+^\dag (t_s)\hat{\mathcal{U}}_-(t_s)
\equiv\rho(t_s,\phi_0)+\imath\tau(t_s,\phi_0)\sigma_1,\\
&&\rho(t_s,\phi_0)=\cos\left( 2E_J\cos
t_s\sin\phi_0\right),\nonumber\\
&&\tau(t_s,\phi_0)=-\sin\left(2E_J\cos t_s\sin \phi_0\right).
\nonumber
\end{eqnarray}
As a result, the state of the system after changing the direction of
the bias voltage takes the following form:
\begin{eqnarray}\label{WF1}
&&\hspace{-1.1cm}|\Psi(t)\rangle=
\sum\limits_\nu\mathbf{e}^\nu_2(t,\phi_0)
\otimes\nonumber\\
&&\otimes\left(\rho A_\nu^-|-\nu \mathrm{z}_+/\sqrt
2\rangle+\imath\tau A_{-\nu}^-|\nu \mathrm{z}_-/\sqrt
2\rangle\right),
\end{eqnarray}
where $\mathrm{z}_\pm=\mathrm{z}_1\pm\mathrm{z}_2$ and
\begin{eqnarray}\label{zz}
&&\mathrm{z}_1=\text{e}^{-\imath(t-\phi_0)}\tilde\varepsilon t_s,\nonumber\\
&&\mathrm{z}_2=\text{e}^{-\imath(t+\phi_0)}\tilde\varepsilon(t-t_s).
\end{eqnarray}
A schematic representation of evolution of the coherent states can be seen in Fig.~\ref{fig:fig3_2}. Equation~(\ref{WF1}) demonstrates that the state of the
system is represented by the entanglement of two qubit state with
two so-called "cat states" (superposition of coherent states) whose
structure is controlled by the parameters $E_J\ (\rho )$ and $\phi_0$. As it
follows from Eqs.~(\ref{WF1}), (\ref{zz}), the bias voltage switching
does not affect the dynamics of the system if $\phi_0= \pi n$ with $n$ standing for an integer number.


\section{Entanglement entropy.} \label{sect3_4}

Let us limit ourselves for simplicity to considering a most interesting case when $\phi_0 = \pi/2$ and put
$\mathbf{e}_{in} = (\mathbf{e}_2^+ + \mathbf{e}^-_2)/\sqrt 2$, that
is, we suppose that immediately before the interaction was switched
on, the qubit was in the eigenstate of the operator $\hat
H_q(t=0-\delta)$. With these assumptions Eq.~(\ref{Anu}) transforms to:
\begin{equation}\label{Anu2}
   A_+^-=A_-^-=\exp(\imath E_J)/\sqrt 2.
\end{equation}
To characterize the entanglement between the qubit states and the
states of the mechanical oscillator, we introduce the reduced
density matrices,
$\hat\varrho_{q(m)}(t)=\mathrm{Tr}_{m(q)}\hat\varrho$, where
\begin{equation}\label{rhofull}
   \hat\varrho=|\Psi(t)\rangle\langle\Psi(t)| 
\end{equation}
is a
complete density matrix of the system and $\mathrm{Tr}_{m(q)}$
denotes the trace over mechanical (qubit) degrees of freedom. Using Eqs.~(\ref{WF}),(\ref{WF1}), one can get the following expression for the reduced qubit density matrix $\hat\varrho_q$,
\begin{equation}\label{54}
\hat\varrho_q(t)=\frac{ I+\lambda(t,t_s)\sigma_1}{2},
\end{equation}
where
\begin{eqnarray} \label{DMQ}
&& \lambda(t,t_s)=\exp{\left(-\tilde{\varepsilon}^2t^2\right)},
\hspace{1.0cm} 0<t\le t_s, \\
&& \lambda(t,t_s)=\rho^2\exp{\left[
-\tilde\varepsilon^2(t-2t_s)^2\right]}+\tau^2\exp{\left( -\tilde{\varepsilon}^2t^2
\right)},\hspace{1.3cm} t>t_s. \label{RHO}
\end{eqnarray}
Note that $\lambda (t,t_s)\geq 0$. When deriving Eq.~(\ref{54}), we took into account relation
$\mathbf{e}_2^+(\mathbf{e}_2^-)^\dag+\mathbf{e}_2^-
(\mathbf{e}_2^+)^\dag=\sigma_1$. The entropy of entanglement (also called the von~Neumann entropy) is defined as:
\begin{equation}\label{667}
S_{en}(t) \equiv -\operatorname{Tr} \hat\varrho_q(t) \log
\hat{\varrho}_q(t) =- \operatorname{Tr} \hat\varrho_m(t) \log
\hat{\varrho}_m(t).
\end{equation}
Since the basis of the coherent state is not orthonormal (quasiorthogonal, in fact) and overcomplite, it is suitable to use Eq.~(\ref{54}). In order to calculate the entanglement entropy, it is convenient to present the matrix in a diagonal form. The reason is if $\lambda_i, i=1,2$ is an eigenvalue of the matrix $\hat{\varrho}_q$, then Eq.~(\ref{667}) can be rewritten as:
\begin{equation}\label{667b}
    S_{en}(t) =-\sum_{i=1,2}\lambda_i \log{\lambda_i}.
\end{equation}
 One can easily find the eigenvalues of the reduced qubit density matrix $\hat{\varrho}_q$, Eq.~(\ref{54}),
\begin{equation}\label{eigenentropy}
    \lambda_{1,2}=\frac{1}{2}\left[ 1-\lambda (t,t_s)\right].
\end{equation}
From these equations one can see that the maximal value of the entanglement is $S_{en}^{(max)}=\log{2}$ (when $\lambda_{1,2}=1/2$), meaning that correlations between the electronic and mechanical subsystems are maximal. In the calculations such a usual way of treating the uncertainty $0 \log{0}=0$ is accepted. In general, entropy of entanglement is a limited function, $0\leq S_{en} \leq \log{N}$, where $N$ stands for the number of subsystems (degrees of freedom) which a whole system consists of.

\begin{figure}
\centering
\includegraphics[width=0.75\columnwidth]{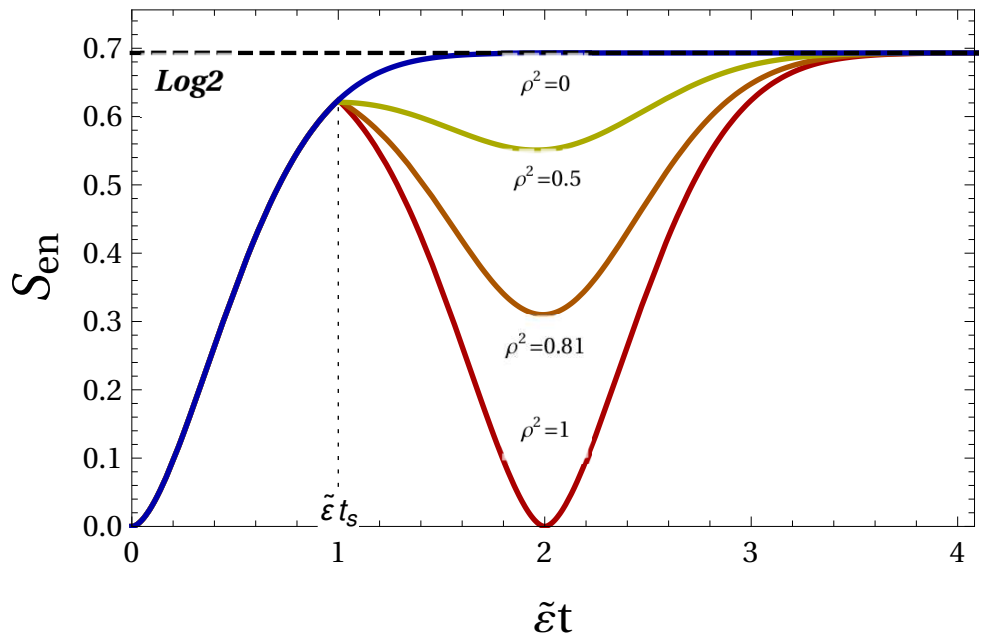}
\caption{\textit{The entanglement entropy dependent on time for different values of
$\rho=0,1/\sqrt{2},0.9,1$ (blue, yellow, orange and red curves). The thin dotted line indicates the bias voltage switching time. The dashed curve corresponds to the maximal value of the entanglement,} log$2$.}
\label{fig:fig3_3}
\end{figure}

A plot of $S_{en}(t)$ for $\tilde{\varepsilon} t_s=1$ and different values of
$\rho$ (equally, for $\tau^2=1-\rho^2$) is presented in Fig.~\ref{fig:fig3_3}. The entanglement entropy monotonically increases in time within
intervals $0<t<t_s$ and $2t_s<t<\infty$ saturating to the maximal
value  $S_{en}^{\text{(max)}}$ at $t\rightarrow \infty$.
Within interval $t_s<t\leq 2t_s$  behavior of the entanglement
entropy depends on the relation between $\rho$ and $\tau$. In
particular, for $\rho^2>\tau^2$ the entanglement entropy $S_{en}(t)$
starts to decrease after switching, reaching some minimal value
(equals zero for the $\rho^2=1$, i.~e., our system is separable) within interval $t_s<t\leq 2t_s$. If
$\rho^2<\tau^2$, the entropy continues to grow just after the switching.
However, its derivative might be also negative within some time
interval whose existence is controlled by the parameters
$\tilde{\varepsilon}t_s$ and $\tau^2/\rho^2$. 

In addition let us also briefly consider a more general case of an arbitrary value of initial superconducting phase difference $\phi_0$. For the time interval $0<t\leq t_s$, the qubit density matrix is given by Eqs.~(\ref{54}),~(\ref{DMQ}). However, for the time after the bias voltage switching, $t>t_s$, one can find the following expression for the density operator of the qubit,
\begin{eqnarray}
&& \varrho_q (t)= \frac{1}{2}I+I\rho\tau\text{e}^{-\tilde{\varepsilon}^2t_s^2}\sin{\left\{ \tilde{\varepsilon}^2t_s (t-t_s) \sin{(2\phi_0)}\right\}}+\nonumber\\
&&\hspace{0.5cm} +\frac{1}{2}\sigma_1\text{e}^{-\tilde{\varepsilon}^2\left(t_s^2+(t-t_s)^2\right)}\left[\rho^2 \text{e}^{-2\tilde{\varepsilon}^2t_s(t-t_s)\cos{(2\phi_0)}}+\tau^2 \text{e}^{2\tilde{\varepsilon}^2 t_s (t-t_s) \cos{(2\phi_0)} }\right]-\nonumber \\
&&\hspace{0.5cm} -\sigma_1\rho\tau\text{e}^{-\tilde{\varepsilon}^2(t-t_s)^2}\sin{\left\{ \tilde{\varepsilon}^2 t_s (t-t_s) \sin{(2\phi_0)}\right\}}.
\end{eqnarray}
From this equation one can find, in particular, following the above-mentioned procedure, that the maximal effect is achieved at $\phi_0=\pi/2$ and one get Eq.~(\ref{54}).

\section{Time evolution of the mechanical subsystem.} \label{sect3_5}

To describe the evolution of the mechanical subsystem, we consider
the reduced density matrix $\hat{\varrho}_m(t)$. From
Eq.~(\ref{WF1}) one gets that at $t>t_s$ the reduced density matrix of the mechanical subsystem takes a form,
\begin{eqnarray}\label{DMM}
&&\hat{\varrho}_m(t)=
\frac{1}{2}\sum\limits_\nu\left[\rho^2|\nu
\mathrm{z}_+/\sqrt 2\rangle \langle\nu \mathrm{z}_+/\sqrt
2|+\tau^2|\nu \mathrm{z}_-/\sqrt 2\rangle\langle\nu
\mathrm{z}_-/\sqrt 2|- \right.\nonumber\\
&&\hspace{1.5cm}\left.-\imath \rho\tau \left(|-\nu \mathrm{z}_+/\sqrt
2\rangle\langle\nu \mathrm{z}_-/\sqrt 2|-
\text{H.c.}\right)\right].
\end{eqnarray}

To visualize the state of the mechanical subsystem, it is convenient
to use the Wigner function representation for the density matrix
$\hat\varrho_m(t)$,
$$W(x,p,t)=\frac{1}{\pi}\int\varrho_m(x+y,x-y,t)\exp(2\imath py)dy,$$
where $\varrho_m(x,x',t)=\langle x|\hat\varrho_m(t)|x'\rangle $.
Using Eq.~(\ref{DMM}), one gets
\begin{equation} \label{WiF}
W(x,p,t)= W_t(x\cos t -p\sin t, p\cos t +x\sin t),
\end{equation}
where the function $W_t(x,p)$ is defined according to the relation,
\begin{eqnarray} \label{76}
&&W_t(x,p)=\frac{1}{2}\sum\limits_\nu\left[\rho^2
W_0(x,p+\nu|\mathrm{z}_+|)
+\tau^2W_0(x,p-\nu|\mathrm{z}_-|)+ \right.\nonumber\\
&&\hspace{2cm}\left. +2\rho\tau \sin\left(2\nu
Z_-x\right)W_0\left(x,p+\nu Z_+\right)\right].
\end{eqnarray}
In Eq.~(\ref{76}) $ Z_\pm= \left(
|\mathrm{z}_-|\pm|\mathrm{z}_+|\right)/2$  and
\begin{equation}\label{89}
W_0(x,p)=\frac{1}{\pi}\exp\left[-(x^2+p^2)\right]
\end{equation}
is the Wigner function corresponding to the ground state of a harmonic oscillator. Plots of $W(x,p,t)$ for $t=2\pi N$, $\rho=0,
\rho=1$ and $\rho=\tau=1/\sqrt 2$ at $|\mathrm{z}_+|=3$ and
$|\mathrm{z}_-|=9$ are presented in Figs.~\ref{fig:fig3_4} and \ref{fig:fig3_5}.

From Equations~(\ref{DMM}),(\ref{76}) one can see that in the case when $ \rho $ is
equal to zero or one (in particular, when $ t_ s = 0$) the Wigner
function is positive and has two maxima, demonstrating the
entanglement between two states of the qubit and two coherent states
(see Fig.~\ref{fig:fig3_4}). In general case $\rho\tau\neq 0$, and the Wigner function
takes both positive and negative values at $t>t_s$, demonstrating
the entanglement of two states of the qubit with the superposition of two quasi-orthogonal coherent states of the
nanomechanical resonator (see Fig.~\ref{fig:fig3_5}).
\begin{figure}
\centering
\begin{minipage}{0.45\columnwidth}
\includegraphics[width=\columnwidth]{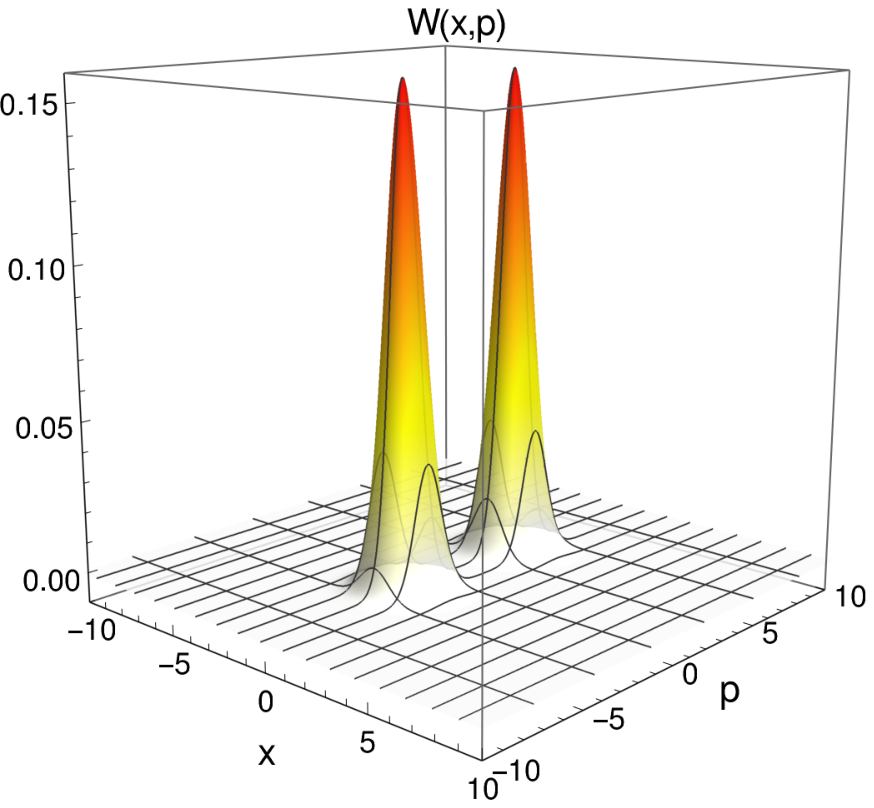}
\end{minipage}
\begin{minipage}{0.45\columnwidth}
\includegraphics[width=1.0\columnwidth]{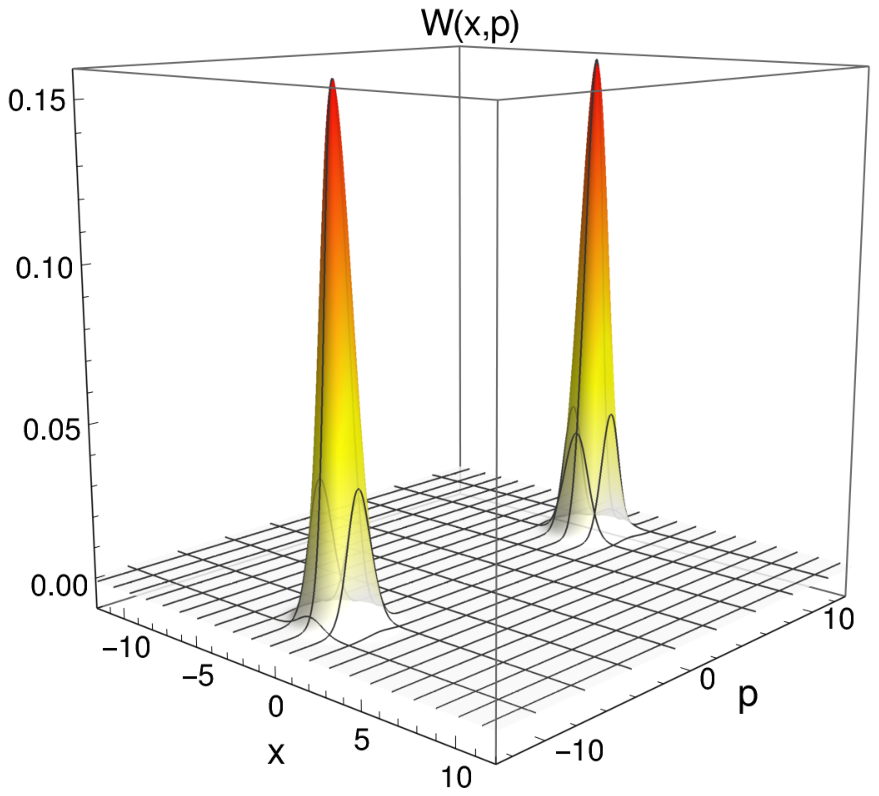}
\end{minipage}
\caption{\textit{The Wigner functions $W(x,p,t=2\pi N)$ for  $\rho=1$ (a)
and $\rho=0$ (b). It takes only positive values and has two maxima
demonstrating the entanglement between two qubit states and two coherent
states of the nanomechanical resonator.}}
\end{figure}\label{fig:fig3_4}

\begin{figure}
\centering
\includegraphics[width=.55\columnwidth]{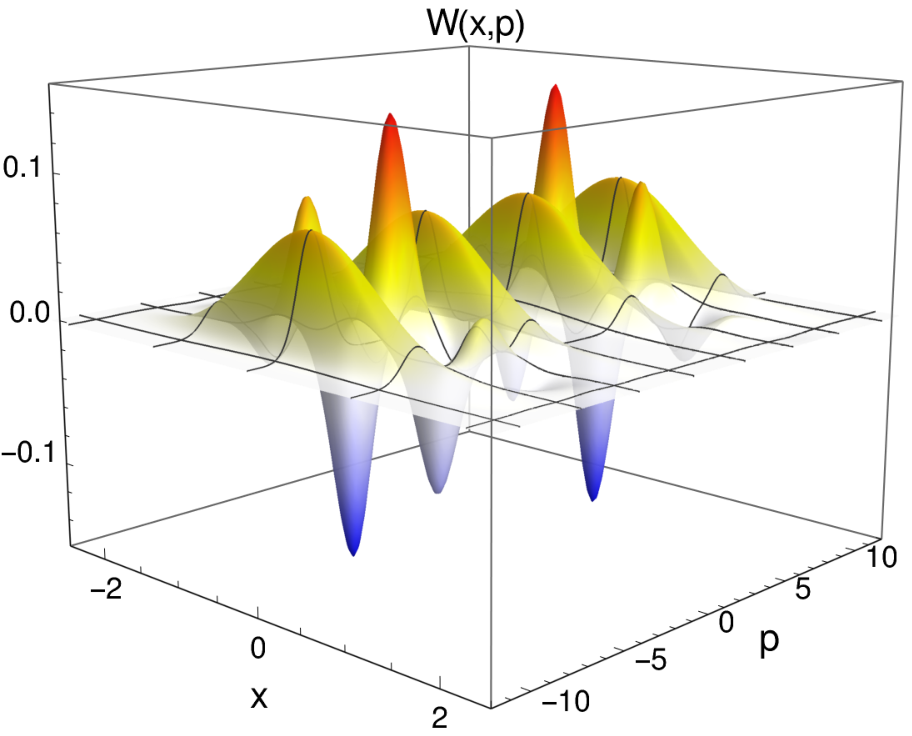}
\caption{\textit{Wigner function $W(x,p,t=2\pi N)$ for $\rho
=1/\sqrt{2}$. It takes both positive and negative values
demonstrating entanglement between the qubit states and "cat states"
of the nanomechanical resonator.}}\label{fig:fig3_5}
\end{figure}

\section{Time-averaged electric current.} \label{sect3_6}

As it follows from the above consideration, the amplitude of
mechanical fluctuations, and therefore the energy stored in the
mechanical subsystem, changes over time.  This energy comes from the
electronic subsystem causing a rectification of ac current. To
analyze this phenomenon, we calculate the dimensionless (normalized
to $I_0=2e/\hbar$) ac Josephson current averaged over the N-th
period of the Josephson oscillations,
$$I_N=\frac{1}{2\pi}\int\limits^{2\pi N}_{2\pi(N-1)}dt
\mathrm{Tr}\left(\frac{\partial \hat H_q(t)}{\partial\phi}\hat
\varrho(t)\right).$$  
Taking into account that $\partial \hat
H_q/\partial\Phi=\eta\partial \hat H/\partial t$ and $\hat
H_q(t=2\pi N)=0$, one gets the following expression for $I_N$,
\begin{eqnarray}\label{current}
&&I_N=\frac{\kappa}{2\pi}\nabla_N\mathrm {Tr}
\left(\hat H_m+\hat H_{int}\right)\hat\varrho(2\pi N) \nonumber \\
&&\hspace{0.6cm}=\frac{\kappa}{2\pi}\nabla_N
\left[E_m(N)+E_{int}(N)\right],
\end{eqnarray}
where $\nabla_N f(N)\equiv f(N)-f(N-1)$ is the first difference.
From this equation, one can see that the average current is given by
the change of the mechanical energy $E_{m}$ and the energy of interaction
$E_{int}$ after N-th period. One can find that at
$N>Ns=[t_s/2\pi]+1$ the functions $E_m(N)$ and $E_{int}(N)$ can be
written as follows,
\begin{eqnarray}
&&E_m(N)=2\pi^2\tilde\varepsilon^2\left(\rho^2(2N_s-N)^2
+\tau^2 N^2\right), \nonumber \\
&&E_{int}(N)=2\pi
\varepsilon\tilde{\varepsilon}\left[\rho^2\left(N-2N_s\right)
\text{e}^{-(2\pi\tilde\varepsilon)^2(N-2 N_{s})^{2}} +\tau^2 N\text{e}^{-(2\pi\tilde\varepsilon
N)^2 }\right].
\end{eqnarray}

The change in the interaction energy contributes to the averaged current
as well as the mechanical energy. However, this contribution is of the
order of $\tilde\varepsilon^2$ and important only for periods for
which $I(N)/\tilde\varepsilon\simeq\tilde\varepsilon^2$. Thus, the
average current is determined by the change of mechanical energy
mainly, and is defined by the following equations,
\begin{eqnarray}\label{I}
&&\frac{I(N)}{\tilde\varepsilon}\approx
I_{m}(N)=-2\pi\tilde\varepsilon N,
\hspace{0.5cm} N\leq N_s-1\\
&&\frac{I(N)}{\tilde\varepsilon}\approx
2\pi\tilde\varepsilon\left( N-2\rho^2N_s\right), \hspace{0.5cm} N>
N_s.
\end{eqnarray}
\begin{figure}
\centering
\includegraphics[width=0.55\columnwidth]{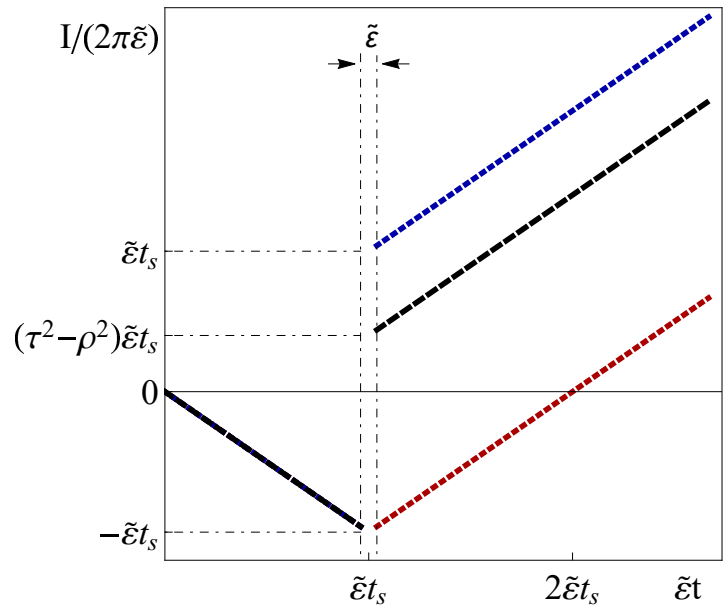}
\caption{\textit{Schematic illustration of the time-averaged Josephson
current as a function of time for different values of $\rho$ (black dashed curve). The dotted lines indicate the limiting cases of $\rho=0$ (top, blue online) and $\rho=1$ (bottom, red online) The current for $t<t_s$ does not depend on $\rho$ ($\rho=1$), see Eqs.~(\ref{I}),(\ref{Sw1}).
The period $N_s$ corresponding to the moment of the bias voltage switching, is out of the consideration.}}
\end{figure}\label{fig:fig3_6}
From Fig.~\ref{fig:fig3_6} one can see that the averaged current exhibits a jump
equal to $-\rho^2 I(N_s)$ after the period during which the bias
voltage is switched. It originates in the fact that when we switch the
sign of the bias voltage (at $t=t_s$) the power, pumped into the
mechanical subsystem, changes depending on the magnitude of $\rho^2$.
For $\rho=1$, the supplied power, $P=I V$, just changes its sign with the bias voltage,
and the current continues to flow in the same direction as it did
before switching. For $\rho=0$ supplied power is not changed and
consequently the current direction changes after switching.


\section*{Conclusions}
\addcontentsline{toc}{section}{Conclusions}

In this chapter the quantum dynamics of the NEMS
comprising the movable CPB qubit, subjected to an electrostatic
field and coupled to two bulk superconductors, controlled by the bias voltage,
via tunneling processes, is analyzed. It is demonstrated 
analytically that if the ac Josephson frequency of superconductors,
controlled by the bias voltage, is in resonance with the mechanical
frequency of the CPB, the initial pure state (direct product of the
CPB state and ground state of the oscillator) evolves in time into
the coherent states of the mechanical oscillator entangled with the
qubit states. Furthermore, we established the protocol of the bias
voltage manipulation which results in the formation of entangled
states incorporating so-called cat-states (the quantum superposition
of the coherent states). The organization of such states is confirmed
by the analysis of the corresponding  Wigner function taking
negative values, while their specific features provide the
possibility for their experimental detection by measuring the
average current. The discussed phenomena may serve as a foundation
for the encoding of quantum information from charge qubits into a
superposition of the coherent mechanical states.  It may constitute
interest  for the field of quantum communications due to the
robustness of such multiphonon states regarding external
perturbation, comparing to a single-phonon Fock state.

The main results of this chapter are published, Refs.~\cite{Bahrova2021, BahrovaC2021}.

\clearpage           
\chapter{NANOMECHANICS DRIVEN BY SUPERCONDUCTING PROXIMITY EFFECT} \label{chapt4}

In this chapter a hybrid nanoelectromechanical weak link based on a nanotube is considered. More precisely, a carbon nanotube is suspended above a trench in a normal metal electrode and positioned in a gap between two superconducting ones. It is shown that under a constant bias voltage, in such a system the mechanical instability, resulting in self-sustained nanotube oscillations, occurs, on the one hand. On the other hand, the system can also operate in cooling regime. The phenomena emerge due to superconducting proximity effect (the hybrid (normal-superconducting) structure of the device). In the first section~(\ref{sect4_1}) bending vibrations of the nanotube are treated semi-classically and details of mechanical instability leading to a self-saturation effect is discussed together with an scheme for an possible experimental detection of the considered effects. In the second section~(\ref{sect4_2}) quantum effects are taken fully into account. It is demonstrated that quantum fluctuations of the nanotube lead to the cooling effect, which can be observed in an experiment due to the electric current measurement discussed in the last subsection.

\section{Self-sustained nanomechanical oscillations.} \label{sect4_1}

In this section the occurrence of the mechanical instability in the hybrid nanoelectromechanical system is discussed. A region and required conditions for it are obtained. Also, a strong enhancement of the electric current through the system in the stationary regime of self-sustained oscillations of the nanotube is found. The latter leads to the possibility for the device to operate as a transistor or a diode.

\subsection{Model of nanoelectromechanical device. Hamiltonian and dynamics.} \label{subsect4_1_1}

\begin{figure}
\centering
\includegraphics[width=.7\columnwidth]{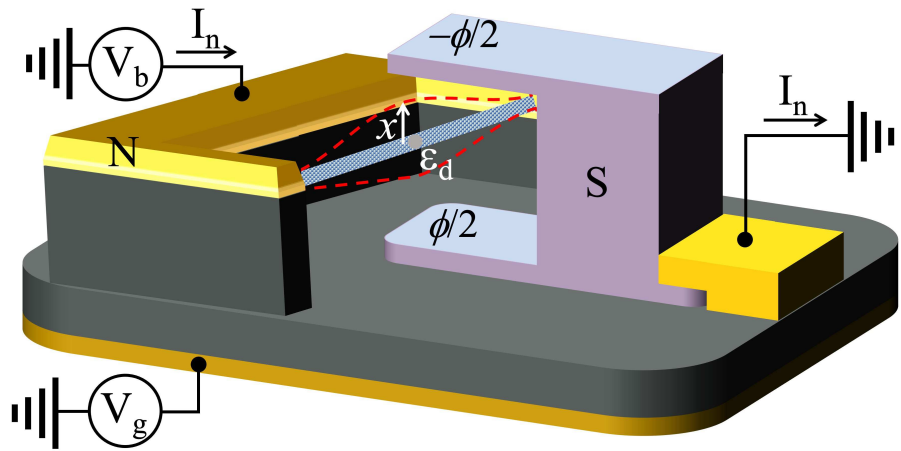}
\caption{\textit{Schematic illustration of the nanoelectromechanical device under consideration. A carbon nanotube (CNT) is suspended in a gap between two edges of a normal electrode ($N$) and tunnel-coupled to it. The electronic energy levels of the CNT are tuned such that only one energy level with energy $\varepsilon_d$, which is well separated from the other levels, is considered. Bending of the CNT in the $x$ direction between two superconducting leads ($S$) affects the values of the tunneling barriers between them. The bias voltage $V_b$ is applied to the normal electrode.
}}\label{fig:fig4_1_1}
\end{figure}

A sketch of the NEMS investigated in this paper is presented in Fig.~\ref{fig:fig4_1_1}. A single-walled CNT is suspended above a trench in a bulk normal metal electrode
biased by a constant voltage $V_b$. Two superconducting
leads with the superconducting phase difference $\phi$ are positioned
near the middle of the nanotube in such a way that the bending of
the nanotube moves it closer to one electrode and further
away from the other. The distance between the quantized electronic
levels inside the nanotube is much greater than the other energy
parameters, allowing one to consider the nanotube as a single-level quantum dot (QD). The bending dynamics of the CNT are reduced to the
dynamics of the fundamental flexural mode. We suppose that the
amplitude of this mode, $x$, is larger than the amplitude of
zero-point oscillations. Thus, we consider it as a classical
mechanical oscillator with mass $m$ and frequency $\omega$. 

The dynamics of the mechanical subsystem is described by Newton's equation,
\begin{equation}\label{xEq1}
    \ddot{x}+\omega^2 x=-\frac{1}{m}\text{Tr}\left\{\hat{\rho}\frac{\partial H(x)}{\partial x}\right\},
\end{equation}
where
\begin{equation}\label{H}
    H=H_d+H_l+H_t
\end{equation}
is the Hamiltonian of the electronic subsystem. The first term
$H_d$ represents the single-level QD,
\begin{equation}\label{Hd}
    H_d=\sum_{\sigma}\varepsilon_d d^\dag_\sigma d_\sigma,
\end{equation}
where $d_\sigma^\dag (d_\sigma)$ is the creation (annihilation) operator of an electron with spin projection $\sigma=\uparrow,\downarrow$ on the dot.
The Hamiltonian $H_l=H_l^n+H_l^s$ describes the normal and superconducting leads, respectively, with
\begin{eqnarray}\label{Hl}
   && H_l^n=\sum_{k\sigma }(\varepsilon_k-eV_b)a_{k \sigma}^\dag a_{k\sigma},\\
    && H_l^s=\sum_{k j\sigma}\left(\varepsilon_{k}c^\dag_{kj\sigma}c_{kj\sigma}-\Delta_{s} (\text{e}^{\imath\phi_j } c^\dag_{kj\uparrow}c^\dag_{-kj\downarrow}+\text{H.c.})\right).
\end{eqnarray}
Here, $a^\dag_{k\sigma} (a_{k\sigma})$, and $c^\dag_{kj\sigma}
(c_{k\sigma})$ are the creation (annihilation) operators of an electron
with quantum number $k$ and spin projection $\sigma$ in the normal
and superconducting $j=1,2$ leads, respectively, and
$\Delta_{s}\text{e}^{\imath\phi_j}$ is the superconducting order
parameter (in the $j$ electrode). Note that the energies
$\varepsilon_{d},\varepsilon_{k}$ are counted from the Fermi energy of
the superconductors. In what follows, we set $\phi_{1}=-\phi_{2}=\phi/2$.

The Hamiltonian $H_t=H_t^n+H_t^s$ describes the tunneling of
electrons between the dot and the leads, where
  \begin{eqnarray}\label{Ht}
  && H_t^n=\sum_{k\sigma}t^n_0 (a_{k\sigma}^\dag d_\sigma +\text{H.c.}), \\
 &&  H_t^s=\sum_{k j\sigma}t^s_{ j}(x) (c^\dag_{kj\sigma}d_\sigma +\text{H.c.}).\label{Hts}
\end{eqnarray}
Here is the position-dependent superconducting tunneling amplitude
\begin{equation}\label{tunnela}
  t^s_{1(2)}(x)=t^s_0 \text{e}^{(-1)^j (x+a)/2\lambda},  
\end{equation}
 where $2\lambda$ is the characteristic tunneling length and $a$ is a
parameter for asymmetry. For a typical CNT-based nanomechanical
resonator, $2\lambda\sim 0.5$ nm~\cite{Morpurgo1999}. We concentrate our attention on the symmetric case
$a=0$ and leave the asymmetric one for a brief discussion in the subsection~\ref{subsect4_1_5} because taking into account the asymmetry does not bring any qualitative result, as we will show. Also, the completely asymmetric case (only one SC electrode is present) is discussed in Ref.~\cite{Parafilo2022}.

\subsection{Density matrix approximation.} \label{subsect4_1_2}

The time evolution of the electronic density matrix $\hat{\rho}$ is described by the Liouville–von Neumann equation ($\hbar=1$),
  \begin{equation}\label{LvN}
      \imath \partial_t \hat{\rho}=[H,\hat{\rho}],
  \end{equation}
which together with Eq.~(\ref{xEq1}) forms a closed system of
equations that describe the nanoelectromechanics of our system. We restrict ourselves to the case $\Delta_{s}\gg|eV_b|
\gg \Delta_d \sim \Gamma_n$, where $\Delta_d=(2\pi)\nu_{s}|t^{s}_0|^2$ and $\Gamma_n=(2\pi)\nu_{n}|t^{n}_0|^2$, with $\nu_{s(n)}$ the density of
states in the superconducting (normal) electrode.

To describe the electronic dynamics of the QD, we use the reduced density
matrix approximation in which the full density matrix of the system
is factorized to the tensor product of the equilibrium density
matrices of the normal and superconducting leads and the density matrix of the dot as
$\hat{\rho}=\hat{\rho}_n\otimes\hat{\rho}_s\otimes\hat{\rho}_d$.
Using the standard procedure, one can trace out the degrees of
freedom of the leads
 and obtain the following equation for the reduced density matrix $\hat{\rho}_d$~\cite{Parafilo2020} (in the deep subgap regime $\Delta_{s}\rightarrow
\infty$),
\begin{equation}\label{EDM}
  \partial_t \hat{\rho}_d=-\imath \left[H_d^{eff},\hat{\rho}_d \right]+\mathcal{L}_n\{\hat{\rho}_d\},
\end{equation}
where
\begin{equation}\label{Hdeff}
    H^{eff}_d=H_d+\Delta_{d}(x,\phi)d_{\downarrow}d_{\uparrow} +\Delta_{d}^* (x,\phi) d_\uparrow^\dag d_\downarrow^\dag,
\end{equation}
\begin{eqnarray}\label{deltad}
  && \Delta_{d}(x,\phi)=\dfrac{1}{2}\Delta_d\sum\limits_{j=1,2}\text{e}^{(-1)^{j}(x/\lambda+\imath\phi/2)}=\Delta'(x,\phi)+i\Delta''(x,\phi)\nonumber\\
 &&  \hspace{2cm}= \Delta_d \cosh(x/\lambda+i\phi/2).
\end{eqnarray}

Above, $\Delta_d(x,\phi)$ is the off-diagonal order parameter induced by the superconducting
proximity effect~\cite{Rozhkov2000, Stadler2016}, and $\Delta^{\prime,\prime\prime}(x,\phi)$ are real functions. The Lindbladian term in
Eq.~(\ref{EDM}) reflects the incoherent electron exchange
between the normal lead and QD. The latter in the high bias voltage
regime, $|eV_b|\gg \varepsilon_0, k_BT$, takes the form:
\begin{equation}\label{Ln}
\mathcal{L}_n\{\hat{\rho}_d\}=\Gamma_n\sum\limits_\sigma
\begin{cases}
2 d^\dag_\sigma\hat{\rho}_d d_\sigma-\left\{ d_\sigma d_\sigma^\dag ,\hat{\rho}_d \right\} , & \kappa =+1; \\
2 d_\sigma\hat{\rho}_d d^\dag_\sigma-\left\{ d^\dag_\sigma d_\sigma ,\hat{\rho}_d \right\} , & \kappa=-1;
\end{cases}
\end{equation}
where $\kappa=\text{sgn}(eV_b)$.

Another way to obtain the effective Hamiltonian Eq.~(\ref{Hdeff}) is to use the equation of motion method within the Green function formalism, see, e.g., Ref.~\cite{Flensberg2004} and the appendix in Ref.~\cite{Stadler2016}. The idea of this well-established method is to obtain a series of coupled differential equations for a desired Green function by differentiating it several times. 

\begin{figure}
    \centering
    \begin{subfigure}[t]{0.45\columnwidth}
        \centering
        \subfloat[\textit{a)}]{\includegraphics[width=\columnwidth, keepaspectratio]
        {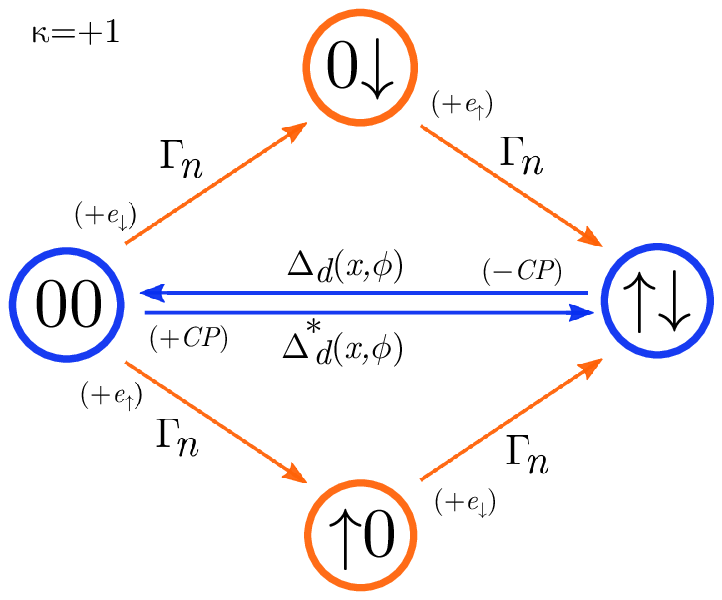}}
     \caption{} \label{fig:subfig4_1_2a}
    \end{subfigure}
    \quad
    \begin{subfigure}[t]{0.45\columnwidth}
        \centering
       \subfloat[\textit{b)}]{\includegraphics[width=\columnwidth, keepaspectratio]
        {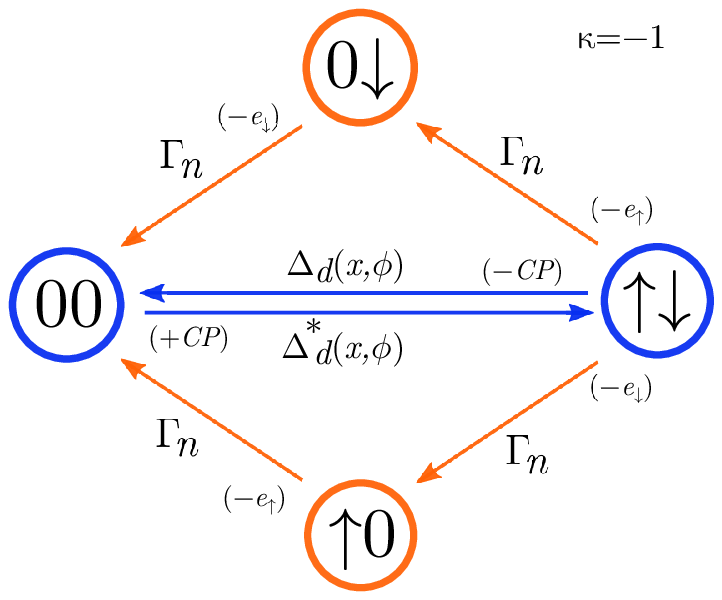}}
       \caption{}\label{fig:subfig4_1_2b}
    \end{subfigure}
    \caption{\textit{Diagrams representing the transitions between electronic states in the quantum dot. The single-electron states change due to transitions from the empty to the single-occupied QD and then from the single-occupied to the double-occupied one (indicated by orange arrows). In the high bias voltage regime, the tunneling of electrons (a) or holes (b) with spin $\downarrow$ or $\uparrow$ is allowed only from the normal lead to the dot and forbidden in the opposite direction. Transitions between the empty and double-occupied quantum dot are due to coupling with the superconducting leads (indicated by blue arrows).} }\label{fig:fig4_1_2}
\end{figure}

Figure~\ref{fig:fig4_1_2} represents the electronic dynamics on the dot for
$\kappa=\pm1$. From Fig.~\ref{fig:fig4_1_2}, one can see that not all electron processes are allowed due to the parameter scales in this work. In the subgap regime, single-electron transitions between the dot and the superconducting leads are prohibited, and thus only an exchange of Cooper pairs occurs. Moreover, because of the high bias voltage, single-electron tunneling between the dot and the normal leads is enabled exclusively in one direction (from the lead to the dot, see Fig.~\ref{fig:fig4_1_2}a, or vice-versa, Fig.~\ref{fig:fig4_1_2}b), establishing that our model is electron-hole symmetric.

As a consequence, the QD density matrix $\hat{\rho}_d$ acts in the Hilbert  space $\mathcal{H}_{4}$, which may be presented as a direct sum of
two $\mathcal{H}_{2}$ spaces via $\mathcal{H}_{4}=\mathcal{H}_{e}\oplus\mathcal{H}_{CP}$ spanned over state vectors
$\vert\uparrow\rangle = d^\dag_\uparrow\vert 0\rangle$, $\vert\downarrow \rangle = d^\dag_\downarrow\vert 0\rangle$,
and $\vert 0\rangle$, $\vert 2\rangle = d^\dag_\uparrow d^\dag_\downarrow\vert 0\rangle$ (with $d_{\uparrow,\downarrow}\vert 0\rangle=0$).
Then, the equations for the dot density matrix are the following:
\begin{eqnarray}\label{EDME}
    && \partial_t\rho_0=-4\Gamma_n \rho_0-\imath\Delta_d(x,\phi)\rho_{20}+\imath\Delta_d^*(x,\phi)\rho_{02}, \\
    &&\partial_t\rho_2=2\Gamma_n(1-\rho_0-\rho_2)+\imath\Delta_d(x,\phi)\rho_{20}-\imath\Delta_d^*(x,\phi)\rho_{02} ,\\
    && \partial_t\rho_{02}=-2\Gamma_n\rho_{02}+\imath\Delta_d(x,\phi)(\rho_0-\rho_2)+2\imath \varepsilon_d\rho_  {02}, \\
    && \partial_t\rho_{20}=-2\Gamma_n\rho_{20}-\imath\Delta_d^*(x,\phi)(\rho_0-\rho_2)-2\imath\varepsilon_d\rho_{20}.
\end{eqnarray}
Here we use the normalization condition $\rho_0+\rho_\uparrow+\rho_\downarrow+\rho_2=1$. The equation for the displacement $x$,~Eq.(\ref{xEq1}), has the form,
\begin{equation}\label{Ex}
    \ddot{x}+\omega x=-\frac{2\Delta_d}{\lambda}\left[\sinh{\left(\frac{x}{\lambda}-\imath\frac{\phi}{2}\right)}\rho_{02}+ \sinh{\left(\frac{x}{\lambda}+\imath\frac{\phi}{2}\right)}\rho_{20}\right].
\end{equation}

The superselection rule, which forbids the superposition of states with integer and half-integer spins, allows us to present the density matrix $\hat{\rho}_d$
as a direct sum of two density matrices $\hat{\rho}_d=\hat{\rho}_e\oplus\hat{\rho}_{CP}$ acting in the
$\mathcal{H}_{2}$ Hilbert space spanned over state vectors $\vert\uparrow\rangle,\vert\downarrow\rangle$ and $\vert 0\rangle,\vert 2\rangle$, respectively.
 Moreover, taking into account spin-rotation symmetry, one can conclude that $\hat{\rho}_e$ should be proportional to the unit matrix, $\hat{\rho}_e =\rho_e\hat{I}$,
 while $\hat{\rho}_{CP}$ can be written in the form
 \begin{equation}\label{rhocp}
     \hat{\rho}_{CP}=\frac{1}{2}R_{0}\hat{I}+\frac{1}{2}\sum_i R_i\sigma_i,
 \end{equation}
where $\sigma_i ,(i=1,2,3)$ are the Pauli matrices.
Here new variables are $R_i=\text{Sp}(\sigma_i\hat{\rho}_{CP})$, with
\begin{equation}\label{matrixrho}
  \hat{\rho}_{CP}=  \begin{pmatrix}
    \rho_0&\rho_{02}\\
    \rho_{20}&\rho_2
    \end{pmatrix},
\end{equation}
(the equation for $R_0=\frac{1}{2}\text{Sp}\hat{I}\hat{\rho}=(\rho_{0}+\rho_2)/2$ is decoupled and is not relevant).

Then by introducing the dimensionless time $\omega t\rightarrow t$ and displacement $x/\lambda \rightarrow x$, and taking into account the normalization condition
 $\text{Tr}\hat{\rho}_d=1$, we get the following closed system of equations for $x(t)$ and $R_{i}(t)$,

\begin{equation}
   \ddot{x}+x=-\xi\left[\sinh{(x)}\cos{\left(\frac{\phi}{2}\right)}R_{1}-\cosh{(x)}\sin{\left(\frac{\phi}{2}\right)}R_{2} \right],\label{xEq}
\end{equation}
\begin{equation}
    \alpha\dot{\vec{R}}=\hat{L}\vec{R}- \kappa\vec{e}_3,\label{EqR}
\end{equation}
where $\vec{R}=(R_{1},R_{2},R_{3})^{T}$, $\vec{e_3}=(0,0,1)^T$, $\xi=\Delta_{d}/( m\lambda^{2}\omega^{2})$ is the
nanoelectromechanical coupling parameter, and $\alpha=\omega /(2\Gamma_n)$ is the adiabaticity parameter. For a typical CNT-based
NEMS, one can estimate $\xi\sim 10^{-3}\ll 1$~\cite{Morpurgo1999, Moser2014}. The
matrix $\hat{L}$ is defined as follows,

\begin{equation}\label{matrixA}
\hat{L}(x)=
\begin{pmatrix}
-1 &  \tilde{\varepsilon}_d & -  \tilde{\Delta}'' (x,\phi) \\
-\tilde{\varepsilon}_d & -1 & -\tilde{\Delta}'(x,\phi) \\
\tilde{\Delta}''(x,\phi) & \tilde{\Delta}'(x,\phi) & -1
\end{pmatrix},
\end{equation}
where $\tilde{\varepsilon}_d\equiv\varepsilon_d/\Gamma_n, \tilde{\Delta}_d\equiv\Delta_d/\Gamma_n$.

\subsection{Equation of motion method for Green functions.}\label{subsect4_1_3}

In this subsection we present another method to obtain the effective Hamiltonian Eq.~(\ref{Hdeff}) as it was pointed out in the previous subsection,~\ref{subsect4_1_2}. It is based on the equation of motion method within the Green function formalism, see, e.g., Ref.~\cite{Flensberg2004} and the appendix in Ref.~\cite{Stadler2016}. The idea of this well-established method is to obtain a series of coupled differential equations for a desired Green function by differentiating it several times. 

Since in the considered system the dot coupled to superconducting and normal leads independently, in the following derivation we can omit the contribution of the normal lead. Let us define the retarded Green function in the 2x2 Nambu space as~\cite{Cuevas1996},
\begin{equation}\label{gr1}
    \hat{G}^r(t)=-\imath \theta(t)
    \begin{pmatrix}
    \langle\{d_\uparrow (t),d_\uparrow^\dag (0)\}\rangle & \langle\{ d_\uparrow(t), d_\downarrow (0)\}\rangle\\
    \langle \{d_\downarrow^\dag (t), d_\uparrow^\dag (0)\}\rangle & \langle\{ d_\downarrow^\dag (t), d_\downarrow (0)\}\rangle
    \end{pmatrix}.
\end{equation}

Then, by differentiating this function two times, we find for the Fourier harmonics
\begin{equation}
 \hat{G}^r (\tau)=\int_{-\infty}^{+\infty} \hat{G}^r (t)\text{e}^{-\imath \tau t},  \qquad \hat{G}^r (t)=\frac{1}{2\pi}\int_{-\infty}^{+\infty} \hat{G}^r (\tau)\text{e}^{\imath \tau t},
\end{equation}
the following closed system of equations:
\begin{eqnarray}
       &&\hat{g}_0^{-1}(\tau)\hat{G}^r(\tau)=\hat{I}+\sigma_z\sum_{kj}t^s_{kj}\hat{G}^{r,1}_j(k,\tau),\label{gr2}\\
       &&\hat{g}_{kj}^{s,-1}(\tau)\hat{G}_j^{r,1}(k,\tau)=\sigma_z t_{kj}^s\hat{G}^r(\tau),\label{gr3}
\end{eqnarray}
where $\hat{G}^{r,1}_j(\tau)$ is Fourier harmonics of the following correlation function:
\begin{equation}\label{gr40}
    \hat{G}^{r,1}_j(k,t)=-\imath\theta(t)
    \begin{pmatrix}
    \langle\{c_{kj\uparrow}(t),d^\dag_\uparrow(0)\}\rangle & \langle \{ c_{kj\uparrow}(t),d_\downarrow(0)\}\rangle \\
    \langle\{ c^\dag_{kj\downarrow}(t),d_\uparrow^\dag (0)\}\rangle & \langle\{ c_{kj\downarrow}^\dag (t), d_\downarrow (0)\}\rangle
    \end{pmatrix}.
\end{equation}
Here also the Green function of a single-level QD with the level energy $\varepsilon_d$,
\begin{equation}\label{gr0}
    \hat{g}_0(\tau)=
    \begin{pmatrix}
    \frac{1}{\tau-\varepsilon_d+\imath 0} & 0\\
    0 & \frac{1}{\tau+\varepsilon_d+\imath 0}
    \end{pmatrix},
\end{equation}
with its inverse matrix,
\begin{equation}\label{gr0inv}
    \hat{g}_0^{-1}(\tau)=
    \begin{pmatrix}
   \tau-\varepsilon_d+\imath 0 & 0\\
    0 & \tau+\varepsilon_d+\imath 0
    \end{pmatrix},
\end{equation}
and the (standard) Green function of the $j$ superconductor:
\begin{equation}\label{grs}
    \hat{g}^s_{kj}(\tau)=\frac{\imath}{\tau^2-E_k^2}
    \begin{pmatrix}
    \tau+\varepsilon_k & -\Delta_s\text{e}^{\imath\phi_j}\\
    -\Delta_s\text{e}^{-\imath\phi_j} & \tau -\varepsilon_k
    \end{pmatrix},
\end{equation}
where the eigenvalues of the Hamiltonian $H_l^s$ are $E_k=\pm\sqrt{\varepsilon_k^2+\Delta_s^2}$ (Andreev level energies).

Then, by substituting Eq.~(\ref{gr3}) into Eq.~(\ref{gr2}), one gets the following Dyson equation for the retarded Green function of the dot $\hat{G}^r(\tau)$ :
\begin{equation}\label{grd}
    \hat{G}^r(\tau)=\hat{g}_0(\tau)+\hat{g}_0(\tau)\hat{\Sigma}^r(\tau)\hat{G}^r(\tau),
\end{equation}
where the self-energy function is determined as
\begin{equation}\label{grsigma}
    \hat{\Sigma}^r(\tau)=\imath \sum_{kj}(t^s_{kj})^2\hat{g}_{kj}^s (-\Delta_s,\tau),
\end{equation}
and after the integrating over the energy variable, we obtain:
\begin{equation}\label{grsigma2}
     \hat{\Sigma}^r(\tau)=\frac{1}{2}\sum_j \frac{\Delta_d(x)}{\sqrt{\Delta_s^2-\tau^2}}
     \begin{pmatrix}
     \tau & \Delta_s\text{e}^{\imath\phi_j}\\
     \Delta_s\text{e}^{-\imath\phi_j} & \tau
     \end{pmatrix}.
\end{equation}
Moreover, in case of symmetric tunneling contacts with the superconducting leads, one can re-write this equation as follows:
\begin{equation}\label{grsigma3}
    \hat{\Sigma}^r(\tau)=\frac{\Delta_d}{\sqrt{\Delta_s^2-\tau^2}}
     \begin{pmatrix}
     \tau \cosh{(x/\lambda)} & \Delta_s\cosh{\left(x/\lambda-\imath\phi/2\right)}\\
     \Delta_s\cosh{\left(x/\lambda+\imath\phi/2\right)} & \tau\cosh{(x/\lambda)}
     \end{pmatrix}.
\end{equation}

Furthermore, from a formal solution of Eq.~(\ref{grd})
\begin{equation}
    \hat{G}^r(\tau)=\left[ \hat{g}_0^{-1}+\hat{\Sigma}^r\right]^{-1},
\end{equation}
one gets the following expression, 
\begin{equation}
    \hat{G}^r(\tau)=\frac{1}{\det (\hat{g}_0^{-1}-\hat{\Sigma}^r)}
    \begin{pmatrix}
    g^{-1}_{22}-\Sigma^r_{22} & g^{-1}_{12}-(-\Sigma^r_{12})\\
   g^{-1}_{21}-(-\Sigma^r_{21}) & g^{-1}_{11}-\Sigma^r_{11}
    \end{pmatrix},
\end{equation}
or, more specifically, see, e.g., Refs.~\cite{Sun2000,Stadler2016},
\begin{equation}\label{gr5}
    \hat{G}^r(\tau)=\frac{1}{D(\tau)}
    \begin{pmatrix}
    g^{-1}_{22}-\Sigma^r_{22} & \Sigma^r_{12}\\
  \Sigma^r_{21} & g^{-1}_{11}-\Sigma^r_{11}
    \end{pmatrix},
\end{equation}
with 
\begin{equation}
    D(\tau)=\left( g^{-1}_{11}-\Sigma^r_{11}\right)\left( g^{-1}_{22}-\Sigma^r_{22}\right)-\Sigma^r_{12}\Sigma^r_{21},
\end{equation}
by definition.
In the deep subgap regime ($\Delta_s\to\infty$) straightforward calculations leads to:
\begin{equation}\label{gr6}
    \hat{G}^r(\tau)=\frac{1}{D(\tau)}
    \begin{pmatrix}
    \tau+\varepsilon_d & \Delta_d^*(x,\phi)\\
    \Delta_d(x,\phi) & \tau-\varepsilon_d,
    \end{pmatrix}
\end{equation}
\begin{equation}\label{Dtau}
    D(\tau)=\tau^2-\varepsilon_d^2-\vert \Delta_d(x,\phi)\vert^2.
\end{equation}

One the other hand, let us consider an effective QD Hamiltonian of such a type of Eq.~(\ref{Hdeff}),
\begin{equation}\label{Heff2}
    H'=\sum_\sigma \varepsilon_d d^\dag_\sigma d_\sigma +\varUpsilon d_\downarrow d_\uparrow +\varUpsilon^\dag d_\uparrow^\dag d_\downarrow^\dag,
\end{equation}
where $\varUpsilon$ is supposed to be unknown for now function, or an operator-function in more general situation, of the dot position $x$ and superconducting phase difference $\phi$. Then, one can find the retarded Green function, Eq.~(\ref{gr1}), of a dot described by the Hamiltonian Eq.~(\ref{Heff2}) using the method of equations of motion which was employed above. By differentiating only one time, one gets the following expression for the desired Green function Fourier component:
\begin{equation}\label{gr10}
    \hat{G}^r(\tau)=\frac{1}{\tau^2-\varepsilon_d^2-\vert\varUpsilon\vert^2}
    \begin{pmatrix}
    \tau+\varepsilon_d & \varUpsilon^\dag\\
    \varUpsilon & \tau -\varepsilon_d
    \end{pmatrix}.
\end{equation}
At this point one can compare Eq.~(\ref{gr10}) with Eq.~(\ref{gr6}) and as a matter of fact justify that $\varUpsilon=\Delta_d(x,\phi)$.

\subsection{Dynamics of the system in adiabatic regime.} \label{subsect4_1_4}

The system of Eqs.~(\ref{xEq}) and (\ref{EqR}) has an obvious static
solution $x_{st}=0+\mathcal{O}(\xi)$,
$\vec{R}_{st}=\kappa
L^{-1}(0)\vec{e}_{3}+\mathcal{O}(\xi)\vec{R}^{(1)}$, here
$\|\vec{R}^{(1)}\|=1$. The stability of this solution can then be
investigated in standard ways, see, for example, Ref.~\cite{Ilinskaya2018} and below.
However, to simplify this procedure, we will consider the adiabatic
case when $\alpha\ll 1$, which corresponds to a
typical experimental situation~\cite{Willick2020} and reduces the
problem to one that allows the use of Poincare analysis. More specifically, this
inequality allows one to find a solution of Eq.~(\ref{EqR}) to the
accuracy $\alpha$,
\begin{equation}\label{Rexpand}
    \vec{R}(x,t)= \kappa L^{-1}(x(t))(1+\alpha
\dot{x}\partial_{x}L^{-1}(x(t))+\mathcal{O}(\alpha^{2}))\vec{e}_{3}.
\end{equation}

In the zeroth order of the perturbation theory over the adiabaticity parameter $\alpha$, from Eq.~(\ref{Rexpand}) one can find:
\begin{eqnarray}
     && R^{(0)}_1(x,t)=\kappa\frac{\tilde{\Delta}''(x,\phi)+\tilde{\varepsilon}_d\tilde{\Delta}'(x,\phi)}{\tilde{\mathcal{D}}^2(x,\phi)},\\ &&R^{(0)}_2(x,t)=\kappa\frac{\tilde{\Delta}'(x,\phi)-\tilde{\varepsilon}_d\tilde{\Delta}''(x,\phi)}{\tilde{\mathcal{D}}^2(x,\phi)},\\ &&R^{(0)}_3(x,t)=-\kappa\frac{1+\tilde{\varepsilon}_d^2}{\tilde{\mathcal{D}}^2(x,\phi)}.
\end{eqnarray}\label{solutionR0}

Here
$\tilde{\mathcal{D}}^{2}\equiv\mathcal{D}^2(x,\phi)/\Gamma_n^2=\tilde{\Delta}_d^2\left[ \sinh^2{x}+\cos^2{(\phi/2)}\right]+\tilde{\varepsilon}^{2}_{d}+1$.

The equation for the first order correction to the re-normalized matrix elements of the dot density operator, using Eq.~(\ref{Rexpand}), takes the form:
\begin{equation}\label{R1}
    \vec{R}^{(1)}(x,t)=\alpha \dot{x}L^{-1}(x(t))\partial_x\vec{R}^{(0)}(x,t).
\end{equation}

The coefficients $R^{(1)}_1$ and $R^{(1)}_2$ can be expressed in terms of $R^{(1)}_3$ which has the following form
\begin{equation}\label{R3}
    R^{(1)}_3(x,t)=\alpha \kappa\dot{x}\frac{\tilde{\Delta}_d^2}{\tilde{\mathcal{D}}^6}\left[\sinh{(2x)}\left\{\left
    (1-\tilde{\varepsilon}_d^2\right)\tilde{\mathcal{D}}^2- 4\left( 1+\tilde{\varepsilon}_d^2\right)\right\}+2\tilde{\varepsilon}_d\tilde{\mathcal{D}}^2\sin{\phi}  \right].
\end{equation}

Then by substituting these solutions into Eq.~(\ref{xEq}), one gets
 (to accuracy $\alpha$) the following nonlinear differential equation
for $x(t)$:
\begin{equation}\label{xEqsol}
    \ddot{x}-\eta(x,\phi)\dot{x}+x=F(x,\phi),
\end{equation}
the solution of which may be analyzed via Poincare's theory. Here, the
nonlinear force $F(x,\phi)$ and friction coefficient $\eta(x,\phi)$, which in what follows we refer to as a pumping coefficient,
are generated by the interaction with the non-equilibrium electronic environment. In terms of $R^{(1)}_3$, the pumping coefficient takes a form:
\begin{eqnarray}
  &&\dot{x}\eta (x,\phi)=\frac{\tilde{\Delta}_d}{1+\tilde{\varepsilon}_d^2}\left[-\left[\sin{\phi}-\tilde{\varepsilon}_d\sinh{(2x)}\right] R_3^{(1)}+\nonumber\right.\\
  &&\hspace{2.cm}\left. +\frac{\alpha\kappa\dot{x}}{\tilde{\mathcal{D}}^4}\left\{2\tilde{\varepsilon}_d\tilde{\mathcal{D}}^2\left(\cosh{(2x)}-\cos{\phi}\right) +\nonumber\right.\right.\\
  &&\hspace{2.cm}\left.\left. +\tilde{\Delta}_d^2\sinh{(2x)}\left[\left(1-\tilde{\varepsilon}_d^2\right)\sin{\phi}-2\tilde{\varepsilon}_d\sinh{(2x)}\right]\right\}\right].
\end{eqnarray}
The straightforward calculations lead to:
\begin{eqnarray}
  &&F(x,\phi)= \kappa\xi\frac{\tilde{\Delta}_d}{2\tilde{\mathcal{D}}^2}\left[ \sin{\phi}-\tilde{\varepsilon}_d\sinh{\left(2x\right)}\right],\label{fxphi}\\
  &&\eta(x,\phi)=\kappa\alpha\xi\frac{\tilde{\Delta}_d}{2\tilde{\mathcal{D}}^6}\left[ \tilde{\Delta}_d^2\sinh{(2x)}\left(\tilde{\mathcal{D}}^2+4\right)
  \left\{ \sin{\phi}-\tilde{\varepsilon}_d\sinh{(2x)}\right\}+\right.\nonumber\\
  &&\hspace{4.cm}+\left. 8\tilde{\varepsilon}_d\tilde{\mathcal{D}}^2\left\{\sin^2{(\phi/2)}+\sinh^2{x}\right\}\right].\label{etaxphi}
\end{eqnarray}

\subsection{Stability of a static solution. Linearization.}\label{subsect4_1_5}

The strongly nonlinear Eq.~(\ref{xEqsol}) in the adiabatic limit has static solutions determined by the equation:
\begin{equation}\label{Exst}
    x_{st}=F(x_{st},\phi),
\end{equation}
which is in the considered limit of small electromechanical coupling, $\xi\ll 1$, has a trivial solution,
\begin{equation}\label{xst}
    x_{st}=\kappa\xi\frac{\tilde{\Delta}_d}{2\tilde{\mathcal{D}}^2}\sin{\phi}\propto \mathcal{O}(\xi).
\end{equation}
It is obvious from Eq.~(\ref{xst}) that $x_{st}$ is strictly equal to zero when $\phi=\pi n$ ($n$ is an integer number). This corresponds to the straightforward configuration of the nanotube. For the time evolution of a small deviations from the equilibrium position $\delta x(t)=x(t)-x_{st}$ one can obtain the following equation (linearized Eq.~(\ref{xEqsol}) near the equilibrium point $x_{st}$),
\begin{equation}\label{deltax}
    \delta\ddot{x}-\eta (0,\phi)\delta\dot{x}+\delta x=0.
\end{equation}
Here we ignore a small shift of the mechanical oscillation frequency due to the re-normalization. And,
\begin{equation}\label{eta0}
    \eta(0,\phi)=+\kappa\alpha\xi\frac{4\tilde{\varepsilon}_d\tilde{\Delta}_d}{\tilde{\mathcal{D}}^4(0,\phi)}\sin^2{\left(\frac{\phi}{2}\right)}.
\end{equation}
Thereby, the static solution $x_{st}$ is \textit{unstable}, that is, an initial small spontaneous fluctuation leads to exponential increasing of the QD vibration amplitude, when $\eta (0,\phi)>0$ and it is stable otherwise. 
It is clearly seen from Eq.~(\ref{eta0}) that the presence of the \textit{mechanical instability} directly depends on the
position of the dot energy level $\varepsilon_d$ and the direction of the applied bias voltage $\kappa$. In case of positive $\kappa\varepsilon_d$ the equilibrium position $x_{st}$ is \textit{unstable} with respect to the increase of the oscillation amplitude for all values of the
superconducting phase difference $\phi$ (except of $\phi=0$ which is so only for the symmetric junction as will be discussed in the subsection~\ref{subsect4_1_7}). The later reveals the reason we have called $\eta (x,\phi)$ as a pumping coefficient. We can also easily note from Eq.~(\ref{eta0}) that the pumping coefficient takes its maximum at $\phi=\pi$ (the force $F(0,\pi)=0$). One should compare it to the case of $\phi=0$ when the force and pumping are equal to zero, $F(0,0)=0$ and $\eta(0,0)=0$. It is rather an unexpected situation because usually a friction/pumping is generated via the force acting on a QD. However, in our case, the force responsible for the nanomechanical pumping is of electronic nature and is sensitive to the fact that Cooper pairs are delocalized between nanotube and superconducting leads. This force emerges when the Cooper pair is in a state of quantum superposition controlled by phase difference, nanotube bending, and other parameters.

\begin{figure}
\centering
\includegraphics[width=.7\columnwidth]{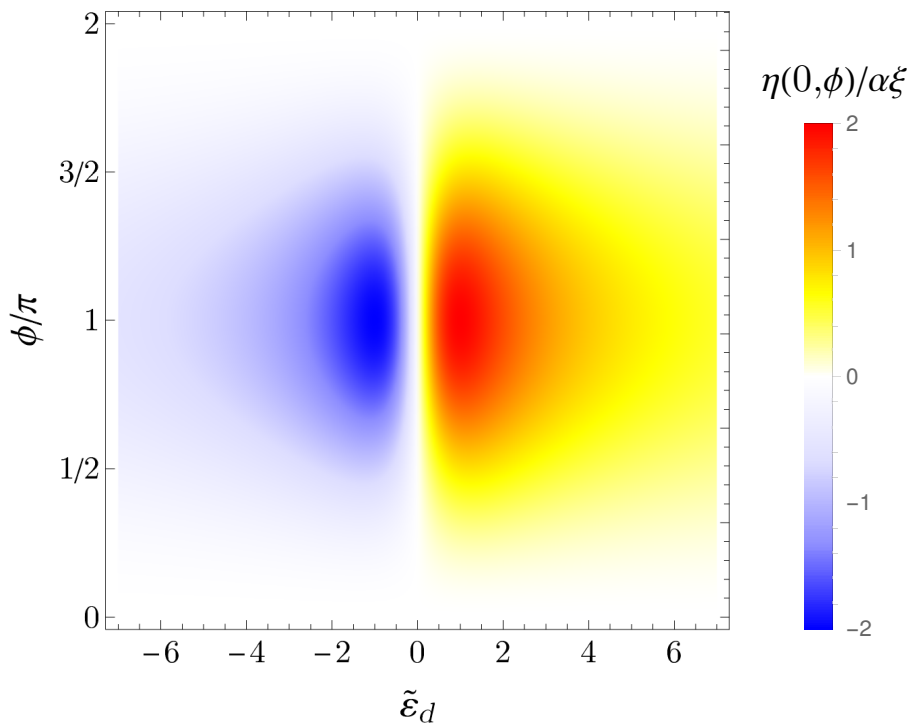}
\caption{\textit{Dependence of the pumping coefficient $\eta (0,\phi)$ (in units of $\alpha\xi$) on the position of the QD energy level $\tilde{\varepsilon}_d$ and superconducting phase difference $\phi$ for the positive direction of the bias voltage $\kappa=1$ and $\tilde{\Delta}_d=1$. The red color scheme indicates the regime where the mechanical instability occurs and blue one corresponds to the damping situation.
}}\label{fig:fig4_1_3}
\end{figure}

The phase diagram of the mechanical instability occurred in the proposed system is presented in Fig.~\ref{fig:fig4_1_3}. The maximal pumping takes place at $\tilde{\varepsilon}_d=1/\sqrt{3}$ (and $\phi=\pi)$.

\subsection{Self-saturation effect.} \label{subsect4_1_6}

In order to analyse what the mechanical instability leads to, we find the stationary solutions $x_{c}(t)$ of
Eq.~(\ref{xEqsol}). It is natural to use the smallness of the parameter
$\xi$ and, then, Eq.(\ref{xEqsol}) can be treated as the one describing an harmonic oscillator slightly perturbed by the interaction with the electronic subsystem,
\begin{equation}
    \ddot{x}+x=\xi f(\dot{x},x).
\end{equation}
In such a case of small electromechanical coupling $\xi\ll 1$, one can use the Krylov-Bogoliubov method of averaging~\cite{Bogoliubov1985} (see also~\cite{Kovalev1989}) to find an approximate solution of the Eq.(\ref{xEqsol}). In the first order of the perturbation theory this method matches the Van der Pol averaging method. Following it, let us assume that in a stationary regime, the dot displacement $x_c$ takes a form (with the accepted accuracy $\xi$),
\begin{equation}\label{xc}
    x_{c}(t)=x_{st}+\sqrt{A}\sin\left(t+\varphi(t)\right)+\mathcal{O}(\xi),
\end{equation}
where the amplitude $\sqrt{A(t)}$ and the phase $\varphi (t)$ vary slowly in time, $x_{st},\dot{A}(t),\dot{\varphi}(t)\sim \xi$. Then, by substituting the ansatz (\ref{xc}) into Eq.(\ref{xEqsol}) and assuming the same functional form for the derivative $\dot{x}_c$, 
\begin{equation}
    \dot{x}_c(t)=\sqrt{A}\cos{(t+\varphi (t))},
\end{equation}
one can find following equations for $A(t)$ and $\varphi (t)$,
\begin{eqnarray}
     &&\dot{A}= 2 A\cos^2{\psi}\eta\left(\sqrt{A}\sin{\psi},\phi\right),\label{dA} \\
     && \dot{\varphi}=- A^{-1/2}\sin{\psi}F\left(\sqrt{A}\sin{\psi},\phi\right).\label{dphi}
\end{eqnarray}
Since the amplitude  and the phase change scarcely during the period of oscillations, the r.h.s. of Eqs.(\ref{dA})-(\ref{dphi}) can be replaced by their average over the period, and, as a result, one obtain
\begin{eqnarray}
      &&\dot{A}=  A\bar{\eta}(A,\phi), \label{dotA} \\
      && \dot{\varphi}=- A^{-1/2}\bar{F}(A,\phi).\label{Ephi}
\end{eqnarray}\label{EA}
Here
\begin{eqnarray}
&&\bar{\eta}(A,\phi)=(\pi)^{-1}\int_{0}^{2\pi}d\psi
\cos^2{(\psi)}
\eta\left(\sqrt{A}\sin\psi,\phi\right)\equiv\kappa\xi\alpha
W(A,\phi),\label{W}\\
&&\bar{F}(A,\phi)=(2\pi)^{-1}\int_{0}^{2\pi}d\psi\sin{(\psi)}F\left(\sqrt{A}\sin\psi,\phi\right).\label{barF}
\end{eqnarray}
The pumping coefficient $\bar{\eta}(A,\phi)$ has an obvious physical
meaning: it gives the ratio between the energy supplied into the
mechanical degree of freedom for one period of mechanical
oscillations with amplitude $\sqrt{A}$ and the total mechanical
energy.

\begin{figure}
    \centering
    \begin{subfigure}[t]{0.45\columnwidth}
        \centering
        \subfloat[\textit{a)}]{\includegraphics[width=\columnwidth, keepaspectratio]
        {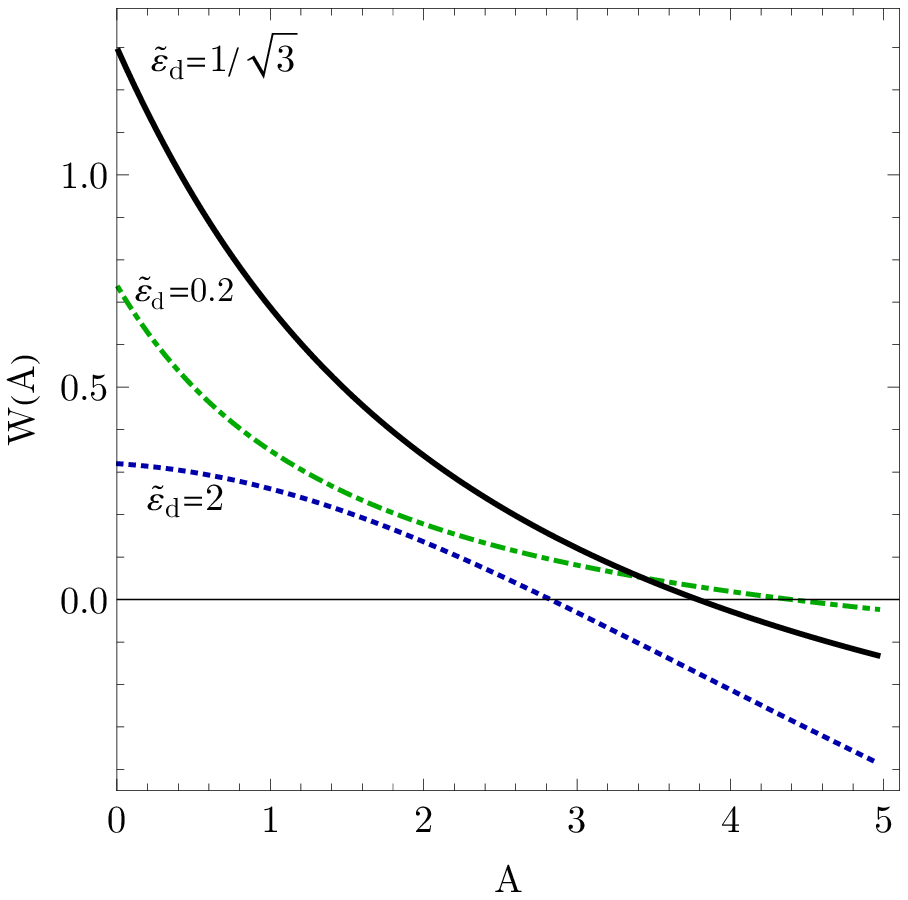 }}
        \caption{}\label{fig:subfig3a}
    \end{subfigure}
    \quad
    \begin{subfigure}[t]{0.45\columnwidth}
        \centering
        \subfloat[\textit{b)}]{\includegraphics[width=\columnwidth, keepaspectratio]
        {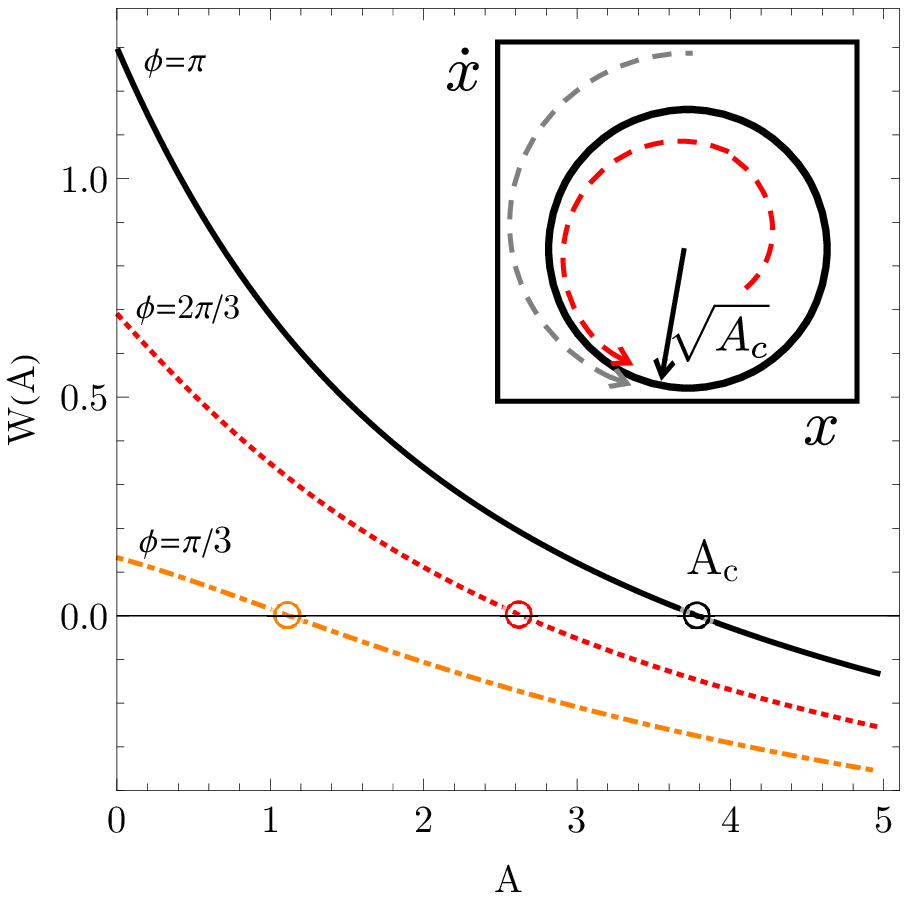}}
        \caption{}\label{fig:subfig3b}
    \end{subfigure}
    \caption{\textit{Plots of the function $W(A)$, proportional to the pumping coefficient $\bar{\eta}(A,\phi)$, for different values of
    (a) the relative position of the dot energy level $\tilde{\varepsilon}_d=0.2; 1/\sqrt{3}; 2$ for $\phi=\pi$, and
    (b) the superconducting phase difference $\phi=\pi/3, 2\pi/3, \pi$ for $\tilde{\varepsilon}_d=1/\sqrt{3}$.
     The zeroes of the functions correspond to the amplitude of the limiting cycle [see the inset in (b)], which strongly depends on the superconducting
     phase difference and reaches its maximum at $\phi=\pi$. The other parameters are $\tilde{\Delta}_d=1, \kappa=+1$.} }\label{fig:fig4_1_4}
\end{figure}

\begin{figure}
    \centering
    \begin{subfigure}[t]{0.45\columnwidth}
        \centering
        \subfloat[\textit{a)}]{\includegraphics[width=\columnwidth, keepaspectratio]
        {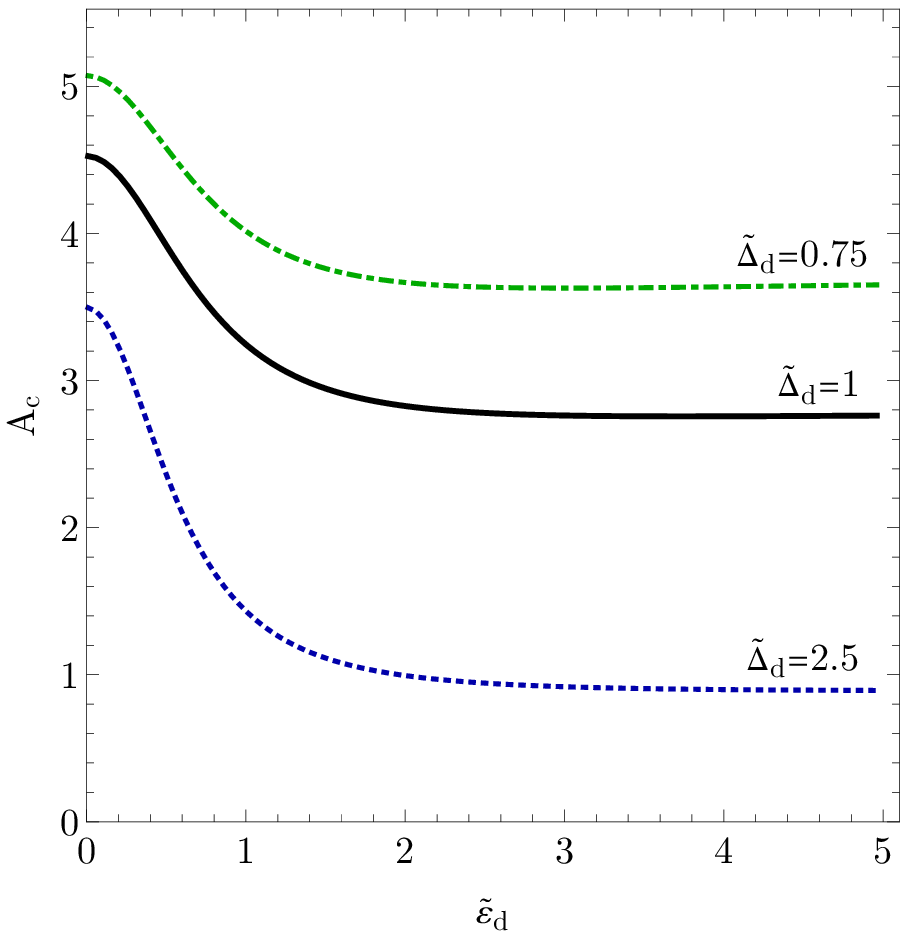}}
        \caption{}\label{fig:subfig4a}
    \end{subfigure}
    \quad
    \begin{subfigure}[t]{0.45\columnwidth}
        \centering
        \subfloat[\textit{b)}]{\includegraphics[width=\columnwidth, keepaspectratio]
        {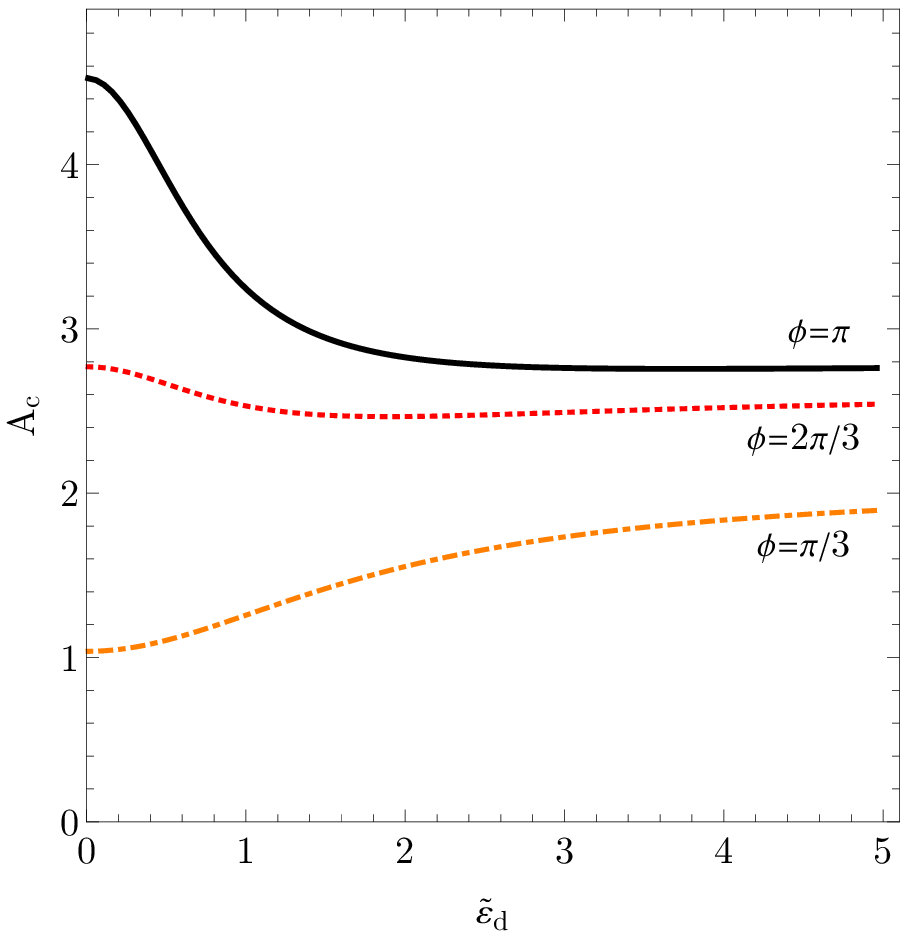}}
        \caption{}\label{fig:subfig4b}
    \end{subfigure}
    \caption{\textit{Dependencies of $A_c$ of the limiting cycles on the relative position of the dot energy level $\tilde{\varepsilon}_d$ (counted from the Fermi energy) for different values of (a) $\tilde{\Delta}_d=0.75; 1; 2.5$ for $\phi=\pi$, and (b) the superconducting phase difference $\phi=\pi/3, 2\pi/3; \pi$ for $\tilde{\Delta}_d=1$.}}\label{fig:fig4_1_5}
\end{figure}

It is evident from Eq.~(\ref{dotA}) that stationary regimes $\dot{A}=0$
are given by equations $A=0$ ($x(t)=x_{st}$) and
$\bar{\eta}(A,\phi)=0$. The first one is a static state of the
nanotube, and the second one corresponds to periodic oscillations with the 
amplitude $\sqrt{A_{c}}$, where $W(A_{c},\phi)=0$. The static regime is stable when $ \bar {\eta} (0, \phi)
<0 $ and unstable otherwise. The stability of the periodic solution
is defined by the sign of the derivative
$\partial_A{\bar{\eta}(A,\phi)}|_{A=A_{c}}$: if it is negative (positive), then the
periodic regime is stable (unstable). Then, analyzing Eqs.~(\ref{xEqsol}) and (\ref{W}), one can conclude that the
pumping coefficient $\bar{\eta}(A,\phi) \propto \kappa$ is an odd
function of $\varepsilon_{d}$ (the first term in the r.h.s. of Eq.~(\ref{etaxphi}) does not give an contribution) and takes the following limit values,
\begin{eqnarray}
   && \bar{\eta}(0,\phi)=\kappa\alpha\xi W(0,\phi)=+\kappa\alpha\xi\frac{4\tilde{\varepsilon}_d\tilde{\Delta}_d}{\tilde{\mathcal{D}}^4(0,\phi)}\sin^2{\left(\frac{\phi}{2}\right)},\label{etaphi}\\
  &&\bar{\eta}(A\rightarrow\infty,\phi)= \kappa\alpha\xi W(A \rightarrow\infty,\phi)=-\kappa\alpha\xi\frac{\tilde{\varepsilon}_d}{2\tilde{\Delta}_d},\label{etaphiInfty}
\end{eqnarray}
from which follows that at $\phi\neq 0$, the solution of the equation
$W(A_{c},\phi)=0$, corresponding to the stationary periodic regime, exists at
any values of the other parameters. However, at small  $\phi\ll 1$ the pumping coefficient, $\eta(x,\phi)$, determined by interaction with the electronic subsystem, is small and can be equalized by the friction coefficient induced by the thermal environment. The case when $\phi=0$ is very
unstable with respect to the small asymmetry parameter $|a|\ll 1$ and will be discussed in the next section,~\ref{subsect4_1_7}. The function
$W(A)$ and $A_c$ at different $\phi\geq 1$ and $\tilde{\varepsilon}_{d}>0$ are
presented in Figs.~\ref{fig:fig4_1_4} and \ref{fig:fig4_1_5}.

It follows from Eq.~(\ref{etaphi})  that if $\kappa \varepsilon_{d}>0$
and $\phi\neq 0$, the static mechanical state $x=x_{st}\ll1$ is
unstable with respect to the appearance of bending oscillations with
amplitudes that exponentially increase in time with the increment
\begin{equation}
  \gamma=\kappa\alpha\xi W(0,\phi).  
\end{equation}
 The latter takes its maximum at
$\phi=\pi$ for the fixed values of all other parameters (notice that
$x_{st}(\pi)=0$). However, the increase saturates at the amplitude
$\sqrt{A_{c}}$, resulting in self-sustained oscillations at this
amplitude. It should be noted that the amplitude saturation in the system
under consideration is a completely internal effect and still takes place
when the friction caused by interaction with a
thermodynamic environment is zero. A "self-saturation" effect was
also reported in~\cite{Ilinskaya2019a} where a special magnetic NEM system was
considered.

\subsection{Influence of the asymmetry and friction generated by a thermodynamic environment.}\label{subsect4_1_7}

In an experiment the presence of the asymmetry of tunnel contacts naturally arises. Following it, let us take into account the asymmetric superconducting tunnel junctions with a parameter $\Delta_{d1}/\Delta_{d2}=\zeta$, where $\zeta=\text{e}^{-a/\lambda}$ (see Eq.~(\ref{tunnela})). Thus, by combining Eqs.~(\ref{tunnela}) and~(\ref{deltad}), one gets that the asymmetry in the superconducting tunneling amplitudes results in a relative shift of a dot equilibrium position and we need to do the following replacement in obtained formulae,
\begin{equation}\label{deltada}
    \Delta_d(x,\phi)\rightarrow \Delta_d(x+a,\phi).
\end{equation}
In what follows it is convenient to introduce the re-normalized parameter of the asymmetry, $a/\lambda\rightarrow a$.
Then, for the criterion for the mechanical instability one obtains $\eta (0+a,\phi)>0$, where
\begin{eqnarray}
    &&\eta(a,\phi)=\kappa\alpha\xi\frac{\tilde{\Delta}_d}{2\tilde{\mathcal{D}}^6}\left[ \tilde{\Delta}_d^2\sinh{(2a)}\left(\tilde{\mathcal{D}}^2+4\right)
  \left\{ \sin{\phi}-\tilde{\varepsilon}_d\sinh{(2a)}\right\}+\right.\nonumber\\
  &&\hspace{4.cm}+\left. 8\tilde{\varepsilon}_d\tilde{\mathcal{D}}^2\left\{\sin^2{(\phi/2)}+\sinh^2{a}\right\}\right].
\end{eqnarray}\label{etaa}
From this inequality one can get that the mechanical subsystem is \textit{unstable} when
\begin{equation}\label{instability1a}
    \begin{cases}
    \vert\tilde{\Delta}_{d}\sinh{a}\vert < \tilde{\Delta}_d^{cr}, & \tilde{\varepsilon}_d>0;\\
     \vert\tilde{\Delta}_{d}\sinh{a}\vert > \tilde{\Delta}_d^{cr}, & \tilde{\varepsilon}_d<0;\\
    \end{cases}
    \qquad \phi=\pi;
\end{equation}
\begin{equation}\label{instability1b}
    \begin{cases}
    \vert\tilde{\Delta}_{d}\cosh{a}\vert < \tilde{\Delta}_d^{cr}, & \tilde{\varepsilon}_d>0;\\
     \vert\tilde{\Delta}_{d}\cosh{a}\vert > \tilde{\Delta}_d^{cr}, & \tilde{\varepsilon}_d<0;\\
    \end{cases}
    \qquad \phi=0;
\end{equation}
or,
\begin{equation}\label{instability2}
    \begin{cases}
    \vert\tilde{\Delta}_{d1}\pm\tilde{\Delta}_{d2}\vert < 2\tilde{\Delta}_d^{cr}, & \tilde{\varepsilon}_d>0;\\
     \vert\tilde{\Delta}_{d1}\pm\tilde{\Delta}_{d2}\vert > 2\tilde{\Delta}_d^{cr}, & \tilde{\varepsilon}_d<0.\\
    \end{cases}
\end{equation}
Here "-" corresponds to the case $\phi=\pi$ and "$+$" -- to $\phi=0$, ($\zeta\ne 1)$. The critical value defined as
\begin{equation}\label{deltacritical}
    (\tilde{\Delta}_d^{cr})^2=\frac{1}{2}\left( \sqrt{(3+\tilde{\varepsilon}_d^2)^2+8(1+\tilde{\varepsilon}_d^2)}-(3+\tilde{\varepsilon}_d^2)\right),
\end{equation}
is localized around 1, $(\tilde{\Delta}_s^{cr})^2\in [(\sqrt{17}-3)/2,2)$, because of $\lim_{\tilde{\varepsilon}_d\to\infty}  (\tilde{\Delta}_d^{cr})^2=2$. 
Another way to obtain the stability criteria is to present displacement in a form
\begin{equation}
    x(t)\sim \text{e}^{\imath\Omega t},
\end{equation}
substitute into Eq.~(\ref{xEqsol}) and look for a solution of the inequality $\text{Im}\Omega <0$ which corresponds to the case of exponential increase of the amplitude of oscillations of the QD. 
One can conclude from the above-mentioned stability criteria, Eqs.~(\ref{instability1a})-(\ref{instability2}), that the presence of "self-saturation" effect is not affected by the asymmetry. 

Note that an infinitesimal value of the asymmetry can result in the emergence of the mechanical instability even in the case of zero superconducting phase difference. However, in the latter case the strength of the pumping is small and need to be in comparison with the damping generated by the influence of an thermodynamic equilibrium bath. In order to do this, we incorporate the phenomenological friction term $+\gamma \dot{x}$ in the l.h.s. of Eq.~(\ref{xEq1}). The friction coefficient $\gamma$ is determined as $\propto Q^{-1}$, where $Q$ is the quality factor determined by interaction with the thermodynamic environment~\cite{Steele2009}. Thus, the mechanical instability occurs if
\begin{equation}
    \gamma -\eta(a,\phi)<0.
\end{equation}

It means that for small values of superconducting phase difference, $\phi <1$, the pumping coefficient $\eta(x+a,\phi)$ determined by interaction with the electronic subsystem, is much smaller than in case of $\phi\approx \pi$ and equalizes a friction coefficient $\gamma$ associated with the influence of a thermodynamic environment. It results in competition between these two processes because the regime of self-sustained oscillations emerges when the pumping is equal to damping. As a consequence, friction induced by the interaction with a thermal bath leads to a decrease in the amplitude of self-sustained oscillations. Nevertheless, as the pumping caused by electronic non-equilibrium environment strongly depends on the superconducting phase difference $\phi$, in the most pronounced case $\phi=\pi$ it dominates over the “thermodynamic” friction for high-quality nanomechanical resonators, $Q\sim 10^{5}$.

\begin{figure}
\centering
\includegraphics[width=.55\columnwidth]{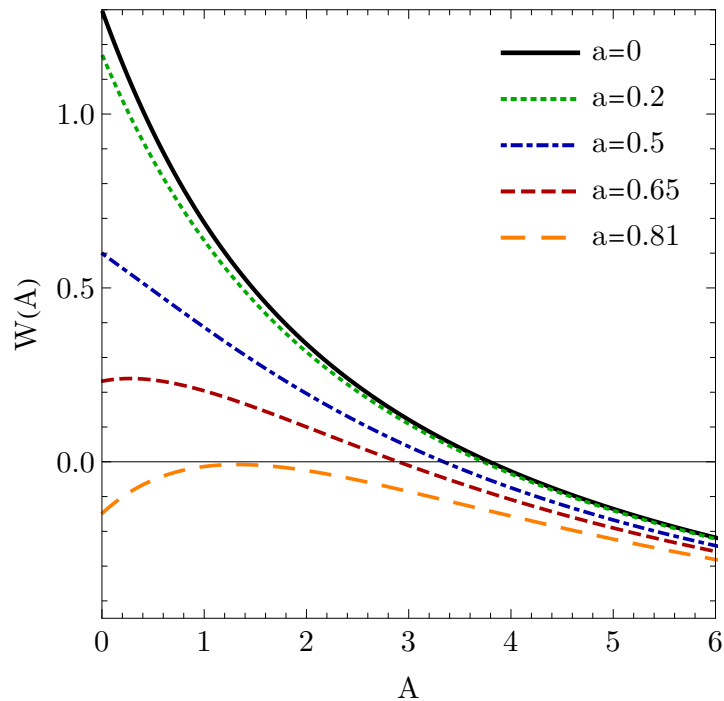}
\caption{\textit{Dependencies of the function $W(A)$, proportional to the pumping coefficient $\bar{\eta}(A,\phi)$, for different values of the asymmetry parameter $\vert a\vert =0$ (black thick curve, symmetric case); $0.2$ (green dotted curve); $0.5$ (blue dotdashed); $0.65$ (red densely dashed); $0.81$ (orange loosely dashed curve) for $\phi=\pi$, and $\tilde{\varepsilon}_d=1/\sqrt{3}$.
     The zeroes of the functions correspond to the amplitude of the limiting cycle. Other parameters are: $\tilde{\Delta}_d=1, \kappa=+1$.
}}\label{fig:fig4_1_6}
\end{figure}

\begin{figure}
\centering
\includegraphics[width=.55\columnwidth]{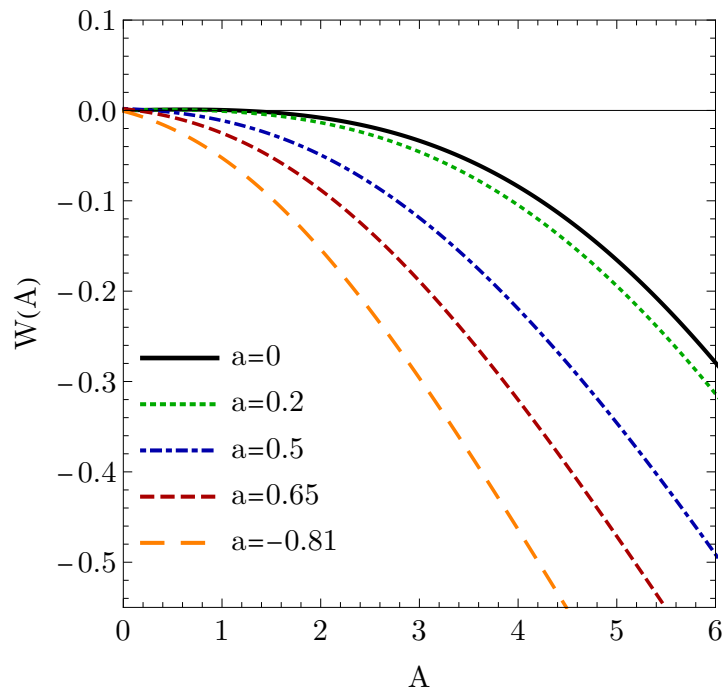}
\caption{\textit{Dependencies of the function $W(A)$, proportional to the pumping coefficient $\bar{\eta}(A,\phi)$, for different values of the asymmetry parameter $\vert a\vert =0$ (black thick curve, symmetric case); $0.2$ (green dotted curve); $0.5$ (blue dotdashed); $0.65$ (red densely dashed); $0.81$ (orange loosely dashed curve) for $\phi=0$, and $\tilde{\varepsilon}_d=1/\sqrt{3}$.
      Other parameters are: $\tilde{\Delta}_d=1, \kappa=+1, \gamma =10^{-5}$. The functions take non-positive values which corresponds to the absent of self-sustained oscillations in a limiting cycle.
}}\label{fig:fig4_1_7}
\end{figure}

Figures~\ref{fig:fig4_1_6},~\ref{fig:fig4_1_7} represent dependencies of function $W(A)$ which is proportional to the pumping coefficient $\bar{\eta}(A,\phi)$ for different values of the asymmetry parameter $a$. Note that $W(A)$ is an even function of $a$. One can see from Fig.~\ref{fig:fig4_1_6}, which corresponds to a case of maximal strength of the pumping, $\phi=\pi, \tilde{\varepsilon}_d=1/\sqrt{3}$, that the value of the amplitude of a limit cycle decreases when the asymmetry increases (shift of the position of a zero of the function to the left). The black curve is associated with the symmetric case ($a=0$) and is the same one as in Fig.~\ref{fig:fig4_1_4}. The orange curve corresponds to the critical value (with the further increase of the asymmetry parameter the pumping is vanishes) obtained from Eq.~(\ref{deltacritical}), $(\tilde{\Delta}_d^{cr})^2=2/3$ for $\tilde{\varepsilon}_d=1/\sqrt{3}$. Figure~\ref{fig:fig4_1_7} demonstrates the case of $\phi=0$ when the effective pumping is negligibly small and, as a consequence, a limit cycle of self-sustained oscillations does not occurs. 

\subsection{Transistor- and diode-like behaviour in electric current.} \label{subsect4_1_8}

The self-sustained oscillations considered above have a very specific
transport signature. This raises the possibility of detecting the
mechanical instability through an electric current measurement. To
explore such a possibility, let us consider the electric current through
the system, $I_{n}$, determined in a standard way,
\begin{equation}\label{in1}
    I_n=e\kappa\text{Tr}\left\{\hat{\dot{N}}\hat{\rho}\right\},
\end{equation}
where
$\hat{\dot{N}}= i[\hat{H},\hat{N}]$ and $\hat{N}=\sum_{k\sigma}a^\dag_{k\sigma} a_{k\sigma}$ is the operator of
the number of electrons in the normal electrode. In the deep subgap regime considered above ($\Delta_s\to\infty$), one can neglect quasiparticle current since it is exponentially small. As a result, in the high bias voltage regime at $\kappa=+1$, where electron tunneling from the QD to the normal leads is forbidden, an expression for $I_{n}$ can
be easily obtained by analyzing Fig.~\ref{fig:fig4_1_2}. From
those diagrams, one can see that a decrease in the number of electrons
in the normal electrode is defined by two different processes. The first
one is the tunneling of an electron with spin up or down into the empty
dot. The rate of this process is $2\Gamma_{n}\rho_{0}$, where
$\rho_{0}=\left( R_{0}+R_{3}\right) /2$ is the probability that the dot is empty.
The second one is the tunneling of an electron into the dot occupied by
 a single electron with spin up or down. The rate of this process is
$2\Gamma_{n}\rho_{e}$. Taking into account the normalization condition
$2\rho_{e}+R_{0}=1$, and using a similar speculation for $eV_b<0$, one gets
from Eq.~(\ref{Rexpand}) the following equation for $I_{n}$,
\begin{equation}\label{Icurrent}
    I_n(t)=\kappa I_{0}\left(1+\kappa R_3\right),
\end{equation}
where $I_0=e\Gamma_n$. In the adiabatic limit one can use the expansion for $\vec{R}$ in the perturbation theory over the adiabaticity parameter $\alpha$, Eq.~(\ref{Rexpand}), and obtain for the current the following equation,
\begin{equation}\label{in2}
    I_n(t)=
    \kappa I_{0}\left[\frac{|\Delta_{d}(x,\phi)|^{2}}{\mathcal{D}^{2}(x,\phi)}+ \alpha \dot{x}f(x)+\mathcal{O}(\alpha^{2})\right].
\end{equation}

\begin{figure}
\centering
\includegraphics[width=.7\columnwidth]{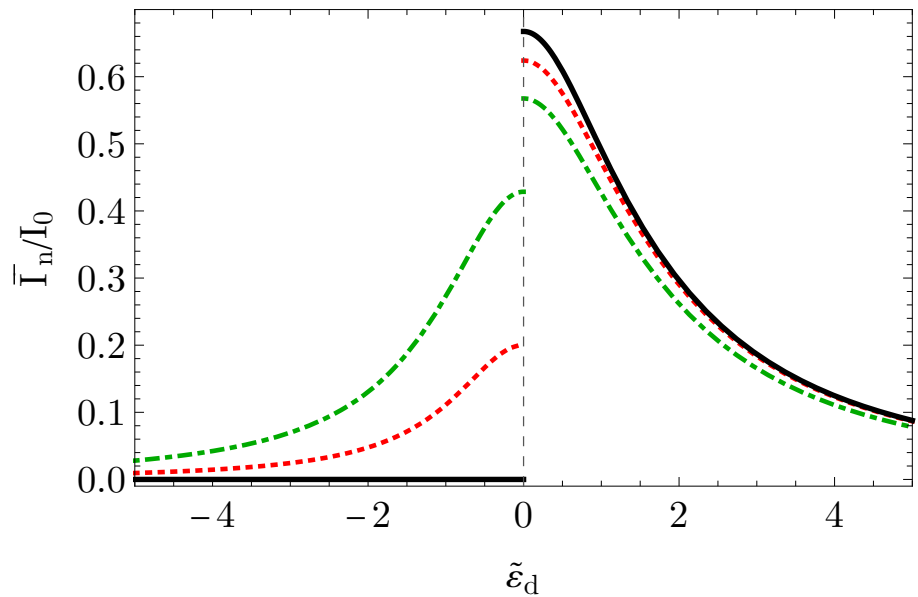}
\caption{\textit{Dependencies of the dc electric current $\bar{I}_n$ (normalized to $I_0=e\Gamma_n$) on the relative position of the QD energy level $\tilde{\varepsilon}_d$ for different values of superconducting phase difference $\phi=\pi/3$ (green dot-dashed curve), $2\pi/3$ (red dotted), and $\pi$ (black solid) for $\tilde{\Delta}_d=1$ and $\kappa=+1$. The maximum effect occurs at $\phi=\pi$ when the dc current is absent in the static regime, while it is close to the maximum one in the stable stationary regime of the self-sustained oscillations.}
}\label{fig:fig4_1_8}
\end{figure}

In the stationary regime
corresponding to the generation of self-sustained oscillations with
amplitude $\sqrt{A_{c}}$, the averaged over a period of the oscillations electric current
$\bar{I}_{n}$ is defined as
\begin{equation}\label{inav}
    \bar{{I}}_n=\frac{1}{2\pi}\int_0^{2\pi} I_n[x(t)] dt.
\end{equation}
After the integrating of Eq.~(\ref{in2}), the main contribution to the current with an accuracy $\alpha ^2$ (since the first-order term is averaging out) is given as
\begin{equation}
    {\bar{I}}_{n}(\kappa,\varepsilon_{d})=\kappa I_0\left[\frac{\Delta_d^2\cos^{2}{(\phi/2})}{\Delta_d^2\cos^{2}(\phi/2)+\Gamma_n^2+\varepsilon_d^2}
+ \theta(\kappa\varepsilon_{d})\delta\bar{I}(A_{c})\right].
\end{equation}\label{Iav}
Here the first term corresponds to the static dc current which crucially depends on the superconducting
 phase difference $\phi$. In particular, the first term is equal to zero at $\phi=\pi$, in contrast to the second term,
\begin{equation}\label{deltaI}
    \delta\bar{I}(A_{c})=\dfrac{1}{2\pi}\dfrac{\Delta_d^2 (\Gamma_n^2+\varepsilon_d^2)}{\mathcal{D}^2(0,\phi)}\int_0^{2\pi}d\psi \dfrac{\sinh^2{\left(\sqrt{A_c}\sin{\psi}\right)}}{\mathcal{D}^2\left( \sqrt{A_c}\sin{\psi},\phi\right)}>0,
\end{equation}
which emerges exclusively due to the self-sustained oscillations and
equals zero if the static state is stable, as indicated by the Heaviside
step function $\theta(\kappa\varepsilon_d)$. The maximal value of the current $\delta\bar{I}_n(A_c)$ can be estimated as
\begin{equation}\label{inmax}
    \delta\bar{I}_n^{max}\approx \frac{\Delta_d^2 \sinh^2{\sqrt{A_c}}}{\Delta_d^2\sinh^2{\sqrt{A_c}}+\Gamma_n^2+\varepsilon_d^2}.
\end{equation}

\begin{figure}
\centering
\includegraphics[width=.7\columnwidth]{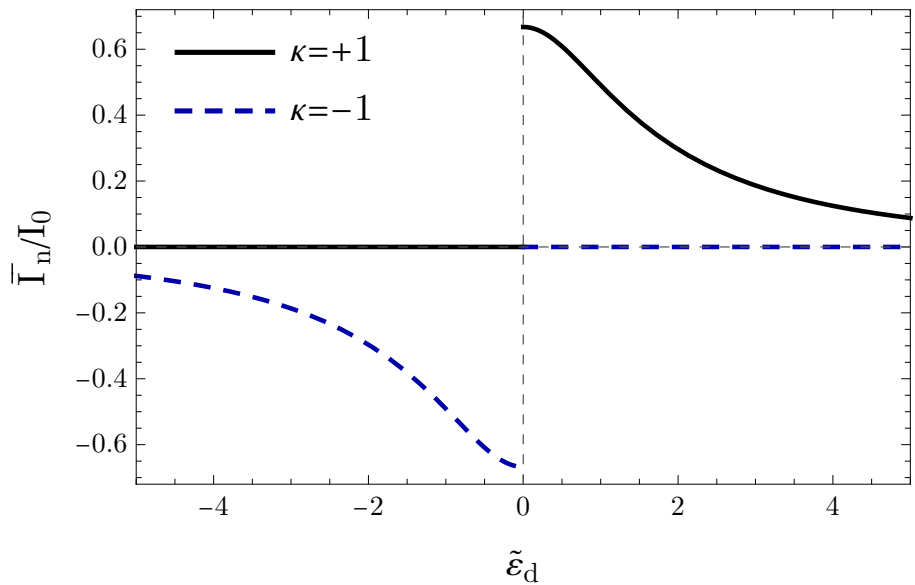}
\caption{\textit{Dependencies of the dc electric current $\bar{I}_n$ (normalized to $I_0=e\Gamma_n$) on the relative position of the QD energy level $\tilde{\varepsilon}_d$ for $\kappa=+1$ (black solid curve, associated with the same one in Fig.~\ref{fig:fig4_1_8}) and for $\kappa=-1$ when the bias voltage is applied in the opposite direction (blue dashed curve), representing a diode-like behaviour of the current. Other parameters: $\phi=\pi$.}
}\label{fig:fig4_1_9}
\end{figure}

Dependencies of the current $\bar{I}_{n}$ as a function of $\tilde{\varepsilon}_{d}$ at
$\tilde{\Delta}_{d}=1$ and $\phi=\pi/3, 2\pi/3, \pi$ are presented in
Figs.~\ref{fig:fig4_1_8},~\ref{fig:fig4_1_9}. These graphs show that the nanomechanical
instability discussed in this article leads to the emergence of
significant diode and transistor effects. The effects are most
pronounced at $\phi=\pi$ when in the static regime $A_{c}=0$ where the Cooper pair exchange between the dot and the superconducting leads is completely blocked. In such a situation, a jump in the average current
from zero to a finite value $\sim I_{0}$ (or vice-versa) occurs if the direction
of the bias voltage changes (diode effect: for one direction of the bias voltage the current is present but for the opposite one --- absent (blocked)), see Fig.~\ref{fig:fig4_1_9} or if the position of the dot energy level $\varepsilon_{d}$ controlled by the gate voltage (third electrode in a transistor), passes zero
(transistor effect), see Fig.~\ref{fig:fig4_1_8}. Note that the discontinuity of the average current
as a function of $\varepsilon_{d}$ must be treated to the accuracy
$\xi$.

\subsection{Numerical results.}\label{subsect4_1_9}
\begin{figure}[ht!!]
    \centering
      \begin{subfigure}[t]{0.45\columnwidth}
        \centering
        \subfloat[\textit{a)}]{\includegraphics[width=\columnwidth, keepaspectratio]
        {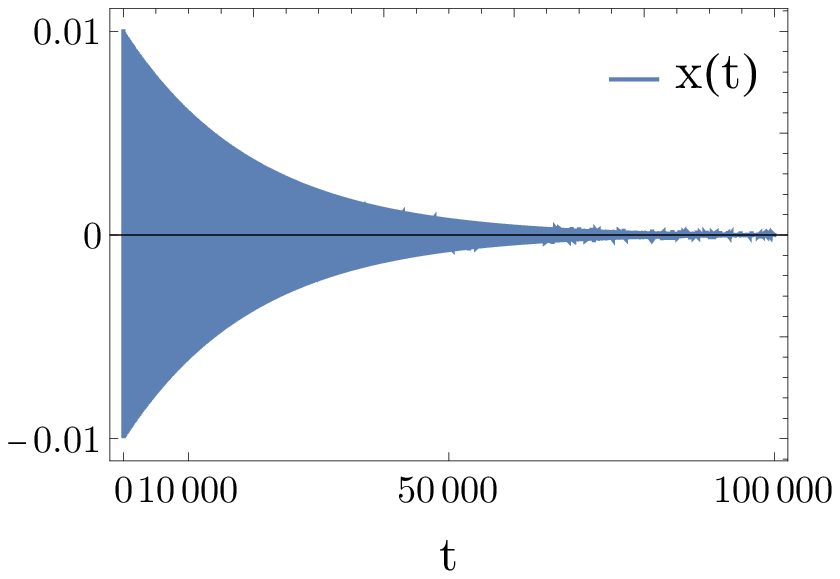}}
        \caption{}\label{fig:subfig4_1_10a}
    \end{subfigure}
    \quad
    \begin{subfigure}[t]{0.45\columnwidth}
        \centering
        \subfloat[\textit{b)}]{\includegraphics[width=\columnwidth, keepaspectratio]
        {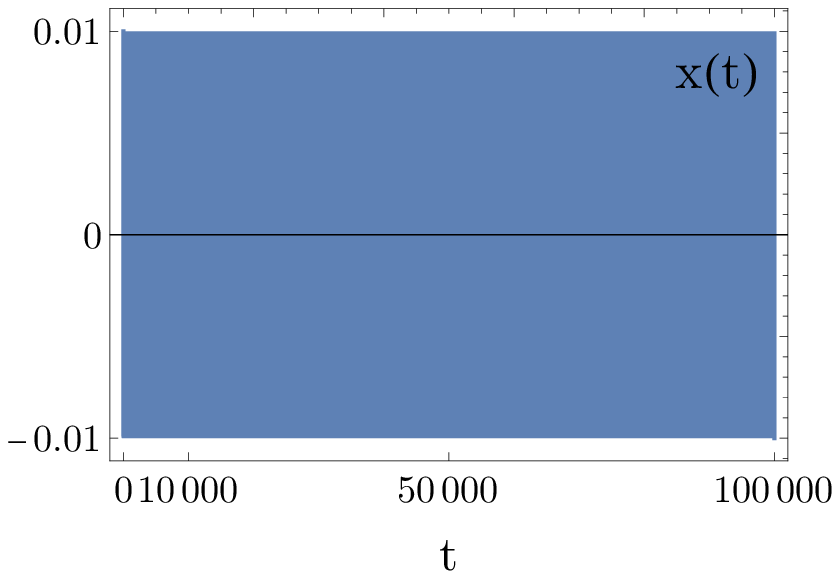}}
        \caption{}\label{fig:subfig4_1_10b}
    \end{subfigure}
    \quad
    \begin{subfigure}[t]{0.5\columnwidth}
        \centering
        \subfloat[\textit{c)}]{\includegraphics[width=\columnwidth, keepaspectratio]
        {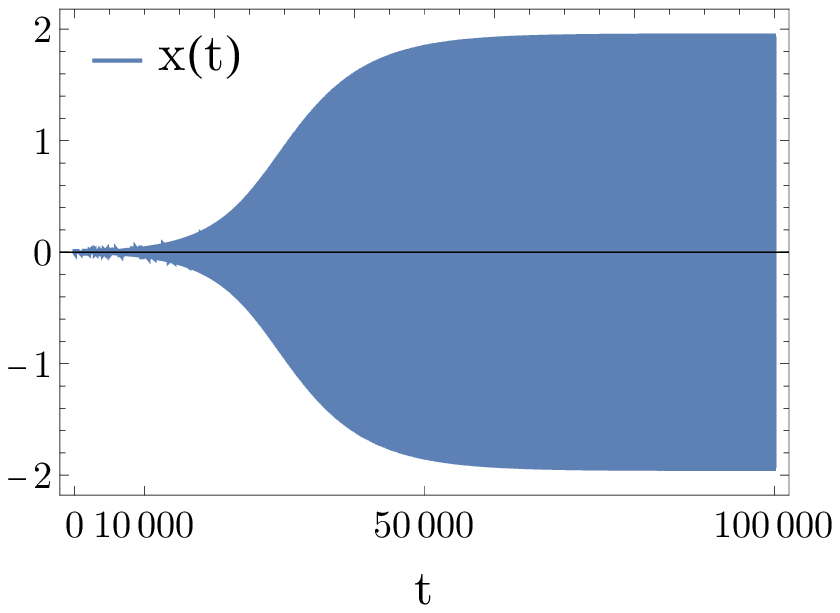}}
        \caption{}\label{fig:subfig4_1_10c}
    \end{subfigure}
    \caption{\textit{Dependencies of the displacement $x$ (in units of the tunneling length $\lambda$) of the QD in three different regimes realized for the system, on time (in units of $\omega$). The figure (a) corresponds to the stable to an initial spontaneous fluctuation ($x_0=0.01$) from an equilibrium position, regime, $\tilde{\varepsilon}_d<0$. The figure (b) corresponds to critical value, $\tilde{\varepsilon}_d=0$ when the state of the mechanical subsystem is not stable or unstable, either. And the figure (c) demonstrates the time evolution of the QD oscillations in the regime when the mechanical instability occurs, $\tilde{\varepsilon}_d=+1/\sqrt{3}$. The latter case is presented in Figs.~\ref{fig:fig4_1_4},~\ref{fig:fig4_1_5} by the black curve. Other parameters used in numerical calculations: $\phi=\pi, \tilde{\Delta}_d=1, \alpha=0.05, \xi=10^{-2}$. }}\label{fig:fig4_1_10}
\end{figure}
\begin{figure}
\centering
\includegraphics[width=.55\columnwidth]{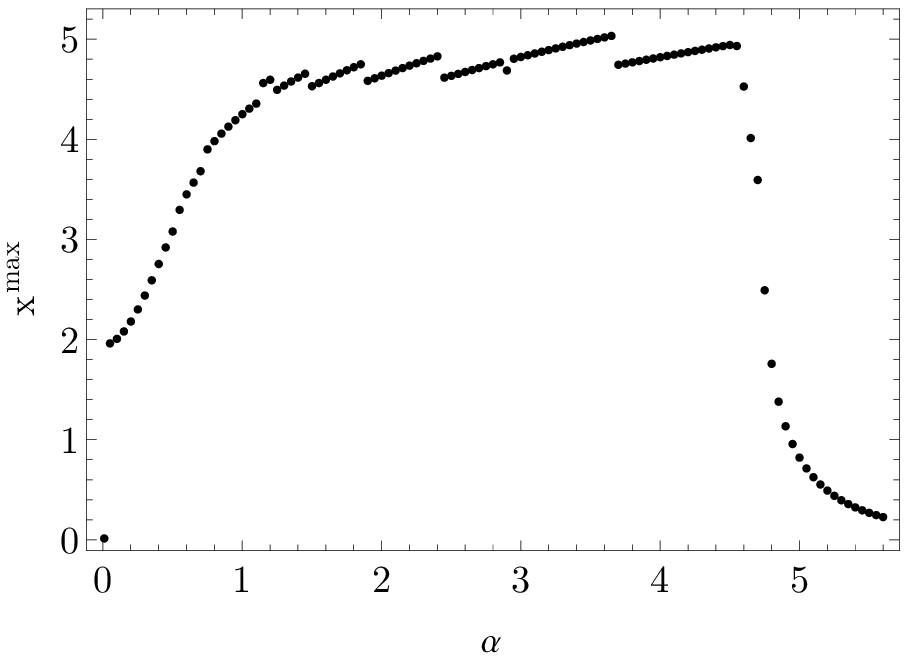}
\caption{\textit{Dependence of the amplitude $\sqrt{A_c}$ of QD self-sustained oscillations (normalized to the tunneling length $\lambda$) on the adiabaticity parameter $\alpha$. One can note that the magnitude of the dot oscillation amplitude is maximal when $\omega\sim\Gamma_n$ and tends to zero in the diabatic regime. Other parameters used in numerical calculations: $\kappa=+1, \tilde{\varepsilon}_d=1/\sqrt{3}, \phi=\pi , \tilde{\Delta}_d=1, \xi=10^{-2}$.}
}\label{fig:fig4_1_11}
\end{figure}
The analytical procedure described above is done in the assumption of the adiabatic limit, $\alpha=\omega/(2\Gamma_n)\ll 1$. However, in order to enlarge the range of parameters for which predicted effects are valid, the system of equations for the dot density matrix elements coupled to the equation for the QD displacement, Eqs.~(\ref{xEq})-(\ref{EqR}), was solved numerically.

Figure~\ref{fig:fig4_1_10} demonstrates numerical solution results for the time evolution of the QD displacement after a small initial spontaneous fluctuation of the nanotube position, $x_0=0.01$, in three different regimes. The stable regime is represented in  Fig.~\ref{fig:fig4_1_10}a. Figure~\ref{fig:fig4_1_10}b corresponds to the critical value of the dot energy level $\tilde{\varepsilon}_d=0$ (the same as for $\phi=2\pi n$, where $n$ is an integer number) when the state of the mechanical subsystem is
neither stable, nor unstable. Lastly, in the regime of presence of the mechanical instability, Fig.~\ref{fig:fig4_1_10}c, the amplitude of the dot oscillations starts to exponentially grow after a small shift from its equilibrium position. However, after some time, magnitude of the dot oscillations saturates and one has the stable regime of self-sustained oscillations of the nanotube. It emerges even without adding an external friction, $\gamma=0$. The latter fact is associated with the self-saturation effect occurred for the considered hybrid system. We should note that Fig.~\ref{fig:fig4_1_10} was obtained for the case of weak electromechanical coupling $\xi\ll 1$, in the adiabatic limit $\alpha\ll 1$, and for other parameters which are the same as for the black curve in Figs.~\ref{fig:fig4_1_4}-\ref{fig:fig4_1_7} came up from analytical predictions. One can see that the amplitude $\sqrt{A_c}$ of the limit cycle given from Fig.~\ref{fig:fig4_1_4} corresponds to the amplitude of oscillations procured from Fig.~\ref{fig:fig4_1_10}c, $\sqrt{A_c}\approx x^{max}\approx 2$. This is a manifestation of strong validity of the analytical calculations.

Figure~\ref{fig:fig4_1_11} demonstrates dependence of the amplitude $\sqrt{A_c}$ ($x^{max}$) of nanotube self-sustained oscillations (normalized to the tunneling length $\lambda$) in the limit cycle on the adiabaticity parameter $\alpha$. It was obtained from numerical solutions of the system of first-order differential equations for the matrix elements of the dot density operator, Eq.~(\ref{EqR}), coupled to the second-order one for the dot displacement, Eq.~(\ref{xEq}). One can note that the amplitude is maximal in the resonant case, when the frequency (in energy units) of the QD vibrations equals the energy difference between Andreev levels.
The later is in an associated agreement with predictions for an electrical~\cite{Isacsson1998} and magnetic NEM system~\cite{Ilinskaya2019a}. In addition, one should take heed that for the case of an electrical electromechanical coupling, in order to obtain a limit cycle, one need to take into account an thermodynamic friction (damping), $\gamma=10^{-4}-10^{-6}$ (for high-quality nanomechanical resonators). Moreover, the maximal values of the amplitude for mechanically \textit{unstable} case is of order of magnitude as for the one in the adiabatic regime, more precisely, approximately twice bigger. Nevertheless, the amplitude is negligible small in the diabatic (anti-adiabatic) limit $\alpha\gg 1$ that means the \textit{stability} of the equilibrium position of the nanotube in this regime. The adiabatic limit corresponds to the case when a nanotube displacement varies very little during one act of electron tunneling which is typically realized in experiments~\cite{Willick2020}.

\section{Ground-state cooling of nanomechanical vibrations by Andreev tunneling.}\label{sect4_2}

In this section a crucial influence of electron tunneling processes on a state of the mechanical subsystem is discussed. The effect of the ground-state cooling of vibrations of the nanotube is found. Also, a possibility to observe the cooling effect in an experiment is demonstrated.
\subsection{Quantum-mechanical description of the mechanical subsystem.} \label{subsect4_2_1}

In the previous section the nanoelectromechanical weak link composed of the carbon nanotube suspended above a trench in a normal metal electrode and positioned in a gap between two superconducting leads, was considered. Such a setup is a generalization of the experimentally implemented one~\cite{Gramich2015}, where a CNT suspended between normal and superconducting electrodes. The nanotube has been treated as a movable single-level quantum dot, in which the position-dependent superconducting order parameter is induced as a result of Cooper pair tunneling. It has been shown that in such a system self-sustained bending vibrations can emerge if a constant bias voltage is applied between normal and superconducting electrodes. Such a process of electron transport, which essentially involves Andreev conversion~\cite{Andreev1964,Kulik1970} of normal electrons into Cooper pairs, we have called by the Andreev injection. As a consequence, the interplay between coherent two-electron (Cooper pair) and incoherent single-electron tunneling into/out of the movable part of the NEMS can result in pumping or cooling effect~\cite{Stadler2016}.

However, one can note from Eq.~(\ref{in2}) that the direct eclectic current has a contribution $\propto x^2$. It means that it is crucial to take into account quantum fluctuations of the nanotube omitted within the semi-classical approach discussed in the previous section, \ref{sect4_1}. In this section we treat the bending vibrations of the nanotube quantum-mechanically which allow an investigation of the operation of such a NEM device in the cooling regime. 

\begin{figure}
\centering
\includegraphics[width=.7\columnwidth]{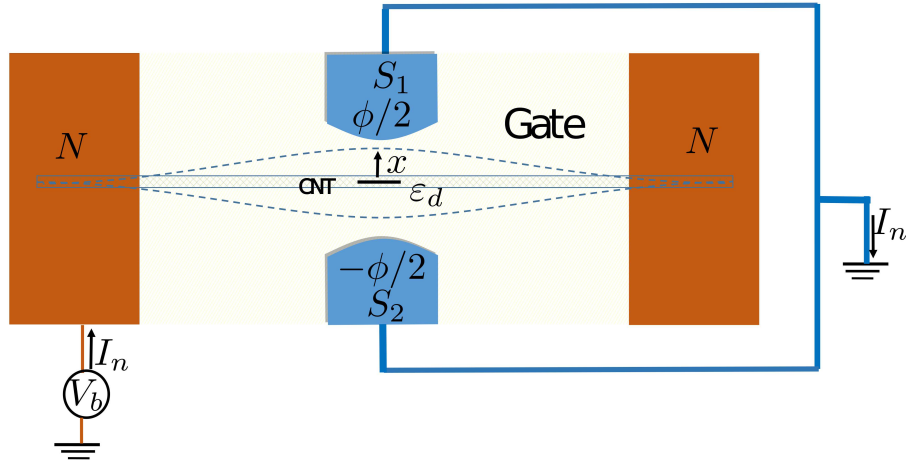}
\caption{\textit{Sketch of the nanoelectromechanical device under consideration. A carbon nanotube (CNT) is suspended in a gap between two edges of a normal electrode ($N$) and tunnel coupled to it. Also, the CNT oscillates in the $x$ direction between two superconducting leads ($S_{1,2}$). This process affects the values of the tunneling barriers between the QD and superconducting electrodes. The normal electrode is biased by voltage $V_b$.}
}\label{fig:fig4_2_1}
\end{figure}

Figure~\ref{fig:fig4_2_1} is an another schematic representation of the system under consideration. 
The Hamiltonian of the system consists of four terms,
\begin{equation}\label{H1}
H=H_d+H_v+H_l+H_t,
\end{equation}
where the Hamiltonian $H_d$ of the single-level QD, the Hamiltonian $H_l=H_l^n+H_l^s$ of the normal and superconducting electrodes, the tunnel Hamiltonian $H_t=H_t^n+H_t^s$ are done by Eqs.~(\ref{Hd})-(\ref{Ht}), respectively. It needs to be stressed that now the displacement $\hat{x}$ is an operator.
The Hamiltonian $\hat{H}_v$,
\begin{equation}\label{0035}
\hat H_v=\frac{p^2}{2m}+\frac{m\omega^2 x^2}{2},
\end{equation}
describes the mechanical dynamic of the dot, $p$ and $x$ are the
canonical conjugated momentum and coordinate,
$\left[p,x\right]=-\imath\hbar; \,m, \omega$ are the mass and
eigenfrequency of the dot, respectively.

\subsection{Equations for Wigner distribution functions.} \label{subsect4_2_2}

The time evolution of the system density matrix $\hat\rho$ is
described by the Liouville-von Neumann equation Eq.~(\ref{LvN}).
We use the reduced density matrix approximation according to which
the full density matrix of the system $\hat{\rho}$ is factorized to the tensor
product of the equilibrium density matrices of the normal and
superconducting leads, and the dot density matrix as:
\begin{equation}\label{reduced}
 \hat\rho=\hat\rho_n\otimes\hat\rho_s\otimes\hat\rho_d.   
\end{equation}
Here the reduced density operator of the QD $\hat \rho_d$ acts in the Hilbert
space which can be presented as the tensor product of the vibrational space of the harmonic
 oscillator and the electronic space of the single-electron level
on the QD.

By following the procedure introduced in subsection~\ref{subsect4_1_2}, we consider the stationary state of the system in the deep subgap case $\Delta_s\gg|eV_b|
\gg \Delta_d , \Gamma_n$, where $\Delta_d=2\pi\nu_s|t^s_0|^2,
\Gamma_n=2\pi\nu_n|t^n_0|^2$ ($\nu_{s(n)}$ is a density of states
in the superconducting (normal) electrode). Using the standard
procedure~\cite{Novotny2003}, one can trace out the lead degrees of freedom and obtain
the following equation for the reduced density matrix $\hat\rho_d$,
\begin{equation}\label{EDM1}
-\imath\left[H_d^{\text{eff}}+H_v,\hat\rho_d \right]+\mathcal
L_n\{\hat\rho_d\}+\mathcal L_\gamma\{\hat\rho_d\}=0,
\end{equation}
where the effective Hamiltonian $H_d^{eff}$ is presented in Eq.~(\ref{Hdeff}) with the off-diagonal dot order parameter, Eq.~(\ref{deltad}), induced by superconducting proximity effect which is now is an operator,
\begin{eqnarray}\label{deltad1}
  && \Delta_{d}(\hat{x},\phi)=\Delta_d \cosh(\hat{x}/\lambda+i\phi/2).
\end{eqnarray}
The Lindbladian term $\mathcal L_n\{\hat\rho_d\}$ in Eq.~(\ref{EDM1}) is induced by the incoherent
electron exchange between the normal lead and the QD and in
 the high bias voltage regime, $|eV_b|\gg \varepsilon_0,\hbar\omega, T$, takes the
form of Eq.~(\ref{Ln}).
Also, in Eq.~(\ref{EDM1}) we phenomenologically introduce the dissipation
term $\mathcal L_\gamma\{\hat\rho_d\}$~\cite{BreuerPetruccione},
\begin{equation}\label{Lgamma}
\mathcal L_\gamma\{\hat\rho\}=-m\omega\gamma\left(n_B+1/2\right)
\left[x, \left[x,\hat\rho\right]\right]-\imath \left(\gamma/2\right) \left[x,
\left\{p,\hat\rho\right\}\right],
\end{equation}
where $ \gamma$ is the damping rate, $n_B$ is the Bose-Einstein
distribution function,
\begin{equation}\label{nb}
 n_B=\frac{1}{e^{\hbar\omega/T}-1},
\end{equation}
and $T$ is temperature of the thermodynamic environment.

The QD density matrix $\hat\rho_d$ acts in the Hilbert space that can be presented as a tensor product of the vibrational space of the harmonic oscillator and the Fock space of the single-level QD which is spanned
on the state vectors $\vert 0\rangle, d^\dag_\uparrow
(d^\dag_\downarrow )\vert 0\rangle =\vert \uparrow (\downarrow)
\rangle, d^\dag_\uparrow d^\dag_\downarrow\vert
0\rangle=\vert\uparrow \downarrow \rangle \equiv \vert 2 \rangle$. It means the dot density matrix can be presented as:
\begin{equation}\label{rhomatrix}
    \hat{\rho}_{d}= \hat{\rho}_{d}(x,x')=\begin{pmatrix} \hat{\rho_0} &\hat{\rho}_{02} &0&0\\ \hat{\rho}_{20}&\hat{\rho}_{2}&0&0\\ 0 & 0& \hat{\rho}_\uparrow &0\\ 0 & 0 &0& \hat{\rho}_\downarrow \end{pmatrix}.
\end{equation}
Thus, we get the following system of equations of
motion for electronic components of the
density matrix $\hat{\rho}_d$ ($\kappa=+1$),
\begin{eqnarray}\label{002}
&&\hspace{-1.8cm}\partial_t\rho_0=-\imath\left[H_v,\rho_0\right]-4\Gamma_n\rho_0-\imath
\Delta_d(x,\phi)\rho_{20}+\imath\rho_{02}\Delta_d^\ast(x,\phi)
+\mathcal L_\gamma\{\hat\rho_0\},\\
&&\hspace{-1.8cm}\partial_t\rho_\uparrow=-\imath\left[H_v,\rho_\uparrow\right]
+2\Gamma_n(\rho_0-\rho_\uparrow)+ \mathcal L_\gamma\{\hat\rho_\uparrow\},\\
&&\hspace{-1.8cm}\partial_t\rho_\downarrow=-\imath\left[H_v,\rho_\downarrow\right]
+2\Gamma_n(\rho_0-\rho_\downarrow)+\mathcal L_\gamma \{\hat
\rho_\downarrow\},\\
&&\hspace{-1.8cm}\partial_t\rho_{02}=
-\imath\left[H_v,\rho_{02}\right]+2\imath\varepsilon_d
\rho_{02}-2\Gamma_n\rho_{02}-\imath
\Delta_d(x,\phi)\rho_2+\imath\rho_0\Delta_d(x,\phi)+\mathcal
L_\gamma\{\hat\rho_{02}\},\\
&&\hspace{-1.8cm}\partial_t\rho_{20}=
-\imath\left[H_v,\rho_{20}\right]-2\imath\varepsilon_d
\rho_{20}-2\Gamma_n\rho_{20}-\imath
\Delta_d^\ast(x,\phi)\rho_0+\imath\rho_2\Delta_d^\ast(x,\phi)
+\mathcal L_\gamma\{\hat\rho_{20}\},\\
&&\hspace{-1.8cm}\partial_t\rho_2=-\imath\left[H_v,\rho_2\right]+2\Gamma_n
(\rho_\uparrow+ \rho_\downarrow)+\imath
\rho_{20}\Delta_d(x,\phi)-\imath\Delta_d^\ast(x,\phi)\rho_{02}
+\mathcal L_\gamma\{\rho_2\}.\label{003}
\end{eqnarray}
Here ordering of the operators $\hat{\rho_{ii}}$ and $\hat{\Delta}_d(\hat{x},\phi)$ is important.
To find the equations in case of the opposite direction of the bias voltage, $\kappa=-1$, one needs to switch $0\rightleftarrows 2$, so that,
\begin{eqnarray}\label{002b}
&&\hspace{-1.8cm}\partial_t\rho_0=-\imath\left[H_v,\rho_0\right]+2\Gamma_n
(\rho_\uparrow+ \rho_\downarrow)+\imath
\rho_{02}\Delta_d(x,\phi)-\imath\Delta_d^\ast(x,\phi)\rho_{20}
+\mathcal L_\gamma\{\rho_0\}.\\
&&\hspace{-1.8cm}\partial_t\rho_\uparrow=-\imath\left[H_v,\rho_\uparrow\right]
+2\Gamma_n(\rho_2-\rho_\uparrow)+ \mathcal L_\gamma\{\hat\rho_\uparrow\},\\
&&\hspace{-1.8cm}\partial_t\rho_\downarrow=-\imath\left[H_v,\rho_\downarrow\right]
+2\Gamma_n(\rho_2-\rho_\downarrow)+\mathcal L_\gamma \{\hat
\rho_\downarrow\},\\
&&\hspace{-1.8cm}\partial_t\rho_{02}=
-\imath\left[H_v,\rho_{02}\right]-2\imath\varepsilon_d
\rho_{02}-2\Gamma_n\rho_{02}-\imath
\Delta_d^\ast(x,\phi)\rho_2+\imath\rho_0\Delta_d^\ast(x,\phi)
+\mathcal L_\gamma\{\hat\rho_{02}\},\\
&&\hspace{-1.8cm}\partial_t\rho_{20}=
-\imath\left[H_v,\rho_{20}\right]+2\imath\varepsilon_d
\rho_{20}-2\Gamma_n\rho_{20}-\imath
\Delta_d(x,\phi)\rho_0+\imath\rho_2\Delta_d(x,\phi)+\mathcal
L_\gamma\{\hat\rho_{20}\},\\
&&\hspace{-1.8cm}\partial_t\rho_2=-\imath\left[H_v,\rho_2\right]-4\Gamma_n\rho_2-\imath
\Delta_d(x,\phi)\rho_{02}+\imath\rho_{20}\Delta_d^\ast(x,\phi)
+\mathcal L_\gamma\{\hat\rho_0\},
\end{eqnarray}
It is convenient to introduce the
linear combinations of the density matrix elements as
follows:
\begin{eqnarray}
&&R_v=\rho_0+\rho_\uparrow+\rho_\downarrow+\rho_2,\nonumber\\
&& R_0=\rho_0+\rho_2,\nonumber\\ 
&& R_1=\rho_{20}+\rho_{02},\nonumber\\
&&R_2=\imath(\rho_{02}-\rho_{20}),\nonumber\\
&& R_3=\rho_0-\rho_2.
\end{eqnarray}\label{005b}
Then the equations for the density matrix elements of the QD, Eqs.~(\ref{002})-(\ref{003}), have the following form:
\begin{eqnarray}\label{EDME2}
    && \partial_t R_0=-\imath[H_v, R_0]+2\Gamma_n (R_v-2R_0-\kappa R_3)+\imath[R_1, \Delta'(x,\phi)]-\imath[R_2, \Delta''(x,\phi)]\nonumber\\
    &&\hspace{2.cm}+\mathcal{L}_\gamma \{ R_0\}, \\
    &&\partial_t R_3=-\imath[H_v,R_3]-2\Gamma_n (\kappa R_v+R_3)+\{ R_1, \Delta''(x,\phi)\}+\{R_2,\Delta'(x,\phi)\}\nonumber\\
    &&\hspace{2cm}+\mathcal{L}_\gamma \{ R_3\},\\
    && \partial_t R_1=-\imath[H_v,R_1]+2 \varepsilon_d R_2-2\Gamma_n R_1+\imath[R_0,\Delta'(x,\phi)]-\{R_3,\Delta''(x,\phi)\}\nonumber\\
    &&\hspace{2cm}+\mathcal{L}_\gamma \{ R_1\}, \\
    && \partial_t R_2=-\imath[H_v,R_2]-2\varepsilon_d R_1-2\Gamma_n R_2-\imath[R_0,\Delta''(x,\phi)]-\{R_3,\Delta'(x,\phi)\}\nonumber\\
    &&\hspace{2cm}+\mathcal{L}_\gamma \{ R_2\}, \\
    &&\partial_t R_v=-\imath[H_v,R_v]+\imath[R_1,\Delta'(x,\phi)]-\imath[R_2,\Delta''(x,\phi)]+\mathcal{L}_\gamma \{ R_v\}.\label{EDME2b}
\end{eqnarray}

The state of the mechanical subsystem is completely
described by the reduced density matrix, 
\begin{equation}\label{rhotrace}
   \hat\rho_v=\text{Tr}
\hat\rho_d, 
\end{equation}
 where the tracing operation is taken over the
electronic degrees of freedom on the dot. It is obvious that in the
 limiting case $\lambda\rightarrow\infty$ the electronic and
vibronic subsystems are independent and the reduced vibronic
density matrix has a form of equilibrium density matrix with
the effective temperature that is determined by an environment
temperature $T$.

It is convenient and natural to use the description
in terms of the Wigner distribution function,
\begin{equation}\label{15w}
W_i (x,p)=\frac{1}{2\pi}\int d\xi e^{-\imath p\xi}\left\langle
x+\frac{\xi}{2}\vert \hat \rho_i\vert x-\frac{\xi}{2}\right\rangle,
\end{equation}
see also section~\ref{sect3_5}. In general, a Wigner function gives probability distributions as:
\begin{eqnarray}
    &&\int dp W(x,p)=\langle x\vert \hat{\rho}\vert x \rangle,\\
    &&\int dx W(x,p)=\langle p\vert \hat{\rho}\vert p \rangle ,
\end{eqnarray}
and is a bounded function,
\begin{equation}
    -2/\hbar \leq W(x,p) \leq 2/\hbar,
\end{equation}
which is an evidence of the uncertainty principle. In addition, Eq.~(\ref{15w}) can be rewritten as follows:
\begin{equation}\label{15b}
W_i (x,p)=\frac{1}{2\pi}\int d\xi e^{\imath x\xi}\left\langle
p+\frac{\xi}{2}\vert \hat \rho_i\vert p-\frac{\xi}{2}\right\rangle.
\end{equation}
Note that here and in what follows we introduce dimensionless variables: $x/x_0\rightarrow x,
px_0/\hbar\rightarrow p$, where $x_0$ is the amplitude of zero-point
oscillations, all energy parameters are measured in units of
$\hbar \omega$, the tunneling length $\lambda$ is measured in
units of $x_0, \gamma/\omega\rightarrow \gamma$.

We are interested in a steady state regime of the
mechanical subsystem in the limit of weak electromechanical coupling $1/\lambda\ll 1$. To find a form of Eqs.~(\ref{EDME2})-(\ref{EDME2b}) in the Wigner-Moyal representation to leading order in this parameter, one can use the following expressions:
\begin{eqnarray}
    && [H_v, \hat{R}_i]\to (x\partial_p-p\partial_x) W_i(x,p),\\
    &&[\hat{x},[\hat{x},\hat{R}_i]]\to -\partial_p^2 W_i(x,p),\\
    &&[\hat{x},\{\hat{p},\hat{R}_i\}]\to 2\imath \partial_ppW_i(x,p);\\
    &&\text{e}^{\hat{x}/\lambda}\hat{R}_i\rightarrow \text{e}^{x/\lambda}W_i(x,p+\imath/(2\lambda))\approx W_i(x,p)+\frac{\imath}{2\lambda}\partial_pW_i(x,p),\\
    &&\hat{R}_i\text{e}^{\hat{x}/\lambda}\rightarrow \text{e}^{x/\lambda}W_i(x,p-\imath/(2\lambda))\approx W_i(x,p)-\frac{\imath}{2\lambda}\partial_pW_i(x,p).
\end{eqnarray}
It results in that the steady-state equation for the Wigner function which describes the mechanical degree of freedom, $W_v(x,p)$, is given by the equation (up to terms of the second order in the parameter $1/\lambda$),
\begin{eqnarray}\label{wv1}
    \left\{ (x\partial_p-p\partial_x)+\mathcal{L}_\gamma\right\}W_v=-\frac{x}{\lambda^2}\Delta_d \cos{(\phi/2)}\partial_p W_1+\frac{1}{\lambda}\Delta_d\sin{(\phi/2)}\partial_p W_2.
\end{eqnarray}
Here we assume $\gamma\sim (1/\lambda^2)$. This equation is coupled to the steady state equation for the
vector-function $\vec{W}=(W_0, W_1,W_2,W_3)^T$ that takes the following form (up to terms of
the first order in the parameter $1/\lambda$),
\begin{eqnarray}\label{010}
&&\hspace{2cm}(x\partial_p-p\partial_x)\vec{W}+2\hat M \vec{W}= \vec{F},\\
\label{A17}
&&\label{011} \hat M=\left(
\begin{array}{cccc}
-2\Gamma_n& 0 & 0 &- \kappa\Gamma_n\\
0 &-\Gamma_n&\varepsilon_d & 0 \\
0 &-\varepsilon_d&-\Gamma_n&-\Delta_d\cos(\phi/2)\\
0&0&\Delta_d\cos(\phi/2)&-\Gamma_n
\end{array}\right),\nonumber\\
&&\hspace{1cm}\label{012}\vec{F}=
-2\Gamma_n W_v\left(\begin{array}{c} 1 \\0\\0 \\-\kappa\\
\end{array}\right)+\frac{\Delta_d}{\lambda}\sin(\phi/2)
\left(\begin{array}{c} \partial_p W_2\\2x W_3\\ \partial_p
W_0\\-2x W_1\\
\end{array}\right).\nonumber
\end{eqnarray}

Furthermore, it is convenient to change from ($x,p$) to polar
coordinates ($A,\varphi$), so that, $x-\bar x=A \sin \varphi$ and $p=A\cos\varphi$,
where $\bar x\sim (1/\lambda)$ stands for an equilibrium
position of the dot. Then, equations~(\ref{wv1})-(\ref{010}) take the following form:
\begin{eqnarray}\label{wv2}
&&\hspace{0cm}-\frac{\partial W_v}{\partial
\varphi}+ \bar x\hat T W_v +\gamma\left(n_B+1/2\right)\hat
T^2W_v+\gamma\left(W_v+A\cos\varphi\hat T W_v\right)-\nonumber\\
&&\hspace{1.5cm}-\frac{\Delta_d}{\lambda}\sin(\phi/2)\hat T W_2+\frac{\Delta_d A}{\lambda^2}\cos(\phi/2)
\sin\varphi\hat T W_1=0;
\end{eqnarray}
\begin{eqnarray}\label{vecw2}
&&\hspace{4cm}-\frac{\partial \overrightarrow
W}{\partial\varphi}+2\hat M \overrightarrow W= \vec{F},\\
&&\hat M=\left(
\begin{array}{cccc}
-2\Gamma_n& 0 & 0 &- \kappa\Gamma_n\\
0 &-\Gamma_n&\varepsilon_d & 0 \\
0 &-\varepsilon_d&-\Gamma_n&-\Delta_d\cos(\phi/2)\\
0&0&\Delta_d\cos(\phi/2)&-\Gamma_n
\end{array}\right),\nonumber\\
&&\hspace{0cm}\vec{F}=-\bar x \hat T \overrightarrow W
-2\Gamma_n W_v\left(\begin{array}{c} 1 \\0\\0 \\-\kappa\\
\end{array}\right)+\frac{\Delta_d}{\lambda}\sin(\phi/2)
\left(\begin{array}{c} \hat T W_2\\2A\sin\varphi W_3\\\hat T
W_0\\-2A\sin \varphi W_1\\
\end{array}\right).\nonumber
\end{eqnarray}
Here the differential operator $\hat T$ is defined
according to the expression:
\begin{equation}\label{hatT}
\hat T\equiv\partial_p=\cos\varphi\frac{\partial}{\partial A}-\frac{\sin\varphi}
{A}\frac{ \partial}{\partial \varphi}.
\end{equation}
Also, one can take a heed that $(x\partial_p-p\partial_x)=-\partial_\varphi$, $\partial_p^2=(1/2A)\partial_AA\partial_A$ and $\partial_pp=(1/2A)\partial_AA^2$.
Equations~(\ref{wv2})-(\ref{vecw2}) have to be solved subject to the
periodic boundary conditions,
\begin{equation}
    W_v(A,\varphi+2\pi)=
W_v(A,\varphi), \qquad \vec W(A,\varphi+2\pi)=\vec
W(A,\varphi).
\end{equation}
We solve these equations by perturbation expansions,
\begin{equation}\label{A181}
W_i(A,\varphi)\rangle=W_i^{(0)}(A,\varphi)+ W_i^{(1)}(A,\varphi)
+...,
\end{equation}
($i=v,0,1,2,3$), where $W_i^{(n)}$ is of
$n$:th order in the parameter $1/\lambda\simeq
10^{-2}-10^{-3}$~\cite{Morpurgo1999} (or the parameter of electromechanical coupling,
$\Delta_d/\lambda\ll 1$). 

It can be easily find from Eqs.~(\ref{wv2})-(\ref{vecw2}) in zeroth order of the perturbation theory that the functions
$W_v^{(0)}(A,\varphi), \vec W^{(0)}(A,\varphi)$ do not
depend on $\varphi$. Hence, $W_v^{(0)}(A,\varphi)=W_v^{(0)} (A)$
and
\begin{eqnarray}\label{A19}
&& W_0^{(0)}=\frac{\varepsilon_d^2+\Gamma_n^2+
(\Delta_d^2/2)\cos(\phi/2)}{D^2}W_v^{(0)},\\
&& W_1^{(0)}=\kappa\frac{\Delta_d\varepsilon_d\cos(\phi/2)}{D^2}W_v^{(0)},\\
&& W_2^{(0)}= \kappa\frac{\Delta_d\Gamma_n\cos(\phi/2)}{D^2}W_v^{(0)},\\
&& W_3^{(0)}=-\kappa\frac{\varepsilon_d^2+\Gamma_n^2}{D^2}W_v^{(0)},
\end{eqnarray}
where 
\begin{equation}
    D^2=\varepsilon_d^2+\Gamma_n^2+\Delta_d^2\cos^2(\phi/2).
\end{equation}
From the requirement, $W_v^{(1)}(A,\varphi)=W_v^{(1)}(A)$, to
first order in the perturbation theory, Eq.~(\ref{wv2}) determines an
equilibrium position of the dot,
\begin{equation}\label{026}
\bar x=\kappa\frac{\Delta_d^2}{\lambda D}\sin(\phi/2)\cos(\phi/2).
\end{equation}
In the second order of the perturbation theory from Eq.~(\ref{wv2}) one gets:
\begin{equation}\label{2order1}
    -\frac{\partial W_v^{(2)}}{\partial\varphi}-\frac{\Delta_d}{\lambda}\sin{(\phi/2)}\hat{T}W_2^{(1)}+\frac{\Delta_d}{\lambda}\cos{(\phi/2)}A\sin{\varphi}\hat{T}W_1^{(0)}+\bar{x}\hat{T}W_v^{(1)}+\mathcal{L}_\gamma\{ W_v^{(0)}\}=0.
\end{equation}
Then, let us average this equation over the $\varphi$ variable in an usual way,
\begin{equation}\label{average}
    \langle W(A,\varphi)\rangle=\frac{1}{2\pi}\int_0^{2\pi}d\varphi W(A,\phi).
\end{equation}
By substituting Eq.~(\ref{hatT}) into Eq.~(\ref{average}), one finds after straightforward calculations the following expression:
\begin{equation}\label{202}
\langle\hat T W(A,
\varphi)\rangle=\frac{1}{A}\frac{\partial}{\partial
A}\left(A\langle\cos\varphi W(A,\varphi)\rangle\right).
\end{equation}
From these equations one can see that the first, third and fourth terms in the r.h.s. of Eq.~(\ref{2order1}) do not give an contribution, so that one obtains the following equation for $W_v^{(0)}$,
\begin{eqnarray}\label{wv3}
&&-\frac{\Delta_0\sin(\phi/2)}{\lambda A} \frac{\partial}{\partial
A} \left( A\left\langle\cos\varphi W_2^{(1)} \right\rangle\right)+\frac{\gamma}
{2A}\frac{\partial}{\partial A}\left(A^2W_v^{(0)}\right)+\nonumber\\
&&\hspace{0.5cm}+\frac{\gamma\left(n_B+1/2\right) }{2A} \frac{\partial}{\partial A}\left(A
\frac{\partial W_v^{(0)}} {\partial A}\right)=0.
\end{eqnarray}
Therefore, to get a closed equation for $W_v^{(0)}(A)$, one needs
to know the function $W_2^{(1)}(A,\varphi)$. This function can be determined from
Eqs.~(\ref{vecw2}) which in the first order in the perturbation theory has a form:
\begin{eqnarray}\label{vecw3}
&&\hspace{0cm}-\frac{\partial \vec{
\tilde{W}}^{(1)}}{\partial\varphi}+2\hat{\tilde{M}} \vec{\tilde{W}}^{(1)}= \vec{\tilde{F}},
\end{eqnarray}
where $\hat{\tilde{W}}$ denotes an reduced vector-function $\hat{\tilde{W}}=(W_1,W_2,W_3)^T$ because of in the first-order approximation the equation for the Wigner function $W_0^{(1)}$ is decoupled from the other ones and is not relevant in what follows. Additionally,
\begin{eqnarray}
&&\hat{\tilde{M}}=\left(
\begin{array}{cccc}
-\Gamma_n&\varepsilon_d & 0 \\
-\varepsilon_d&-\Gamma_n&-\Delta_d\cos(\phi/2)\\
0&\Delta_d\cos(\phi/2)&-\Gamma_n
\end{array}\right),\nonumber\\
&&\hspace{0cm}\vec{\tilde{F}}=-\bar x \hat T \vec{\tilde{W}}^{(0)}
-2\Gamma_n W_v^{(1)}\left(\begin{array}{c} 0\\0 \\-\kappa\\
\end{array}\right)+\frac{\Delta_d}{\lambda}\sin(\phi/2)
\left(\begin{array}{c} 2A\sin\varphi W_3^{(0)}\\\hat T
W_0^{(0)}\\-2A\sin \varphi W_1^{(0)}\\
\end{array}\right).\nonumber
\end{eqnarray}
The system of first-order differential equations, Eq.~(\ref{vecw3}), can be solved exactly by using the Fourier representation  since this equation contains functions which are periodic with the period of $2\pi$, $W(A,\varphi+2\pi)=W(A,\varphi)$,
\begin{equation}\label{Fourier}
    W^{(1)}_i(A,\varphi)=\sum\limits_{n=-\infty}^{+\infty}w(n)\text{e}^{\imath n \varphi},\qquad w(n)=\frac{1}{2\pi}\int_0^{2\pi}W^{(1)}_i(\varphi)\text{e}^{-\imath n\varphi}d\varphi.
\end{equation}
Furthermore, because of the structure of Eq.~(\ref{wv3}), one can note that only the first harmonics ($n=\pm 1$) of the Fourier series, Eq.~(\ref{Fourier}), give an contribution,
\begin{eqnarray}\label{w2}
      && <\cos{\varphi} W_2^{(1)}(A,\varphi)>=\sum_n w_2(n)\left[ <\cos{\varphi}\cos{(n\varphi)}>+\imath <\cos{\varphi}\sin{(n\varphi)}> \right]=\nonumber\\
    &&\hspace{1cm}=\sum_n w_2(n)
     \begin{cases}
     0, & n\ne \pm 1,\\
     1/2, & n=\pm 1;
     \end{cases}=\frac{1}{2}(w_2(1)+w_2(-1))=\text{Re}w_2(1).
\end{eqnarray}
Then, straightforward calculations leads to the following form of Eq.~(\ref{wv3}),
\begin{equation}\label{204}
{\cal D}_1\frac{\partial}{\partial
A}\left(A^2W_v^{(0)}\right)+{\cal D}_2\frac{\partial}{\partial
A}\left(A\frac{\partial W_v^{(0)}}{\partial A}\right)=0,
\end{equation}
that is a stationary Fokker-Planck equation for the Wigner function $W_v^{(0)}(A)$, which describes the state of the mechanical subsystem, with the drift $\mathcal{D}_1$ and diffusive $\mathcal{D}_2$ coefficients, 
\begin{eqnarray}\label{203}
&&{\cal D}_1= -\kappa\frac{\Delta_d^2\Gamma_n\varepsilon_d}
{\lambda^2 D_1}\sin^2(\phi/2)+\gamma,\\ \label{2040} &&{\cal
D}_2=\frac{\Delta_d^2\Gamma_n C}{\lambda^2
D_1}\sin^2(\phi/2)+\gamma\left(n_B+1/2\right).
\end{eqnarray}
Here
\begin{eqnarray}\label{205}
&&D^2=\varepsilon_d^2+\Gamma_n^2+\Delta_d^2\cos^2(\phi/2),\\
&&D_1=\left(D^2-1/4\right)^2+\Gamma_n^2,\\
&&C=\frac{\left(D^2+1/4\right)\left(D^2+\varepsilon_d^2+
\Gamma_n^2 \right)-4\Delta_d^2\Gamma_n^2\cos^2(\phi/2)}{4D^2}.
\label{68}
\end{eqnarray}
The solution of Eq.~(\ref{204}) at small (in comparison to $\lambda$) values of the amplitude has
a form of the Boltzmann distribution function,
\begin{equation}\label{210}
W_v^{(0)}(x,p)=(\beta/\pi)\exp\left[-\beta\left(x^2+p^2\right)\right],
\end{equation}
where  the coefficient $\beta={\cal D}_1/2{\cal D}_2$.

The expressions, Eqs.~(\ref{203}), (\ref{2040}), define the
framework of validity of our consideration. It follows from Eqs.~(\ref{203})-(\ref{68}) that in the region which is related to the maximal cooling effect as we will show in the next section,~\ref{subsect4_2_3},
(the range of the values of parameters ($\phi, \varepsilon_d)$
near the point $\varepsilon_d=1/2, \phi =\pi$) the value of the level
width is restricted from below, $\Gamma_n\geq
\Gamma_n^{(0)}=\Delta_d^2/\lambda^2$.

\subsection{Ground-state cooling of nanomechanical vibrations.} \label{subsect4_2_3}

Nowadays, nanomechanical resonators with a significant value of the quality factor are achieved in experiments~\cite{Moser2014,Laird2011}. For such a case, the electromechanical coupling dominates the coupling with the thermodynamic environment, $1/\lambda\gg \gamma$. Thus, let us consider the case $\gamma\rightarrow 0$. From
Equations~(\ref{203})-(\ref{2040}) it follows  that the sign of the coefficient $\beta$ is determined by the sign of $\kappa\varepsilon_d$. If $\kappa\varepsilon_d$ is positive,
$\beta$ becomes negative. This situation corresponds to the \textit{mechanical
instability} of the system and it was discussed in Ref.~\cite{Bahrova2022}, see section~\ref{sect4_1}. In what
follows we consider the vibronic (stable) regime, when
$\kappa=-1,\,\varepsilon_d>0$ (the same for $\kappa=+1,\,\varepsilon_d<0$).

The coefficient $\beta$ determines the probability $P_0$ that the
system is in its ground state. In terms of Wigner distribution
functions this probability takes a form:
\begin{equation}\label{56}
P_0=2\pi\int dx dp W_v^{(0)}(x,p)W_0(x,p)=\frac{2\beta}{\beta+1},
\end{equation}
where
\begin{equation}\label{w0}
    W_0(x,p)=(1/\pi)\exp[-(x^2+p^2)],
\end{equation}
is the Wigner function of the harmonic oscillator
ground state. Note that according to Heisenberg's uncertainty
principle the maximal value of parameter $\beta$ is equal to
unity, $\beta\leqslant\beta_{\text{max}}=1$.

\begin{figure}
\centering
\includegraphics[width=0.55\columnwidth]{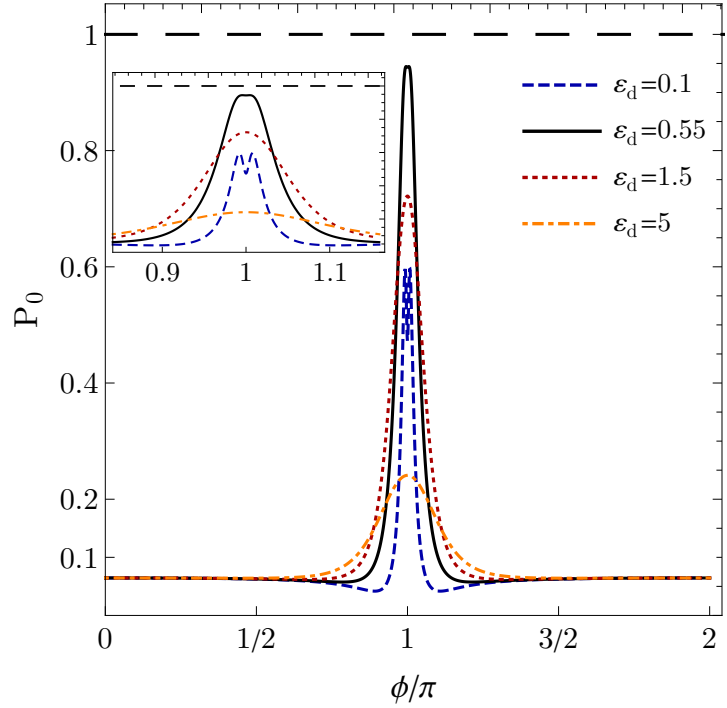}
\caption{\textit{The ground state occupation probability $P_0$ versus the
superconducting phase difference $\phi$ for different values of
the quantum dot energy level: $\varepsilon_d=0.1$ (blue
dashed curve), $0.56$ (black thick), $1.5$ (red dotted), $5$
(orange dot-dashed). The black dotted line indicates the maximal
value of the occupation probability. Inset: zoomed central region
where the cooling reaches its maximum at $\phi=\pi$. Other
parameters: $\Gamma_n=0.2; \Delta_d=25;\lambda=100; \gamma=
10^{-5}, T=15\hbar\omega$.}}\label{fig:fig4_2_3}.
\end{figure}

Dependencies of the probability $P_0$ as
a function of the superconducting phase difference $\phi$ for
different values of the quantum dot energy level
$\varepsilon_d$ are demonstrated in Fig.~\ref{fig:fig4_2_3}.

We can see that the  maximal effect of cooling takes place in the region
$\phi\simeq \pi, \varepsilon_d\simeq 1/2$, the degree of cooling
reaches the significant values, $P_0\simeq 0.95$. One can estimate it as
\begin{equation}\label{p0estimation}
    P_0=\frac{2\varepsilon_d}{\varepsilon_d+\varepsilon_d^2+\Gamma_n^2+1/4},
\end{equation}
and, therefore, note that the
maximal cooling effect occurs in the anti-adiabatic regime,
$\Gamma_n\simeq 0.2<1$.
More precisely, the extrema of the function $\beta$ are the following:
\begin{eqnarray}
 &&\phi_{extr,n}=\pi n;\qquad n\in \mathbb{Z},\\
      &&\phi_{extr,\pm}=2\arccos{\frac{\pm\sqrt{\sqrt{(\varepsilon_d^2+\Gamma_n^2)(16\Gamma_n^2+1)}-2(\varepsilon_d^2+\Gamma_n^2)}}{\sqrt{2}\Delta_d}},
\end{eqnarray}
and associated with the ones seen in Fig.~\ref{fig:fig4_2_3}.
\begin{figure}
\centering
\includegraphics[width=0.55\columnwidth]{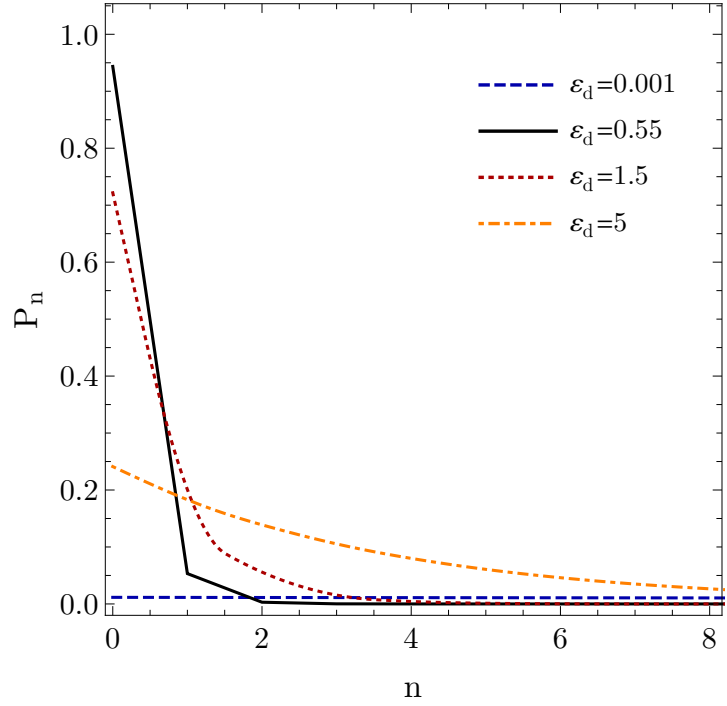}
\caption{\textit{The probability distribution $P_n$ for different values of
the quantum dot energy level: $\varepsilon_d=0.001$ (blue
dashed curve), $0.56$ (black thick), $1.5$ (red dotted), $5$
(orange dot-dashed) for $\phi=\pi$. The black dotted line indicates the maximal
value of the ground-state occupation probability. Other
parameters are the same as in Fig.~\ref{fig:fig4_2_3}: $\Gamma_n=0.2; \Delta_d=25;\lambda=100; \gamma=
10^{-5}, T=15\hbar\omega$.}}\label{fig:fig4_2_4}.
\end{figure}

Additionally, we calculate the probability distribution $P_n$. The probability to find the mechanical subsystem being in a state $n$ is defined as:
\begin{equation}\label{pn}
    P_n=2\pi\int dx dp W_v^{(0)}(x,p)W_n(x,p),
\end{equation}
where now $W_n(x,p)$ stands for the Winger function of the harmonic oscillator corresponded to $n$th Fock state,
\begin{equation}\label{wn}
    W_n(x,p)=\frac{(-1)^n}{\pi}\text{e}^{-(x^2+p^2)}L_n[2(x^2+p^2)],
\end{equation}
where $L_n(x)$ denotes an $n$th Laguerre polynomial. Thus, $n=0$ ($L_0(x)=1$) gives an special case of Eq.~(\ref{w0}).
Straightforward calculations leads to the following expression for the probability $P_n$,
\begin{equation}\label{pn1}
    P_n=\frac{2\beta (1-\beta)^n}{(1+\beta)^{n+1}}=\left(\frac{1-\beta}{1+\beta}\right)^n P_0,
\end{equation}
where $P_0$ is defined by Eq.~(\ref{56}).

\subsection{Non-monotonic behaviour of electric current.} \label{subsect4_2_4}

The effect of cooling of the mechanical vibrations can
be explored by dc current measurements. The Wigner distribution
function gives the possibility to calculate various physical
quantities. The electric current through the system can be defined in a standard way as a change of the number of electrons in the normal lead, see Eq.~(\ref{in1}). Thus, one can obtain the following expression for the electric current in terms of Wigner functions, analogous to Eq.~(\ref{Icurrent}),
\begin{equation}\label{inn}
    I/I_0=\kappa\int_0^{2\pi}d\varphi\int_0^\infty dA A [W_v(A,\varphi)+\kappa W_3(A,\varphi)],
\end{equation}
where $I_0=e\Gamma_n/\hbar$. 
In the zeroth order of perturbation theory over the parameter of the electromechanical coupling, using Eqs.~(\ref{A19}) and (\ref{210}), we get the expression for the static current (see the first term in the r.h.s. of Eq.~(\ref{Iav})),
\begin{equation}\label{currall}
I_n^{(0)}=I_0\frac{\Delta_d^2\cos^2{(\phi/2)}}{\Gamma_n^2+\varepsilon_d^2+
\Delta_d^2\cos^2(\phi/2)}.
\end{equation}
From this equation one can see that the current $I_n^{(0)}$ is equal to zero at $\phi=\pi$ despite the fact of maximal effect of cooling in this regime we are interested the most. Therefore, we need to consider next-order corrections. From Equation~({\ref{inn}}) one notes that in order to find the next perturbation order terms, one need to know the functions $W_v^{(1),(2)}$ and $W_3^{(1),(2)}$, at least. This fact leads to the requirement to include in Eqs.~(\ref{wv1}),(\ref{010}) a contribution from the next order of the perturbation theory. 

However, one can avoid to do that because of the fact that due to the geometry of our system, the normal current is equal to
the sum of the partial currents corresponding to the superconducting
electrodes, $I_n=I_1^{(s)}+I_2^{(s)}$. Here the supercurrent in the $j$ superconducting lead is
determined by the change of the number of Cooper pairs and can be
presented as:
\begin{equation}\label{currj}
I_j^{(s)}=\frac{2e}{\hbar}\text{Tr}\left(\frac{\partial
H_d^{\text{eff}}}{\partial \phi_j}\hat\rho_d\right),
\end{equation}
 where one should take into account that in more general case the expression for the dot order parameter is following:
 \begin{equation}\label{dotd}
     \Delta_d(x,\phi)=\frac{1}{2}\Delta_d (\text{e}^{-x/\lambda-\imath \phi_1}+\text{e}^{x/\lambda-\imath\phi_2}).
 \end{equation}
 Then,
 \begin{equation}
    I_j=\imath e \omega\Delta_d\text{Tr}\left\{\text{e}^{(-1)^j x/\lambda}\left( \text{e}^{\imath\phi_j}\rho_{02}-\text{e}^{-\imath\phi_j}\rho_{20}\right)\right\}.
\end{equation}\label{ij1}
 In
terms of Wigner functions 
the expression for the electric current takes a form:
\begin{eqnarray}\label{curr}
&&I_n=e\omega\int dx dp
\left[\Delta_d\sin(\phi/2)\sinh(x/\lambda)W_1+\right.\nonumber\\
&&\hspace{2cm}+\left.\Delta_d
\cos(\phi/2)\cosh(x/\lambda)W_2\right],
\end{eqnarray}
which is convenient to be re-written in the polar coordinates (angle-action representation) up to the second-order terms in the perturbation theory as
\begin{eqnarray}
       && I_n=e\omega\Delta_d\int_0^{2\pi}d\varphi \int_0^\infty dA A\left\{\sin{(\phi/2)}\left[\frac{A\sin{\varphi}}{\lambda}W_1^{(0)}+\frac{A\sin{\varphi}}{\lambda}W_1^{(1)}\right]+\right.\nonumber\\
       &&\hspace{1.5cm}+\left.\cos{(\phi/2)}\left[W_2^{(0)}+W_2^{(2)}+\frac{A^2\sin^2{\varphi}}{2\lambda^2}W_2^{(0)}\right]\right\}.
\end{eqnarray}\label{inn3}
From this equation one can find that zeroth-order terms give Eq.~(\ref{currall}), the contribution of the first-order corrections equals to zero, and the non-zero second-order contribution (the second term in the integrand) at $\phi=\pi$ is
\begin{equation}
    I_n^{(2)}=e\omega 2\pi \frac{\Delta_d}{\lambda}\int_0^\infty dA A^2 \text{Im}w_1(1).
\end{equation}
 Thus, at $\phi=\pi$ the current is determined
by the mechanical fluctuations and in the leading order (second) of the
electromechanical coupling parameter it reads as
\begin{equation}\label{curr2}
I_n=I_0\left(\frac{\Delta_d}{\lambda}\right)^2
\frac{\left(\Gamma_n^2+\varepsilon_d^2+1/4\right)\langle
x^2\rangle+\varepsilon_d/2}
{\left(\Gamma_n^2+\varepsilon_d^2-1/4\right)^2+\Gamma_n^2},
\end{equation}
where the $\langle ...\rangle$ denote the
average value in the phase space with $W_v^{(0)}(x,p)$ and
\begin{equation}\label{x2average}
    \langle x^2\rangle=(2\beta)^{-1}.
\end{equation}

\begin{figure}
\centering
\includegraphics[width=0.55\columnwidth]{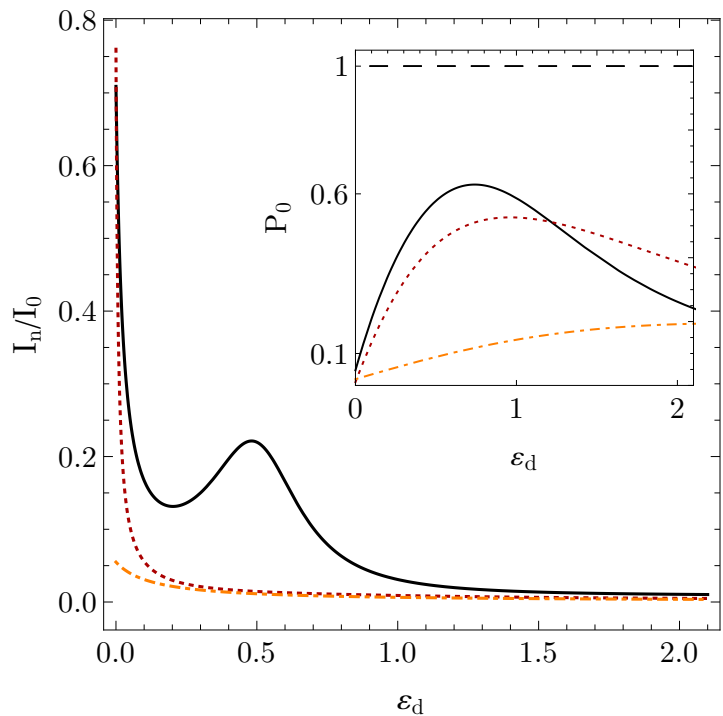}
\caption{\textit{The dependence of the electric current (normalized to
$I_0$) on the quantum dot level energy $\varepsilon_d$ at $\phi=\pi$ for
different values of $\Gamma_n:\,\Gamma_n=0.2$ (black
thick curve), $\Gamma_n=1$ (red dotted), $\Gamma_n=3$ (orange
dot-dashed). Inset: the ground state occupation probability versus
the QD level energy. Other parameters: $\Delta_d=5,\lambda=50, \gamma=5 \times
10^{-5}, T=15$.}}\label{fig:fig4_2_5}
\end{figure}

Figure~\ref{fig:fig4_2_5} presents the  dependence of the electric current on the
quantum dot level energy $\varepsilon_d$ for different values of
$\Gamma_n$ at $\phi=\pi$.  We can see the pronounced drop of the current which corresponds to the cooling regime as one can note from the inset where the ground-state probability increase occurs in this case. This maximum-minimum structure disappears in the heating regime ($P_0\lesssim 0.5$). Also, one can see the maximum of the current in the case of $2\varepsilon_d=\Gamma_n$ at $\Gamma_n\to 0$. It is a resonant peak and has therefore a different from cooling nature and is discussed, for example, in Ref.~\cite{Stadler2016} for a hybrid N-QD-S device. Thus, the above-mentioned facts can
serve as a criterion that the system is in the cooling regime.

\section*{Conclusions}
\addcontentsline{toc}{section}{Conclusions}

In this chapter the nanomechanical weak link that involves a carbon nanotube suspended between two normal leads and biased by a constant voltage is considered. The nanotube, which is treated as a single-level quantum dot, performs bending vibrations in a gap between two superconducting electrodes. The coupling between the electronic and mechanical degrees of freedom is induced due to the superconducting proximity effect which exhibits in the appearance of the position-dependent dot order parameter. 

On the one hand, in the first section we demonstrate that in such a system, the static, straight configuration
of the nanotube is \textit{unstable} regarding the occurrence of
self-sustained bending vibrations in a wide range of parameters if a
bias voltage is applied between the normal and superconducting leads. It is shown that the occurrence of this
mechanical instability crucially depends on the direction of the bias voltage
and the relative position of the QD level. We have also shown that the appearance of
self-sustained mechanical vibrations strongly affects the dc current
through the system, leading to transistor and diode effects. The
latter can be used for the direct experimental observation of the
predicted phenomena.

On the other hand, in the second section, using the density matrix approximation, we find that at certain direction of the applied bias voltage, the stationary state of the mechanical subsystem has a Boltzmann form. Moreover, the probability to find the system in the ground state has been demonstrated to be $P_0\lesssim 1$. The latter is related to the cooling regime of the considered system. Additionally, the probability depends on the superconducting phase difference and the relative position of the QD energy level in a key manner. Also, we have discussed that the direct electric current behaviour mirrors the stationary state of the system. 

Thus, the clear possibility to
govern the operating mode of the device by changing the bias and
gate voltages is demonstrated and the schemes for an experimental detection of the predicted effects are proposed.

The main results of this chapter are published in Refs.~\cite{BahrovaC2021b, Bahrova2022, Bahrova2022b, BahrovaC2022a}.

\clearpage            
\chapter*{CONCLUSIONS}						
\addcontentsline{toc}{chapter}{CONCLUSIONS}	

In the works this dissertation based on, phenomena which emerge due to an electromechanical coupling in nano-structures
based on a movable quantum dot, are investigated.

The main results are the following:

\begin{enumerate}
    \item The electron transport through an single-molecule transistor with coherent vibrons was theoretically investigated. It has been shown that current-voltage characteristics of such a nanoscale transistor are step-like functions of the bias voltage similarly to the polaronic ones. However, the lifting of this coherent-oscillation induced blockade occurs at voltages much lower than the ones predicted within Franck-Condon theory.
    \item The possibility to generate quantum entanglement between charge qubit states and mechanical coherent ones in a nanoelectromechanical system was shown. The experimentally simple protocol of the bias voltage manipulation, which results in the formation of entangled states incorporating "cat states", was proposed. 
    \item The non-trivial behavior of a hybrid nanoelectromechanical device, which emerges due to a fundamentally new type of electromechanical coupling based on the quantum delocalization of Cooper pairs, has been theoretically investigated. The range of existence of the mechanical instability in such a system was theoretically found. Moreover, the instability results in the generation of self-sustained mechanical oscillations under the self-saturation effect.
    \item The regime when the stationary state of the mechanical subsystem of the hybrid nanoelectomechanical system has a Boltzmann form determined by parameters of the device, was investigated. In this case the probability to find the system in the ground state was shown to be sufficiently large which corresponds to the ground-state cooling effect in the system.
    \item It has been theoretically demonstrated that the mechanical vibrations in the hybrid nanoelectromechanical device strongly affect the direct electric current through the one. It allows to probe experimentally the presence and characteristics of the predicted self-sustained oscillations as well as the cooling effect.
\end{enumerate}

\chapter*{\text{Acknowledgements}}

The candidate deeply acknowledges the very best supervision of Sergei I. Kulinich and \fbox{Ilya V. Krive}.
Also, the candidate greatly acknowledges all the support and advisement of Leonid Y. Gorelik.

Moreover, the candidate's warm thanks are to Sergey N. Shevchenko for the all-round support.

Nevertheless, the candidate says a hundred of thanks to O.A.~Ilinskaya, A.V.~Parafilo, A.D.~Shkop, H.C.~Park and R.I.~Shekhter for the helpful discussions and collaboration together with the useful advises.

The candidate sincerely thanks Sergei V. Koniakhin for the all-round help, too.


In addition, the candidate appreciates the support of the Department of Theoretical Physics headed by Dr.~Victor V. Slavin, in particular, special candidate's thanks go to A.N. Kalinenko, of B.~Verkin Institute for Low Temperature Physics and Engineering of the NAS of Ukraine (Kharkiv, Ukraine) as well as the Center for Theoretical Physics of Complex Systems headed by Prof.~Sergej Flach, of the Institute of Basic Sciences (Daejeon, Republic of Korea).








\clearpage                                  
\phantomsection
\addcontentsline{toc}{chapter}{\bibname}	
\urlstyle{rm}    

\nocite{Baas2004,Amthor2015,Avriller2018,Belzig2003,Bimberg1999,Blais2020,Cassell1999,Cleland2003,Cottet2004,Davies1997,Deng2016,Gorelik2005,Grassl2009,Gu2019,Harrison2005,Hartle2018,Hayakawa2017,Holmqvist2018,Hwang2009,Imry1997,Koch2006,Krantz2019,Krause2015,Laird2015,Liu2021,Micchi2015,Moghaddam2012,Nielsen2002,Parafilo2015,Park2013,Perrin2015,Peskin2019,Popov2017,Rastelli2019,Sapmaz2006,Scheible2004,Shekhter2003,Stadler2015,Houten1996,Delft2001,Wang2016,Zhang2020,Svidzinskii}


\printbibliography


\appendix

\renewcommand*\chaptertitlename{\appendixname~}
\addtocontents{toc}{\def\protect\cftchappresnum{\appendixname{} }%
\setlength{\cftchapnumwidth}{\widthof{\cftchapfont\appendixname~\quad\cftchapaftersnum}}
}





\chapter{\authorbibtitle} \label{AppendixA}

\begin{enumerate}
    \item \textbf{O.M. Bahrova}, S.I. Kulinich, I.V. Krive, Polaronic effects induced by non-equilibrium vibrons in a single-molecule transistor, \textit{Low Temp. Phys.}~\textbf{46}, No. 7, 671, (2020) [\textit{Fiz. Nizk. Temp.}, \textbf{46}, 799 (2020)],  DOI: 10.1063/10.0001362
    \item \textbf{O.M. Bahrova}, L.Y. Gorelik, S.I. Kulinich, Entanglement between charge qubit states and coherent states of nanomechanical resonator generated by ac Josephson effect, \textit{Low Temp. Phys.,} \textbf{47}, No. 4, 287, (2021) [\textit{Fiz. Nizk. Temp.}, \textbf{47}, 315 (2021)], DOI: 10.1063/10.0003739
    \item \textbf{O.M. Bahrova}, L.Y. Gorelik, S.I. Kulinich, R.I. Shekhter, H.C. Park, Nanomechanics driven by the superconducting proximity effect, \textit{New J. Phys.}, \textbf{24}, 033008 (2022), DOI:  10.1088/1367-2630/ac5758
    \item \textbf{O.M. Bahrova}, L.Y. Gorelik, S.I. Kulinich, R.I. Shekhter, H.C. Park, Cooling of nanomechanical vibrations by Andreev injection, \textit{Low Temp. Phys.}, \textbf{48}, No. 6, 476 (2022) [\textit{Fiz. Nizk. Temp.}, \textbf{48}, 535 (2022)], DOI: 10.1063/10.0010443
    %
    %
    %
    \item \textbf{O.M. Bahrova}, I.V. Krive, How to control transport of spin-polarized electrons via magnetic field in a molecular transistor, Physics and Scientific\&Technological progress: student scientific conference, p.3, (2018).
    \item \textbf{O. M. Bahrova}, S. I. Kulinich, I. V. Krive, Polaronic effects induced by coherent vibrons in a single-molecule transistor, I International Advanced Study Conference Condensed matter \& Low Temperature Physics, June 8-14, 2020, Ukraine, Kharkiv, Abstracts, p. 183, (2020).
    \item A.D. Shkop, \textbf{O.M. Bahrova}, Coulomb and vibration effects in spin-polarized current through a single-molecule transistor, XI Conference of Young Scientists “Problems of Theoretical Physics”, December 21-23, 2020, Ukraine, Kyiv, Abstracts, p.15-16, (2020).
    \item \textbf{O.M. Bahrova}, L.Y. Gorelik, S.I. Kulinich, Schrödinger-cat states generation via mechanical vibrations entangled with a charge qubit, II International Advanced Study Conference Condensed matter \& Low Temperature Physics, June 6–12, 2021, Ukraine, Kharkiv, Abstracts, p.201, (2021).
    \item \textbf{O.M. Bahrova}, L.Y. Gorelik, S.I. Kulinich, H.C. Park, R.I. Shekhter, Self-sustained mechanical oscillations promoted by superconducting proximity effect, The International Symposium on Novel maTerials and quantum Technologies, December 14–17, 2021, Abstracts, p.134, (2021).
    %
    %
    \item \textbf{O.M. Bahrova}, L.Y. Gorelik, S.I. Kulinich, H.C. Park, R.I. Shekhter, Nanomechanics provoked by Andreev injection, 29th International Conference on Low Temperature Physics, August 18-24, 2022, Abstracts, p.1554 {\&} 1771, (2022).
\end{enumerate}

\chapter{Information on the approbation of the dissertation results} \label{AppendixB}

\renewcommand{\labelitemi}{$\bullet$}
\begin{itemize}
    \item Physics and Scientific\&Technological progress: student scientific conference (Kharkiv, Ukraine, April 10-12, 2018);
    \item I International Advanced Study Conference Condensed matter \& Low Temperature Physics (Kharkiv, Ukraine, June 8-14, 2020);
    \item XI Conference of Young Scientists “Problems of Theoretical Physics” (Kyiv, Ukraine (online), December 21-23, 2020);
    \item II International Advanced Study Conference Condensed matter \& Low Temperature Physics (Kharkiv, Ukraine, June 6–12, 2021);
    \item The International Symposium on Novel maTerials and quantum Technologies, (Kanagawa, Japan (online), December 14–17, 2021);
    \item Quantum Thermodynamics Conference 2022, (Belfast, United Kingdom (online), June 27-July 1, 2022);
    \item 29th International Conference on Low Temperature Physics, (Sapporo, Japan (online), August 18-24, 2022);
    
\end{itemize}

\end{document}